\documentclass{article}
\usepackage{amssymb}
\usepackage{amsfonts}
\usepackage{amsmath}

\input{tcilatex}
\begin{document}

\begin{center}
\bigskip

{\large PSEUDO-CONFORMAL FIELD\ THEORY}

{\large \ }\bigskip

C. N. Ragiadakos*

\bigskip

\bigskip

\bigskip

\bigskip

\textbf{ABSTRACT}
\end{center}

\begin{quote}
The fundamental structure of the 4-dimensional spacetime is assumed to be
the lorentzian CR-structure (LCR-structure), which contains two correlated
3-dimensional CR-structures. It is defined by explicit Frobenius integrable
relations characterized by "left" and "right" CP(3) points. This
LCR-structure is invariant under a very restrictive tetrad-Weyl symmetry,
which permits a unique special second order partial differential equation
applied to a Yang-Mills field, identified with the gluon field. A class of
metrics with the corresponding self-dual forms are defined. After partially
fixing the tetrad-Weyl symmetry, the electroweak connection is also defined,
which is directly related with the class of LCR-tetrads of the structure.
The "free electron" LCR-structure is identified, which has gravitational and
electroweak potentials (dressings) with fermionic gyromagnetic ratio g=2.
The corresponding massless "neutrino" LCR-structure is also found. These two
solitonic configurations constitute the first leptonic generation identified
with the Petrov type D LCR-structure. The muon and tau leptonic generations
are identified with the Petrov type II and I respectively. Using the
electron LCR-structure, I compute the corresponding quark having an
additional gluonic potential and providing the explication of the
lepton-quark correspondence. The standard model is implied via the causal
Bogoliubov, Epstein-Glaser and Scharf procedure viewed as a targeted
harmonic analysis in the rigged Hilbert-Fock space of precise Poincare
representations of the computed geometric structures.
\end{quote}

\bigskip

\newpage

\bigskip

\bigskip

\bigskip

{\LARGE Contents}

\textbf{1. INTRODUCTION}

\textbf{\qquad 1.1 A brief review of BEGS procedure}

\textbf{2. LORENTZIAN CR-STRUCTURE}

\textbf{\qquad 2.1 Three-dimensional CR-structure}

\textbf{\qquad 2.2 Definition of lorentzian CR-structure}

\textbf{\qquad 2.3 A G(4,2)\ embedding of the LCR-structure}

\textbf{\qquad 2.4 LCR-structures determined with complex trajectories}

\textbf{3. EMERGENCE OF EINSTEIN'S METRIC}

\textbf{4. LEPTONIC LCR-STRUCTURES}

\textbf{\qquad 4.1 Free electron (and positron) LCR-structure}

\textbf{\qquad 4.2 Solving the electron naked singularity problem}

\textbf{\qquad 4.3 Free neutrino LCR-structure}

\textbf{\qquad 4.4 The muon and tau generations}

\textbf{5. EMERGENCE OF ELECTROWEAK\ FIELD}

\textbf{6. GLUONIC FIELD\ AND\ QUARKS}

\textbf{\qquad 6.1 Emergence of quarks}

\textbf{\qquad 6.2 The confinement problem}

\textbf{\qquad 6.3 The gluonic dressing of the "natural U(2)" LCR-structure}

\textbf{7. STANDARD\ MODEL\ DERIVATION}

\textbf{\qquad 7.1 From SU(2,2) down to Poincar\'{e} rigged Hilbert space}

\textbf{\qquad 7.2 BEGS formulation of the standard model}

\textbf{\qquad 7.3 On self consistency conditions}

\textbf{8. DISCUSSION}

\bigskip

\newpage

\renewcommand{\theequation}{\arabic{section}.\arabic{equation}}

\section{INTRODUCTION}

\setcounter{equation}{0}

The standard model (SM) actually provides an experimentally well established
description of elementary particles, except the graviton. The discovery of
the Higgs particle made the SM renormalizable and hence well defined in the
context of quantum field theory (QFT). But the recent experimental results
of ATLAS and CMS from the LHC of CERN show that the minimal supersymmetric
SM does not describe nature. No supersymmetric particles have been observed.
On the other hand highly sensitive experiments did not find any sign of
weakly interacting massive particles. Hence string theory has to be
abandoned, and elementary particle physics has to look for other quantum
models, which do not need supersymmetry to describe nature. The present work
provides such an alternative by simply replacing Einstein's metric with a
special totally real (Cauchy-Riemann) CR-structure in the tangent space of a
differential manifold. The first investigators (E. Cartan, Tanaka, Severi
etc) of CR-structures used the term "pseudo-conformal transformations"\cite%
{Cartan}, therefore the present model is called pseudo-conformal field
theory (PCFT) and its fundamental CR-structure is called lorentzian
Cauchy-Riemann (LCR) structure. In fact it generalizes the metric
independence of the 2-dimensional Polyakov action into a four dimensional
highly symmetric model which derives all the particles of the SM. The
present work systematically describes the emergence of the leptonic and
hadronic sector with the amazing lepton-quark correspondence. These Poincar%
\'{e} representations (free fields) are the basis of the Bogoliubov\cite%
{BoLoTo}-Epstein-Glaser\cite{EpGl} perturbative causal approach\cite{Sch1}
as reformulated by Scharf\cite{Sch2} into a self-consistent (BEGS)
procedure, which is briefly reviewed in the following subsection.

The metric independence is usually related with topological field theories,
which are essentially dynamically "empty". The 2-dimensional generally
covariant Polyakov action 
\begin{equation}
\begin{array}{l}
I_{S}=\frac{1}{2}\int d^{2}\!\xi \ \sqrt{-\gamma }\ \gamma ^{\alpha \beta }\
\partial _{\alpha }X^{\mu }\partial _{\beta }X^{\nu }\eta _{\mu \nu } \\ 
\end{array}
\label{i1}
\end{equation}%
is also metric independent, but it is very rich, because it is not
topological. In the light-cone coordinates $(\xi _{-},\ \xi _{+})$, the
action takes the metric independent form 
\begin{equation}
\begin{array}{l}
I_{S}=\int d^{2}\!z\ \partial _{-}X^{\mu }\partial _{+}X^{\nu }\eta _{\mu
\nu } \\ 
\end{array}
\label{i2}
\end{equation}%
The light-cone coordinates $(\xi _{-},\ \xi _{+})$ may be viewed as the real
part of the structure coordinates $(z^{0},\ z^{\widetilde{0}})$ of the
following trivial totally real CR-manifold of $%
\mathbb{C}
^{2}$%
\begin{equation}
\begin{array}{l}
\rho _{11}(\overline{z^{0}},z^{0})=0\quad ,\quad \rho _{22}(\overline{z^{%
\widetilde{0}}},z^{\widetilde{0}})=0 \\ 
\end{array}
\label{i3}
\end{equation}%
because its corresponding normal form\cite{BaEbRo} is%
\begin{equation}
\begin{array}{l}
\func{Im}z^{0}=0\quad ,\quad \func{Im}z^{\widetilde{0}}=0 \\ 
\end{array}
\label{i4}
\end{equation}

The CR-structures\cite{Jaco} are metric independent. In general relativity,
the geodetic and shear free condition of a null congruence coincides\cite%
{Traut}\cite{ANK} with the CR-structure of conventional mathematics. In
order to make the present work self-consistent, a very brief review of the
3-dimensional CR-structure will be presented in section II, where the
4-dimensional LCR-structure will be defined\cite{Rag} without using the
notion of the metric. It essentially consists of two 3-dimensional
CR-substructures with a common complex tangent subspace. The emergence of
Einstein's gravity is described in section III and the emergence of
electroweak gauge field is described in section V. The solitonic
LCR-structures of the free electron and its neutrino are explicitly computed
in section IV.

\subsection{A brief review of BEGS procedure}

After the Wightman axioms it became clear that the fundamental structures of
QFT are the Poincar\'{e} group and the Schwartz distributions\cite{Schwa}.
The later are functionals applied over vector spaces of functions with
infinite number of successive finer norms (nuclear spaces) called test
functions. The useful kinds of test functions on $%
\mathbb{R}
^{4}$ are: 1) Smooth functions $\phi (x)$ with compact support and infinite
number of derivatives ($\in C_{c}^{\infty }$) denoted with $\mathcal{D}$.
The corresponding functionals $f(\phi )\in 
\mathbb{C}
$ are denoted $\mathcal{D}^{\prime }$. 2) Smooth functions $\phi (x)$
decreasing faster than any polynomial of $x$ denoted with $\mathcal{S}$
usually called Schwartz space. The corresponding functionals $f(\phi )\in 
\mathbb{C}
$ are denoted $\mathcal{S}^{\prime }$. 3) Real analytic functions $\phi (x)$
denoted with $\mathcal{A}$. The corresponding functionals $f(\phi )\in 
\mathbb{C}
$ are denoted $\mathcal{A}^{\prime }$ usually called Sato's distributions.
4) General smooth functions $\phi (x)\in C^{\infty }$ denoted with $\mathcal{%
E}$. The corresponding functionals $f(\phi )\in 
\mathbb{C}
$ are denoted $\mathcal{E}^{\prime }$ usually called distributions with
compact support. Apparently the following set relations 
\begin{equation}
\begin{array}{l}
\mathcal{D\subset S\subset A\subset E} \\ 
\multicolumn{1}{c}{\Downarrow} \\ 
\mathcal{E}^{\prime }\mathcal{\subset A}^{\prime }\mathcal{\subset S}%
^{\prime }\mathcal{\subset D}^{\prime } \\ 
\end{array}
\label{i5}
\end{equation}%
are valid. The Schwartz distributions permit the definition of rigged
Hilbert space (Gelfand triple)\cite{BoLoTo} as the completion of the
test-function normed nuclear spaces. The Gelfand-Maurin theorem permits the
proper definition of the self-adjoint and unitary operators with their
combined discrete and continues eigenvalues. The conventional QFT is based
on the rigged Hilbert space 
\begin{equation}
\begin{array}{l}
\mathcal{S}(%
\mathbb{R}
^{4})\subset L^{2}(%
\mathbb{R}
^{4})\subset \mathcal{S}^{\prime }(%
\mathbb{R}
^{4}) \\ 
\end{array}
\label{i6}
\end{equation}%
where the space $\mathcal{S}(%
\mathbb{R}
^{4})$ is dense in $L^{2}(%
\mathbb{R}
^{4})$ of square integrable functions. In section VI we will find that the
quark gluonic dressing has line singularities implying that the physical
hadrons are colorless distributions with compact support i.e. they belong to 
$\mathcal{E}^{\prime }(%
\mathbb{R}
^{4})$, which is interpreted as confinement. The Schwartz test functions $%
\mathcal{S}$ and the corresponding distributions $S^{\prime }$ decay as \ 
\begin{equation}
\begin{array}{l}
|\varphi (x)|\leq C_{N}|x|^{-N}\quad as\quad |x|\rightarrow \infty \\ 
\\ 
|f(x)|\leq C_{N}|x|^{N}\quad as\quad |x|\rightarrow \infty%
\end{array}
\label{i7}
\end{equation}%
This decaying difference essentially distinguishes the wavefront
singularities\cite{Horm} which characterize the representatives of the
(proper) distributions. The above conditions may be replaced with the
condition that the distributions are represented with functions which are
derivatives of locally integrable "potentials" with singularities. The
typical example is the potential ($\frac{1}{|\overrightarrow{x}|}$) of a
point charge, which is locally integrable but its grad, the corresponding
field strength ($\frac{1}{|\overrightarrow{x}|^{2}}$), is not locally
integrable.

The free fields are operator valued distributions $\mathcal{S}^{\prime }$
which belong to precise representations of the Poincar\'{e} group. The
Bogoliubov expansion of the S-matrix into the coupling $c(x)$ constant
considered as a test function is 
\begin{equation}
\begin{array}{l}
S=1+\underset{n\geq 1}{\sum }\frac{1}{n!}\int
S_{n}(x_{1},x_{2}...x_{n})c(x_{1})c(x_{2})...c(x_{n})[dx] \\ 
\\ 
S_{n}(x_{1},x_{2}...x_{n})=T\{\mathcal{L}_{I}(x_{1})\mathcal{L}_{I}(x_{2})...%
\mathcal{L}_{I}(x_{n})\}%
\end{array}
\label{i8}
\end{equation}%
where $T$ denotes the time-ordering of the interaction lagrangian $\mathcal{L%
}_{I}(x)$, which depends on the \emph{free field} operators. It has the
normal product form of operator valued distributions. In conventional QFT
time-ordering is imposed using the step distribution, which causes
mathematical problems, because its multiplication with the distribution $%
\mathcal{L}_{I}(x)$ is not permitted, implying\cite{Sch1} the appearance of
the well known infinities. Epstein and Glaser\cite{EpGl} found a procedure
to bypass these infinities. They multiply with a $C^{\infty }$ regulator and
after separate the advanced from the retarded parts. After, they find the
degree of the divergence at $x=0$. These divergencies are subtracted. Grang%
\'{e} et.al.\cite{Grange} use a more effective method introducing test
functions as partitions of unity.

Up to now the procedure is equivalent to the ordinary renormalization
procedure. The final essential step was the use by Scharf and his
collaborators\cite{Sch2} a variation of the BRS transformation procedure to
establish order by order the complete interaction lagrangian without the
unphysical modes of the free fields. Considering the free fields as operator
valued distributions, they make the infinitesimal gauge transformation%
\begin{equation}
\begin{array}{l}
A^{\prime \mu }(x)=A^{\mu }(x)+\lambda \partial ^{\mu }u(x)+O(\lambda ^{2})
\\ 
\end{array}
\label{i9}
\end{equation}%
as an operator transformation 
\begin{equation}
\begin{array}{l}
A^{\prime \mu }(x)=e^{-i\lambda Q}A^{\mu }(x)e^{i\lambda Q}\simeq A^{\mu
}(x)-i\lambda \lbrack Q,A^{\mu }(x)]+O(\lambda ^{2}) \\ 
\multicolumn{1}{c}{\Downarrow} \\ 
\lbrack Q,A^{\mu }(x)]=i\partial ^{\mu }u(x)%
\end{array}
\label{i10}
\end{equation}%
implying the commutator of the charge $Q$ with the field operator $A^{\mu
}(x)$. It determines $Q$ up to a complex number 
\begin{equation}
\begin{array}{l}
Q=\tint\limits_{x^{0}=c}d^{3}x(\partial _{\mu }A^{\mu }\partial
_{0}u-(\partial _{0}\partial _{\mu }A^{\mu })u) \\ 
\end{array}
\label{i11}
\end{equation}%
The gauge charge is nilpotent $Q^{2}=\{Q,Q\}=0$ and the field $u(x)$ is a
ghost massless "fermionic" field. Its partner is denoted $\widetilde{u}(x)$.
The purpose of these ghost fields is to eliminate\cite{Sch2} the unphysical
longitudinal and scalar components determined by the gauge condition $%
\partial _{\mu }A^{\mu }=0$. That is, a state $\Phi $ of the Fock space is
physical if 
\begin{equation}
\begin{array}{l}
\{Q^{\dag },Q\}\Phi =0 \\ 
\end{array}
\label{i12}
\end{equation}%
This procedure was generalized for the non-abelian groups and applied\cite%
{SchAsDu} to the derivation of the SM of elementary particles. It
demystified the spontaneous symmetry breaking and provided an understanding
of the "naturalness problems" and the neutrino and quark mixing problems.
The application of the algorithm to the coordinate invariance of the spin-2
field $h^{\mu \nu }$ implies the proper Einstein-Hilbert action with a
cosmological constant, which we should expect, because it is well known that
higher order derivatives generate negative norm (unphysical) states\cite%
{Pais}. Hence the entire BEGS procedure provides a well defined
("renormalizable") quantum general relativity.

Concluding this brief review I point out that BEGS procedure essentially
separates QFT into two parts. The first part where the experimentally
observed particles and symmetries (gravity, electroweak and gluon gauge
fields) are assumed and the second part the mathematical harmonic analysis
properly done in the rigged Hilbert-Fock space of the Poincar\'{e}
representations. The assumed symmetries and particles of the first part are
implied by PCFT. The second pure mathematical part is implied by the proper
treatment of the Schwartz distributions and the rigged Hilbert space
(Gelfand triple) described by the BEGS procedure as an operational harmonic
analysis in the asymptotic Poincar\'{e} representations found in PCFT. This
interplay of PCFT and the BEGS procedure provides the known SM of elementary
particles as described in section VI.

\section{THE\ LORENTZIAN\ CR-STRUCTURE}

\setcounter{equation}{0}

In the complex plain $%
\mathbb{C}
$, the Riemann mapping theorem states that two lines (real hypersurfaces)
are holomorphically equivalent. Poincar\'{e} has showed that it is not valid
in higher dimensions. That is, the real surfaces $\rho \left( \overline{%
z^{\alpha }},z^{\alpha }\right) =0$ (for any real function $\rho $) on the
complex plane $%
\mathbb{C}
^{n}\ ,n>1$ cannot be transformed to each other with an \emph{holomorphic}
transformation. These different structures of the real surfaces are called
Cauchy-Riemann structures (CR-structures) and the manifolds endowed with
such structures are called CR-manifolds. Surfaces of $%
\mathbb{C}
^{n}$ determined by one \emph{real} function are called CR-manifolds of the
hypersurface type and those determined by $n$ independent real functions are
called totally real CR-manifolds. In the following subsections I review the
3-dimensional CR-manifolds of the hypersurface type and a special form of
4-dimensional totally real CR-manifolds, which I call lorentzian
CR-manifolds (LCR-manifolds). These two kinds of CR-structures are
sufficient for the reader to understand the fundamental geometric structure
of pseudo-conformal field theory (PCFT).

Throughout this mathematical review, the reader should notice that the
notion of the CR-structure does not need the notion of the metric to be
defined. CR-structure is not a notion of the riemannian geometry, while in
general relativity the CR-structure emerged\cite{Traut} through the notion
of a geodetic and shear free null congruence.

\subsection{Three-dimensional CR-structures}

The simple three-dimensional CR-manifold $M$ is determined by the
annihilation of a real function $\rho \left( \overline{z^{\alpha }}%
,z^{\alpha }\right) =0$ in $%
\mathbb{C}
^{2}$. The condition $d\rho \left( \overline{z^{\alpha }},z^{\alpha }\right)
|_{M}=0$ implies that this surface admits the following cotangent real
1-form \ 
\begin{equation}
\begin{array}{l}
\omega _{0}=2i(\partial \rho )|_{M}=i((\partial -\overline{\partial })\rho
)|_{M} \\ 
\end{array}
\label{c1}
\end{equation}%
If we use the regular coordinates ($u,\zeta ,\overline{\zeta }$)\cite{BaEbRo}%
, which provide the following graph form for the surface \ 
\begin{equation}
\begin{array}{l}
z^{0}=w=u+iU\quad ,\quad z^{1}=\zeta \\ 
\\ 
U=\frac{z^{0}-\overline{z^{0}}}{2i}=\phi \left( u,\zeta ,\overline{\zeta }%
\right) \quad ,\quad \phi (0)=0\quad ,\quad d\phi (0)=0%
\end{array}
\label{c2}
\end{equation}%
and the real surface takes the simple form \ 
\begin{equation}
\begin{array}{l}
\rho \left( \overline{z^{\alpha }},z^{\alpha }\right) =\frac{z^{0}-\overline{%
z^{0}}}{2i}-\phi (\frac{z^{0}+\overline{z^{0}}}{2},z^{1},\overline{z^{1}})
\\ 
\\ 
\phi (0)=0\quad ,\quad d\phi (0)=0%
\end{array}
\label{c2a}
\end{equation}%
We find the following basis of the cotangent space \ 
\begin{equation}
\begin{array}{l}
\omega _{0}=du-i(\frac{\partial \phi }{\partial z^{1}})dz^{1}+i(\frac{%
\partial \phi }{\partial \overline{z^{1}}})d\overline{z^{1}} \\ 
\\ 
\omega _{1}=dz^{1}\quad ,\quad \overline{\omega }_{1}:=\overline{\omega _{1}}%
=d\overline{z^{1}}%
\end{array}
\label{c3}
\end{equation}%
which admit arbitrary multiplication factors. The dual basis of the tangent
space\ is \ 
\begin{equation}
\begin{array}{l}
k_{0}=\frac{\partial }{\partial u} \\ 
\\ 
k_{1}=\frac{\partial }{\partial z^{1}}+2i\frac{\partial _{z^{1}}U}{%
1-i\partial _{u}U}\frac{\partial }{\partial u} \\ 
\\ 
\overline{k}_{1}:=\overline{k_{1}}%
\end{array}
\label{c4}
\end{equation}%
which is normalized by the (interior product $\lrcorner $) relations $%
k_{0}\lrcorner \omega _{0}=1$ and $k_{1}\lrcorner \omega _{1}=1$ and the
other interior products vanish. Do not confuse the interior product with the
inner product defined by a metric tensor!

If the defining function $\rho \left( \overline{z^{\alpha }},z^{\alpha
}\right) $ is real analytic, the surface is called real analytic. This
surface is \emph{diffeomorphically} equivalent to the "flat" hyperplane $U=0$%
. But there is not always a \emph{holomorphic} transformation which performs
this transformation.

In its abstract definition\cite{Jaco} a 3-dimensional CR-manifold is a
(real) differentiable manifold which admits a complex field $%
k_{1}=k_{1}^{\beta }\partial _{\beta }$ in the tangent space $T^{\ast }(M)$,
which is linearly independent to its complex conjugate. The CR-structure is
invariant under the transformation 
\begin{equation}
\begin{array}{l}
k_{1}^{\prime }=bk_{1} \\ 
\\ 
with\ b\neq 0,\infty \ complex\ function%
\end{array}
\label{c5}
\end{equation}%
The CR-structure may be equivalently defined by a real covector field $%
\omega _{0}=\omega _{0\alpha }d\xi ^{\alpha }$ and a complex covector $%
\omega _{1}=$ $\omega _{1\alpha }d\xi ^{\alpha }$ of the cotangent space $%
T_{\ast }(M)$ such that 
\begin{equation}
\begin{array}{l}
k_{1}\lrcorner \omega _{0}=0\quad ,\quad k_{1}\lrcorner \omega _{1}=1\quad
,\quad k_{1}\lrcorner \overline{\omega _{1}}=0 \\ 
\\ 
\omega _{0}\wedge \omega _{1}\wedge \overline{\omega _{1}}\neq 0%
\end{array}
\label{c6}
\end{equation}%
The corresponding equivalence transformation in the cotangent space is 
\begin{equation}
\begin{array}{l}
\omega _{0}^{\prime }=a\omega _{0}\quad ,\quad \omega _{1}^{\prime }=\frac{1%
}{b}\omega _{1}+c\omega _{0} \\ 
\\ 
a\neq 0\ real,\ and\ b\neq 0,\infty \ and\ c\ complex\ functions%
\end{array}
\label{c7}
\end{equation}

The CR-structures have some local invariants (called relative invariants),
which take the discrete values $0$ or $1$. The first relative invariant
emerges from the observation that under a general CR-structure preserving
transformation (\ref{c7}) the relation 
\begin{equation}
\begin{array}{l}
d\omega _{0}=iA\omega _{1}\wedge \overline{\omega }_{1}+B\omega _{0}\wedge
\omega _{1}+\overline{B}\omega _{0}\wedge \overline{\omega }_{1} \\ 
\end{array}
\label{c8}
\end{equation}%
takes the form 
\begin{equation}
\begin{array}{l}
d\omega _{0}^{\prime }=iAab\overline{b}\omega _{1}^{\prime }\wedge \overline{%
\omega }_{1}^{\prime }\func{mod}[\omega _{0}^{\prime }] \\ 
\end{array}
\label{c9}
\end{equation}%
Hence, if at a point $y$ the function does not vanish $A(y)\neq 0$, it will
not vanish for any other set of representative 1-forms of the CR-structure,
which is then called non-degenerate at $y$. On the other hand if the
CR-structure at the point $y$ has $A(y)=0$, it will vanish for any other set
of representative 1-forms of the CR-structure and it is called degenerate at 
$y$. Notice that the non-vanishing condition of the CR-structure defining
1-forms [$\omega _{0}\ ,\ \omega _{1}\ ,\ \overline{\omega _{1}}$] is
related to its degeneracy. Vanishing points of $A(x)$ may be interpreted as
a different CR-manifold, because the transformation 
\begin{equation}
\begin{array}{l}
\omega _{0}^{\prime }=\frac{1}{A}\omega _{0}\quad ,\quad \omega ^{1\prime
}=\omega ^{1} \\ 
\end{array}
\label{c10}
\end{equation}%
at any point of a neighborhood of the zeros of $A(x)$ removes the zeros, but
it makes $\omega _{0}^{\prime }$ not well defined at $x=y$. This is the
reason that the transformation (\ref{c8}) must respect the condition $A(x)$
related to a coordinate atlas of the CR-manifold. We can generally define
non-degenerate CR-manifolds with non-vanishing coefficient $A(x)$, $\forall
x\in M$. A degenerate CR-manifold has $A(x)=0\ ,\ \forall x\in M$. Hence if
in the form (\ref{c8}) $\frac{\partial ^{2}U}{\partial \zeta \partial 
\overline{\zeta }}|_{u=0}=0\ ,\ \forall \ \zeta $ ,\ the CR-manifold is
degenerate. A degenerate CR-structure is equivalent to the trivial one 
\begin{equation}
\begin{array}{l}
\omega _{0}=du\quad ,\quad \omega ^{1}=d\zeta \quad ,\quad \overline{\omega }%
^{1}=d\overline{\zeta } \\ 
\end{array}
\label{c11}
\end{equation}%
A non-degenerate CR-structure (also called pseudoconvex) on a smooth
manifold can always take the form $d\omega _{0}=i\omega _{1}\wedge \overline{%
\omega _{1}}(\func{mod}\omega _{0})$, which is invariant under the
transformation \ 
\begin{equation}
\begin{array}{l}
\omega _{0}^{\prime }=|\lambda |^{2}\omega _{0}\quad ,\quad \omega
_{1}^{\prime }=\lambda (\omega _{0}+\mu \omega _{1}) \\ 
\end{array}
\label{c12}
\end{equation}%
with $\lambda (x)\neq 0$ and $\mu (x)$ arbitrary complex functions.

Moser used the holomorphic transformations to restrict the real function $%
\phi \left( u,\zeta ,\overline{\zeta }\right) $ to the following form up to
special linear transformations. \ 
\begin{equation}
\begin{array}{l}
\phi =\frac{1}{2}\zeta \overline{\zeta }+\tsum\limits_{k\geq 2,j\geq
2}N_{jk}(u)\zeta ^{j}\overline{\zeta }^{k} \\ 
\\ 
N_{22}=N_{32}=N_{33}=0%
\end{array}
\label{c13}
\end{equation}%
The functions $N_{jk}(u)$ characterize the CR-structure. By their
construction these functions belong into representations of the isotropy
subgroup of $SU(1,2)$ symmetry group of the hyperquadric.

The classical domains are usually described as regions\cite{Pyat} of
projective spaces. The $SU(1,2)$ symmetric classical domain is the region of 
$CP^{2}$\ determined by the relation \ 
\begin{equation}
\begin{array}{l}
\overline{Z^{m}}C_{mn}Z^{n}>0 \\ 
\end{array}
\label{c14}
\end{equation}%
where $Z^{n}$ are the homogeneous coordinates of $CP^{2}$\ and $C_{mn}$ are $%
SU(1,2)$ symmetric matrices. The matrix \ 
\begin{equation}
\begin{array}{l}
C_{B}=%
\begin{pmatrix}
1 & 0 & 0 \\ 
0 & -1 & 0 \\ 
0 & 0 & -1%
\end{pmatrix}
\\ 
\end{array}
\label{c15}
\end{equation}%
gives the bounded realization of the classical domain and the matrix \ 
\begin{equation}
\begin{array}{l}
C_{S}=%
\begin{pmatrix}
0 & 0 & -i \\ 
0 & -1 & 0 \\ 
i & 0 & 0%
\end{pmatrix}
\\ 
\end{array}
\label{c16}
\end{equation}%
gives the unbounded realization, which is also called Siegel domain. The
boundary of the classical domain is a $SU(1,2)$ symmetric real submanifold
of $CP^{2}$.\ In the bounded realization it takes the form of $S^{3}$ and in
the unbounded realization it takes the form of the hyperquadric. The unitary
transformation of the hermitian matrices and $C_{S}=U^{\dag }C_{B}U$ with \ 
\begin{equation}
\begin{array}{l}
U=%
\begin{pmatrix}
\frac{1}{\sqrt{2}} & 0 & \frac{-i}{\sqrt{2}} \\ 
0 & -1 & 0 \\ 
\frac{-i}{\sqrt{2}} & 0 & \frac{1}{\sqrt{2}}%
\end{pmatrix}
\\ 
\end{array}
\label{c17}
\end{equation}%
implies the holomorphic transformation between the $S^{3}$ and the
hyperquadric CR-structure coordinates.

The boundary of the classical domain is invariant under the $SU(1,2)$.\ The
action of the group on the boundary is transitive, because for any two
points of the boundary there is a group element which transforms the one
point to the other. But the group action is not effective (faithful),
because there are many group elements, which transform one point to the
other. The boundary is the coset space $SU(1,2)/P$,\ where $P$\ is the
isotropy group (subgroup of $SU(1,2)$) which preserves a point of the
boundary. Hence the hyperquadric and $S^{3}$ may be viewed as a base
manifold of a 8-dimensional bundle with the background Cartan connection of
the group $SU(1,2)$. If we define the connection with $\omega =g^{-1}dg$,\
where $g\in SU(1,2)$,\ we find that its curvature $\Omega =d\omega +\omega
\wedge \omega =0$ vanishes.

Now it is trivial to see that the Moser form (\ref{c13}) is a deformation of
the projective form \ 
\begin{equation}
\begin{array}{l}
X^{\dag }C_{S}X=%
\begin{pmatrix}
1 & \overline{z^{1}} & \overline{z^{0}}%
\end{pmatrix}%
\begin{pmatrix}
0 & 0 & -i \\ 
0 & -1 & 0 \\ 
i & 0 & 0%
\end{pmatrix}%
\begin{pmatrix}
1 \\ 
z^{1} \\ 
z^{0}%
\end{pmatrix}
\\ 
\end{array}
\label{c17a}
\end{equation}%
i.e. of the boundary of the hyperquadric, the unbounded realization of the $%
SU(1,2)$ classical domain. The unbounded realization is needed because of
the local translation is needed to formulate the expansion of the
holomorphic transformation. This point permits the application of the Cartan
method to classify the non-degenerate 3-dimensional CR-structures.

Cartan generalized the Klein geometry by osculating a general manifold with
a coset space. The forms of a general non-degenerate CR-manifold are
extended to the following relations\cite{Jaco} 
\begin{equation}
\begin{array}{l}
d\omega _{0}=i\omega _{1}\wedge \overline{\omega _{1}}-\omega _{0}\wedge
(\omega _{2}+\overline{\omega _{2}}) \\ 
d\omega _{1}=-\omega _{1}\wedge \omega _{2}-\omega _{0}\wedge \omega _{3} \\ 
d\omega _{3}=2i\omega _{1}\wedge \overline{\omega _{3}}+i\overline{\omega
_{1}}\wedge \omega _{3}-\omega _{0}\wedge \omega _{4}-R\omega _{0}\wedge 
\overline{\omega _{1}} \\ 
d\omega _{4}=i\omega _{3}\wedge \overline{\omega _{3}}+\omega _{4}\wedge
(\omega _{2}+\overline{\omega _{2}})-S\omega _{0}\wedge \omega _{1}-%
\overline{S}\omega _{0}\wedge \overline{\omega _{1}}%
\end{array}
\label{c18}
\end{equation}%
where\ $R$ and $S$ are the components of the curvature because for $R=S=0$
we find the $SU(1,2)$ structure equations. If we assume the following
normalization and notation 
\begin{equation}
\begin{array}{l}
d\omega _{0}=i\omega _{1}\wedge \overline{\omega _{1}}+b\omega _{0}\wedge
\omega _{1}+\overline{b}\omega _{0}\wedge \overline{\omega _{1}} \\ 
\omega _{1}=d\zeta \quad ,\quad \overline{\omega }_{1}=d\overline{\zeta } \\ 
\end{array}
\label{c19}
\end{equation}%
we find\cite{Jaco} 
\begin{equation}
\begin{array}{l}
R=\frac{k(x)}{6\lambda \overline{\lambda }^{3}} \\ 
\\ 
k(x):=\overline{e_{1}}-2\overline{b}\overline{a}\quad ,\quad
e:=c_{1}-bc-2ib_{0}\quad ,\quad c:=b_{\overline{1}} \\ 
df=:f_{0}\omega _{0}+f_{1}\omega _{1}+f\overline{\omega _{1}}%
\end{array}
\label{c20}
\end{equation}%
where $k(x)$ depends on the coordinates of the CR-manifold. If $R=0$ we find 
$S=0$ and the CR-manifold is holomorphically equivalent to the hyperquadric.
From the above form of $R\neq 0$, we see that we can always find a function $%
\lambda (x)$ such that $R=1$. Hence $R$ is a relative invariant of the
CR-structure, which may take the values $R=0$ or $R=1$. In the latter case
the CR-structure may be characterized by the sections $\omega _{2}\ ,\
\omega _{3}\ ,\ \omega _{4}\ $where $\omega _{2}$ is imaginary.

I will now apply the Cartan extension to the following CR-manifold 
\begin{equation}
\begin{array}{l}
U=-2a\frac{\zeta \overline{\zeta }}{1+\zeta \overline{\zeta }}\quad ,\quad
a\neq 0 \\ 
\end{array}
\label{c21}
\end{equation}%
Then 
\begin{equation}
\begin{array}{l}
\omega _{0}=du+\frac{2ia}{(1+\zeta \overline{\zeta })^{2}}(\overline{\zeta }%
d\zeta -\zeta d\overline{\zeta })\quad ,\quad \omega _{1}=d\zeta \\ 
\end{array}
\label{c22}
\end{equation}%
and 
\begin{equation}
\begin{array}{l}
d\omega _{0}=-\frac{4ia(1-\zeta \overline{\zeta })}{(1+\zeta \overline{\zeta 
})^{3}}\omega ^{1}\wedge \overline{\omega ^{1}}\quad ,\quad \omega
^{1}=d\zeta \\ 
\end{array}
\label{c23}
\end{equation}%
We see that at the points $\zeta \overline{\zeta }=1$ the CR-structure is
degenerate. The normalized (real) 1-form $\omega _{0}$ is \ 
\begin{equation}
\begin{array}{l}
\omega _{0}=-\frac{(1+\zeta \overline{\zeta })^{3}}{4a(1-\zeta \overline{%
\zeta })}[du+\frac{2ia}{(1+\zeta \overline{\zeta })^{2}}(\overline{\zeta }%
d\zeta -\zeta d\overline{\zeta })] \\ 
\\ 
\omega _{1}=d\zeta \\ 
\end{array}
\label{c24}
\end{equation}%
which is not defined at\ $\zeta \overline{\zeta }=1$. After some
calculations I find $R\neq 0$. Hence I conclude that this CR-manifold is not
holomorphically equivalent neither to the degenerate one nor to the
hyperquadric.

Concluding the present introductory subsection I point out that the
important notion is the existence of the partial complex structure in the
tangent (or equivalently the cotangent) space which is inherited from the
embedding in an ambient complex manifold. The differences between these two
definitions\cite{Jaco} are the dubbed "realizability problems", which will
not concern us. But the reader should realize that the notion of the metric
is not needed to define the CR-structure. On the other hand a related metric
may also be defined. In the present 3-dimensional CR-manifold we can define
the following class of Kaehler metrics and symplectic forms \ 
\begin{equation}
\begin{array}{l}
ds^{2}=2\frac{\partial ^{2}\rho ^{2}}{\partial z^{\alpha }\partial \overline{%
z^{\beta }}}dz^{\alpha }d\overline{z^{\beta }}\quad ,\quad \omega =2i\frac{%
\partial ^{2}\rho ^{2}}{\partial z^{\alpha }\partial \overline{z^{\beta }}}%
dz^{\alpha }\wedge d\overline{z^{\beta }} \\ 
\end{array}
\label{c24a}
\end{equation}%
It is essentially a class of metrics, because the CR-structure does not
uniquely determine $\rho $. Notice that $\rho ^{\prime }=f\rho $, with $%
f\neq 0$ determines the same CR-structure but it gives different ambient
metrics.

\subsection{Definition of the lorentzian CR-structure}

The lorentzian CR-manifold (LCR-manifold) is defined as a 4-dimensional
totally real submanifold of $%
\mathbb{C}
^{4}$ determined by three special functions, with $z^{b}:=(z^{\alpha },z^{%
\widetilde{\alpha }}),\ \alpha =0,1$. 
\begin{equation}
\begin{array}{l}
\rho _{11}(\overline{z^{\alpha }},z^{\alpha })=0\quad ,\quad \rho
_{12}\left( \overline{z^{\alpha }},z^{\widetilde{\alpha }}\right) =0\quad
,\quad \rho _{22}\left( \overline{z^{\widetilde{\alpha }}},z^{\widetilde{%
\alpha }}\right) =0 \\ 
\end{array}
\label{c25}
\end{equation}%
where $\rho _{11}\ ,\ \rho _{22}$ are real functions and $\rho _{12}$ is a
complex function. Notice the special dependence of the defining functions on
the structure coordinates.

Because of $d\rho _{ij}|_{M}=0$ and the special dependence of each function
on the structure coordinates $\left( z^{\alpha },z^{\widetilde{\alpha }%
}\right) $, we find\cite{BaEbRo} the following real 1-forms in the cotangent
space of the manifold \ 
\begin{equation}
\begin{array}{l}
\ell :=2i\partial \rho _{11}|_{M}=2i\partial ^{\prime }\rho
_{11}|_{M}=i(\partial ^{\prime }-\overline{\partial ^{\prime }})\rho
_{11}|_{M}=-2i\overline{\partial ^{\prime }}\rho _{11}|_{M} \\ 
\\ 
n:=2i\partial \rho _{22}|_{M}=2i\partial ^{\prime \prime }\rho
_{22}|_{M}=i(\partial ^{\prime \prime }-\overline{\partial ^{\prime \prime }}%
)\rho _{22}|_{M}=-2i\overline{\partial ^{\prime \prime }}\rho _{22}|_{M} \\ 
\\ 
m_{1}:=2i\partial \frac{\rho _{12}+\overline{\rho _{12}}}{2}|_{M}=i(\partial
^{\prime }+\partial ^{\prime \prime }-\overline{\partial ^{\prime }}-%
\overline{\partial ^{\prime \prime }})\frac{\rho _{12}+\overline{\rho _{12}}%
}{2}|_{M} \\ 
\\ 
m_{2}:=2i\partial \frac{\overline{\rho _{12}}-\rho _{12}}{2i}%
|_{M}=i(\partial ^{\prime }+\partial ^{\prime \prime }-\overline{\partial
^{\prime }}-\overline{\partial ^{\prime \prime }})\frac{\overline{\rho _{12}}%
-\rho _{12}}{2i}|_{M} \\ 
\end{array}
\label{c26}
\end{equation}%
where the accented symbols of the partial exterior derivatives are defined
as follows

\begin{equation}
\begin{array}{l}
d=\partial +\overline{\partial }=(\partial ^{\prime }+\partial ^{\prime
\prime })+(\overline{\partial ^{\prime }}+\overline{\partial ^{\prime \prime
}}) \\ 
\\ 
\partial ^{\prime }f:=\frac{\partial f}{\partial z^{\alpha }}dz^{\alpha
}\quad ,\quad \partial ^{\prime \prime }f:=\frac{\partial f}{\partial z^{%
\widetilde{\alpha }}}dz^{\widetilde{\alpha }} \\ 
\\ 
A_{b}dx^{b}=:A_{\alpha }^{\prime }dz^{\alpha }+A_{\widetilde{\alpha }%
}^{\prime \prime }dz^{\widetilde{\alpha }}%
\end{array}
\label{c27}
\end{equation}%
In order to familiarize the reader with this new formalism we make the
transcription $A\rightarrow A^{\prime }+A^{\prime \prime }$ in details 
\begin{equation}
\begin{array}{l}
A_{\mu }dx^{\mu }=A_{\mu }\delta _{\nu }^{\mu }dx^{\nu }=A_{\mu }(\ell ^{\mu
}n_{\nu }+n^{\mu }\ell _{\nu }-\overline{m}^{\mu }m_{\nu }-m^{\mu }\overline{%
m}_{\nu })dx^{\nu }= \\ 
\\ 
=[(n^{\mu }A_{\mu })\ell _{\alpha }-(\overline{m}^{\mu }A_{\mu })m_{\alpha
}]dz^{\alpha }+[(\ell ^{\mu }A_{\mu })n_{\widetilde{\alpha }}-(m^{\mu
}A_{\mu })\overline{m}_{\widetilde{\alpha }}]dz^{\widetilde{\alpha }}= \\ 
\\ 
=A_{\alpha }^{\prime }dz^{\alpha }+A_{\widetilde{\alpha }}^{\prime \prime
}dz^{\widetilde{\alpha }} \\ 
\end{array}
\label{c28}
\end{equation}%
where the induced symbols are neglected.

The 1-forms are real, because we consider them restricted on the defined
submanifold. The relations become simpler, if we use the complex form \ 
\begin{equation}
\begin{array}{l}
m=m_{1}+im_{2}=2i\partial ^{\prime }\overline{\rho _{12}}=-2i\overline{%
\partial ^{\prime \prime }}\overline{\rho _{12}}=i(\partial ^{\prime }-%
\overline{\partial ^{\prime \prime }})\overline{\rho _{12}} \\ 
\end{array}
\label{c29}
\end{equation}%
We see that a proper LCR-manifold is characterized by a pair of
3-dimensional CR-submanifolds with a common complex tangent (and cotangent)
vector $m$.

The corresponding tangent basis with the real vectors $\ell ^{\mu }\partial
_{\mu }\ ,\ n^{\mu }\partial _{\mu }$\ and the complex one $m^{\mu }\partial
_{\mu }$ is defined via the contractions 
\begin{equation}
\begin{array}{l}
(\ell ^{\prime \mu }\partial _{\mu })\lrcorner (n_{\nu }dx^{\nu })=1\quad
,\quad (n^{\prime \mu }\partial _{\mu })\lrcorner (\ell _{\nu }dx^{\nu })=1
\\ 
(m^{\prime \mu }\partial _{\mu })\lrcorner (\overline{m}_{\nu }dx)=-1\quad
,\quad (\overline{m}^{\prime \mu }\partial _{\mu })\lrcorner (m_{\nu
}dx^{\nu })=1 \\ 
\\ 
all\ the\ other\ vanish%
\end{array}
\label{c30}
\end{equation}%
where the accent distinguishes the tangent vectors (directional derivatives)
from the cotangent vectors (differential forms). Recall that the symbol $%
\lrcorner $\ defines the interior product (antiderivation) which is not
related to any metric! No metric has been introduced \cite{Chand}.

An abstract LCR-manifold is defined by two real $\ell \ ,\ n$\ , and a
complex $m$ 1-forms, such that $\ell \wedge n\wedge m\wedge \overline{m}\neq
0$ and 
\begin{equation}
\begin{array}{l}
d\ell =Z_{1}\wedge \ell +i\Phi _{1}m\wedge \overline{m} \\ 
\\ 
dn=Z_{2}\wedge n+i\Phi _{2}m\wedge \overline{m} \\ 
\\ 
dm=Z_{3}\wedge m+\Phi _{3}\ell \wedge n%
\end{array}
\label{c31}
\end{equation}%
where the vector fields $Z_{1\mu }\ ,\ Z_{2\mu }\ $ are real, the vector
field$\ Z_{3\mu }$ is complex, the scalar fields $\Phi _{1}\ ,\ \Phi _{2}$
are real and the scalar field$\ \Phi _{3}$ is complex. Using the relations (%
\ref{c30}) one can prove that the above definition of the LCR-structure in
the cotangent space is equivalent to the following commutation relations

\begin{equation}
\begin{tabular}{l}
$\lbrack (\ell ^{\prime \mu }\partial _{\mu }),(m^{\prime \mu }\partial
_{\mu })]=f_{0}(\ell ^{\prime \mu }\partial _{\mu })+f_{1}(m^{\prime \mu
}\partial _{\mu })$ \\ 
\\ 
$\lbrack (n^{\prime \mu }\partial _{\mu }),(\overline{m}^{\prime \mu
}\partial _{\mu })]=h_{0}(n^{\prime \mu }\partial _{\mu })+h_{1}(\overline{m}%
^{\prime \mu }\partial _{\mu })$ \\ 
\end{tabular}
\label{c32}
\end{equation}%
of the corresponding tangent directional derivatives. That is, the
LCR-structure conditions constitute an integrable system where the
application of the (holomorphic) Frobenius theorem in the cotangent and the
tangent space implies the embedding relations (\ref{c25}), which will be
explicitly proved below.

The above conditions are invariant under the transformation

\begin{equation}
\begin{tabular}{l}
$\ell _{\mu }^{\prime }=\Lambda \ell _{\mu }\quad ,\quad \ell ^{\prime
\prime \mu }=\frac{1}{N}\ell ^{\prime \mu }$ \\ 
\\ 
$n_{\mu }^{\prime }=Nn_{\mu }\quad ,\quad n^{\prime \prime \mu }=\frac{1}{%
\Lambda }n^{\prime \mu }$ \\ 
\\ 
$m_{\mu }^{\prime }=Mm_{\mu }\quad ,\quad m^{\prime \prime \mu }=\frac{1}{%
\overline{M}}m^{\prime \mu }$ \\ 
\end{tabular}
\label{c33}
\end{equation}%
with non-vanishing $\Lambda \ ,\ N\ ,\ M$ functions and which imply the
following transformations of the vector and scalar fields \ 
\begin{equation}
\begin{array}{l}
Z_{1\mu }^{\prime }=Z_{1\mu }+\partial _{\mu }\ln \Lambda \quad ,\quad
Z_{2\mu }^{\prime }=Z_{2\mu }+\partial _{\mu }\ln N \\ 
\\ 
Z_{3\mu }^{\prime }=Z_{3\mu }+\partial _{\mu }\ln M \\ 
\\ 
\Phi _{1}^{\prime }=\frac{\Lambda }{M\overline{M}}\Phi _{1}\quad ,\quad \Phi
_{2}^{\prime }=\frac{N}{M\overline{M}}\Phi _{2}\quad ,\quad \Phi
_{3}^{\prime }=\frac{M}{\Lambda N}\Phi _{3}%
\end{array}
\label{c33a}
\end{equation}%
Hence $\Phi _{1}\ ,\ \Phi _{2}\ ,\ \Phi _{3}$ are LCR-structure relative
invariants and the differential forms \ 
\begin{equation}
\begin{array}{l}
F_{1}=dZ_{1}\quad ,\quad F_{2}=dZ_{2}\quad ,\quad F_{3}=dZ_{3} \\ 
\end{array}
\label{c34}
\end{equation}%
are LCR-invariants. These relative invariants act as topological invariants
characterizing the different sectors of the LCR-structure solitons. Hence a
LCR-structure defines a class of LCR-tetrads, which will imply a class of
Einstein metrics.

Notice that this definition permits us to apply the holomorphic Frobenius
theorem for ($\ell \ ,\ m$) and ($n\ ,\ \overline{m}$), which is always
possible, if the tetrad 1-forms are real analytic functions. For that, we
have to complexify the coordinates $x^{\mu }$, considering the basis
covectors real analytic. This theorem implies that there are two sets of
generally complex coordinates $(z^{\alpha }(x),\;z^{\widetilde{\alpha }}(x))$%
,\ $\alpha =0,\ 1$ such that%
\begin{equation}
\begin{array}{l}
dz^{\alpha }=f_{\alpha }\ \ell _{\mu }dx^{\mu }+h_{\alpha }\ m_{\mu }dx^{\mu
}\quad ,\quad dz^{\widetilde{\alpha }}=f_{\widetilde{\alpha }}\ n_{\mu
}dx^{\mu }+h_{\widetilde{\alpha }}\ \overline{m}_{\mu }dx^{\mu } \\ 
\\ 
dz^{0}\wedge dz^{1}\wedge dz^{\widetilde{0}}\wedge dz^{\widetilde{1}}\neq 0
\\ 
\\ 
\ell =\ell _{\alpha }dz^{\alpha }\quad ,\quad m=m_{\alpha }dz^{\alpha }\quad
,\quad n=n_{\widetilde{\alpha }}dz^{\widetilde{\alpha }}\quad ,\quad 
\overline{m}=\overline{m}_{\widetilde{\alpha }}dz^{\widetilde{\alpha }} \\ 
\end{array}
\label{c35}
\end{equation}%
After their computation we make $x^{\mu }$ real again, but real analyticity
condition assures that the matrices $\partial _{\mu }z^{b}(x)$ and its
inverse $\partial _{b}x^{\mu }(z^{a})$ do not vanish on the LCR-manifold.
The reality conditions of the tetrad imply that these structure coordinates
satisfy the relations

\begin{equation}
\begin{array}{l}
dz^{0}\wedge dz^{1}\wedge d\overline{z^{0}}\wedge d\overline{z^{1}}=0 \\ 
\\ 
dz^{\widetilde{0}}\wedge dz^{\widetilde{1}}\wedge d\overline{z^{0}}\wedge d%
\overline{z^{1}}=0 \\ 
\\ 
dz^{\widetilde{0}}\wedge dz^{\widetilde{1}}\wedge d\overline{z^{\widetilde{0}%
}}\wedge d\overline{z^{\widetilde{1}}}=0%
\end{array}
\label{c36}
\end{equation}%
that is, there are two real functions $\rho _{11}$ , $\rho _{22}$ and a
complex one $\rho _{12}$, such that the abstract LCR-structure is realized
in $%
\mathbb{C}
^{4}$ via the totally real surface (\ref{c25}). Hence in the case of real
analytic differentiable manifolds, the abstract LCR-structure is embeddable
in $%
\mathbb{C}
^{4}$ through real functions of special form and the two definitions of
LCR-structure coincide.

\subsection{A G(4,2)\ embedding of the LCR-structures}

The permitted (restricted) holomorphic transformations $z^{\prime
a}=f^{\alpha }(z^{\beta }),\ z^{\prime \widetilde{a}}=f^{\alpha }(z^{%
\widetilde{\beta }})$ may be used\cite{BaEbRo} to find structure coordinates
(called regular coordinates) such that (\ref{c25}) take the forms 
\begin{equation}
\begin{array}{l}
\rho _{11}\left( \overline{z^{\alpha }},z^{\widetilde{\alpha }}\right) =%
\func{Im}z^{0}-\phi _{11}(\overline{z^{1}},z^{1},\func{Re}z^{0}) \\ 
\rho _{12}\left( \overline{z^{\alpha }},z^{\widetilde{\alpha }}\right) =z^{%
\widetilde{1}}-\overline{z^{1}}-\phi _{12}(\overline{z^{a}},z^{\widetilde{%
\beta }}) \\ 
\rho _{22}\left( \overline{z^{\widetilde{\alpha }}},z^{\widetilde{\alpha }%
}\right) =\func{Im}z^{\widetilde{0}}-\phi _{22}(\overline{z^{\widetilde{1}}}%
,z^{\widetilde{1}},\func{Re}z^{\widetilde{0}}) \\ 
\\ 
\phi _{ij}(0)=0\quad ,\quad d\phi _{ij}(0)=0%
\end{array}
\label{c37}
\end{equation}%
where $z^{1},z^{\widetilde{1}}$, are the complex coordinates of $CP^{1}$,
because this regular form of the LCR-structure continues to permit the
following $SL(2,%
\mathbb{C}
)$ transformation%
\begin{equation}
\begin{array}{l}
z^{\prime 1}=\frac{c+dz^{1}}{a+bz^{1}}\quad ,\quad z^{\prime \widetilde{1}}=%
\frac{\overline{c}+\overline{d}z^{\widetilde{1}}}{\overline{a}+\overline{b}%
z^{\widetilde{1}}} \\ 
\\ 
ad-bc=1%
\end{array}
\label{c38}
\end{equation}%
That is, the corresponding spinors transform relative to the conjugate
representations of $SL(2,%
\mathbb{C}
)$%
\begin{equation}
\begin{array}{l}
\begin{pmatrix}
\lambda ^{\prime } \\ 
\lambda ^{\prime }z^{\prime 1}%
\end{pmatrix}%
=%
\begin{pmatrix}
a & b \\ 
c & d%
\end{pmatrix}%
\begin{pmatrix}
\lambda \\ 
\lambda z^{1}%
\end{pmatrix}
\\ 
\\ 
\begin{pmatrix}
-\widetilde{\lambda }^{\prime }z^{\prime \widetilde{1}} \\ 
\widetilde{\lambda }^{\prime }%
\end{pmatrix}%
=%
\begin{pmatrix}
a & b \\ 
c & d%
\end{pmatrix}%
^{\dag -1}%
\begin{pmatrix}
-\widetilde{\lambda }z^{\widetilde{1}} \\ 
\widetilde{\lambda }%
\end{pmatrix}
\\ 
\\ 
ad-bc=1%
\end{array}
\label{c39}
\end{equation}

We saw that the non-degenerate realizable 3-dimensional CR-structures can be
osculated with the boundary of the $SU(1,2)$ classical domain. In the
present case of the LCR-structure the convenient 4-dimensional symmetric
space is the boundary of the $SU(2,2)$ classical domain. For that we first
projectivize (\ref{c25}) to the form

\begin{equation}
\begin{array}{l}
\rho _{11}(\overline{Z^{m1}},Z^{n1})=0\quad ,\quad \rho _{12}\left( 
\overline{Z^{m1}},Z^{n2}\right) =0\quad ,\quad \rho _{22}(\overline{Z^{m2}}%
,Z^{n2})=0 \\ 
\\ 
K(Z^{m1})=0=K(Z^{m2})%
\end{array}
\label{c40}
\end{equation}%
where $Z^{ni}\in CP^{3}$ and $K(Z^{n})$ is a homogeneous function in $%
\mathbb{C}
^{4}$. We will call it Kerr function because it is related to the Kerr
theorem in Minkowski space.

The grassmannian projective manifold $G(4,2)$ is the set of the $4\times 2$
complex matrices \ 
\begin{equation}
\begin{array}{l}
X=%
\begin{pmatrix}
X^{01} & X^{02} \\ 
X^{11} & X^{12} \\ 
X^{21} & X^{22} \\ 
X^{31} & X^{32}%
\end{pmatrix}
\\ 
\end{array}
\label{c48}
\end{equation}%
of rank-2 with the equivalence relation $X\sim X^{\prime }$ if there exists
a $2\times 2$ regular ($\det S\neq 0$) matrix $S$ such that%
\begin{equation}
X^{\prime }=XS  \label{c49}
\end{equation}%
Its typical coordinates are the projective coordinates, which are defined in
every coordinate chart determined by every $2\times 2$\ submatrix with
non-vanishing determinant. In the coordinate chart with \ 
\begin{equation}
\begin{array}{l}
\det 
\begin{pmatrix}
X^{01} & X^{02} \\ 
X^{11} & X^{12}%
\end{pmatrix}%
\neq 0 \\ 
\end{array}
\label{c50}
\end{equation}%
the projective coordinates $\widehat{z}$ are defined by the relation \ 
\begin{equation}
\begin{array}{l}
X=%
\begin{pmatrix}
X_{1} \\ 
\widehat{z}X_{1}%
\end{pmatrix}
\\ 
\end{array}
\label{c51}
\end{equation}%
The projective form (\ref{c40}) is projectively invariant under the general
linear $SL(4,%
\mathbb{C}
)$ transformations. But in order to apply the Cartan expansion we need the
precise form of the $SU(2,2)$ classical domain.

The $SU(2,2)$ symmetric classical domain is the following region\cite{Pyat}
of $G(4,2)$ \ 
\begin{equation}
\begin{array}{l}
X^{\dagger }EX>0 \\ 
\end{array}
\label{c52}
\end{equation}%
which means that the $2\times 2$\ matrix is positive definite. $E$\ is a $%
SU(2,2)$ symmetric $4\times 4$\ matrix. The bounded realization of the
classical domain is achieved with the matrix \ 
\begin{equation}
\begin{array}{l}
E_{B}=\left( 
\begin{array}{cc}
I & 0 \\ 
0 & -I%
\end{array}%
\right) \\ 
\end{array}
\label{c53}
\end{equation}%
and it is \ 
\begin{equation}
\begin{array}{l}
\begin{pmatrix}
Y_{1}^{\dagger } & Y_{2}^{\dagger }%
\end{pmatrix}%
\left( 
\begin{array}{cc}
I & 0 \\ 
0 & -I%
\end{array}%
\right) \left( 
\begin{array}{c}
Y_{1} \\ 
Y_{2}%
\end{array}%
\right) >0\quad \Longleftrightarrow \quad I-\widehat{w}^{\dagger }\widehat{w}%
>0 \\ 
\end{array}
\label{c54}
\end{equation}%
where $\widehat{w}\equiv w_{ij}$\ is the symbol we will use for the
projective coordinates in the bounded realization. It is invariant under the 
$SU(2,2)$\textbf{\ }transformations which take the following explicit form%
\begin{equation}
\begin{array}{l}
\begin{pmatrix}
Y_{1}^{\prime } \\ 
Y_{2}^{\prime }%
\end{pmatrix}%
=%
\begin{pmatrix}
A_{11} & A_{12} \\ 
A_{21} & A_{22}%
\end{pmatrix}%
\left( 
\begin{array}{c}
Y_{1} \\ 
Y_{2}%
\end{array}%
\right) \\ 
\\ 
\widehat{w}^{\prime }=\left( A_{21}+A_{22}\ \widehat{w}\right) \left(
A_{11}+A_{12}\ \widehat{w}\right) ^{-1} \\ 
\\ 
A_{11}^{\dagger }A_{11}-A_{21}^{\dagger }A_{21}=I\quad ,\quad
A_{11}^{\dagger }A_{12}-A_{21}^{\dagger }A_{22}=0 \\ 
A_{22}^{\dagger }A_{22}-A_{12}^{\dagger }A_{12}=I%
\end{array}
\label{c55}
\end{equation}%
The characteristic (Shilov) boundary of this domain is the $S^{1}\times
S^{3} $ [$\simeq U(2)$] manifold with $\widehat{w}^{\dagger }\widehat{w}=I$.

The unbounded (Siegel) realization of the classical domain is determined
with the matrix \ 
\begin{equation}
\begin{array}{l}
E_{U}=\left( 
\begin{array}{cc}
0 & I \\ 
I & 0%
\end{array}%
\right) \\ 
\end{array}
\label{c56}
\end{equation}%
and it has the form \ 
\begin{equation}
\begin{array}{l}
\begin{pmatrix}
X_{1}^{\dagger } & X_{2}^{\dagger }%
\end{pmatrix}%
\left( 
\begin{array}{cc}
0 & I \\ 
I & 0%
\end{array}%
\right) \left( 
\begin{array}{c}
X_{1} \\ 
X_{2}%
\end{array}%
\right) >0\quad \Longleftrightarrow \quad \frac{-i}{2}(\widehat{r}-\widehat{r%
}^{\dagger })=\widehat{y}>0 \\ 
\\ 
\widehat{r}=\widehat{x}+i\widehat{y}=iX_{2}X_{1}^{-1} \\ 
\end{array}
\label{c57}
\end{equation}%
where the projective coordinates in the Siegel realization $\widehat{r}%
\equiv r_{A^{\prime }A}$ are defined with an additional factor $i$\ for
convenience. The fractional transformations, which preserve the unbounded
domain, are%
\begin{equation}
\begin{array}{l}
\begin{pmatrix}
X_{1}^{\prime } \\ 
X_{2}^{\prime }%
\end{pmatrix}%
=%
\begin{pmatrix}
B_{11} & B_{12} \\ 
B_{21} & B_{22}%
\end{pmatrix}%
\left( 
\begin{array}{c}
X_{1} \\ 
X_{2}%
\end{array}%
\right) \\ 
\\ 
\widehat{r}^{\prime }=\left( B_{22}\ \widehat{r}+iB_{21}\right) \left(
B_{11}-iB_{12}\ \widehat{r}\right) ^{-1} \\ 
\\ 
B_{11}^{\dagger }B_{22}+B_{21}^{\dagger }B_{12}=I\quad ,\quad
B_{11}^{\dagger }B_{21}+B_{21}^{\dagger }B_{11}=0 \\ 
B_{22}^{\dagger }B_{12}+B_{12}^{\dagger }B_{22}=0 \\ 
\end{array}
\label{c57a}
\end{equation}%
Notice that the linear part of these transformations ($B_{12}=0$), which
preserves the infinity of the Siegel domain, are the Poincar\'{e}$\times $%
Dilation transformations.

The unitary transformation \ 
\begin{equation}
\begin{array}{l}
\left( 
\begin{array}{cc}
0 & I \\ 
I & 0%
\end{array}%
\right) =\frac{1}{2}\left( 
\begin{array}{cc}
I & I \\ 
I & -I%
\end{array}%
\right) \left( 
\begin{array}{cc}
I & 0 \\ 
0 & -I%
\end{array}%
\right) \left( 
\begin{array}{cc}
I & I \\ 
I & -I%
\end{array}%
\right) \\ 
\end{array}
\label{c58}
\end{equation}%
implies the following Cayley transformation of the projective coordinates%
\begin{equation}
\begin{array}{l}
\widehat{r}=i(I-\widehat{w})(I+\widehat{w})^{-1}=i(I+\widehat{w})^{-1}(I-%
\widehat{w}) \\ 
\\ 
\widehat{w}=(iI-\widehat{r})(iI+\widehat{r})^{-1}=(iI+\widehat{r})^{-1}(iI-%
\widehat{r})%
\end{array}
\label{c59}
\end{equation}

Restricted on the boundary it becomes $U(2)\rightarrow 
\mathbb{R}
^{4}$, which is not bijective. The $SU(2)$\ group manifold implied by
exponentiation of its Lie algebra is%
\begin{equation}
\begin{array}{l}
SU(2)\ni U_{0}=e^{i\psi _{j}\frac{\sigma ^{j}}{2}}=\cos \frac{\psi }{2}+i%
\widehat{\psi }_{j}\sigma ^{j}\sin \frac{\psi }{2}\ ,\quad \psi _{j}=:\psi 
\widehat{\psi }_{j}\ ,\ \psi =\sqrt{\tsum\limits_{j=1}^{3}(\psi _{j})^{2}}
\\ 
U_{0}(\ \widehat{\psi }_{j},\psi +4\pi )=U_{0}(\ \widehat{\psi }_{j},\psi
)=-U_{0}(\ \widehat{\psi }_{j},\psi +2\pi ) \\ 
\end{array}
\label{c60}
\end{equation}%
where $\widehat{\psi }_{j}$\ is the unit vector. Its most convenient image
is the unit ball with radius $2\pi $, center at $U_{0}=I$ and its surface
identified with the point $U_{0}=-I$. When $\psi =:2\rho $ is in this
domain, the cartesian coordinates of the $w=I$ chart%
\begin{equation}
\begin{array}{l}
\widehat{x}_{+}=i(I-\widehat{w})(I+\widehat{w})^{-1}=i(I+\widehat{w})^{-1}(I-%
\widehat{w}) \\ 
\widehat{w}^{\dagger }=\widehat{w}^{-1}%
\end{array}
\label{c61}
\end{equation}%
is found assuming 
\begin{equation}
\begin{array}{l}
\widehat{\psi }_{j}=(-\sin \sigma \cos \chi ,\ -\sin \sigma \sin \chi ,\
\cos \sigma ) \\ 
\\ 
\widehat{w}=e^{i\tau }\left( 
\begin{array}{cc}
\cos \rho +i\sin \rho \cos \sigma & -i\sin \rho \sin \sigma \ e^{-i\chi } \\ 
-i\sin \rho \sin \sigma \ e^{i\chi } & \cos \rho -i\sin \rho \cos \sigma%
\end{array}%
\right) \\ 
\tau \in (-\pi ,\pi )\quad ,\quad \rho \in (0,\pi )\quad ,\quad \sigma \in
\lbrack 0,\pi )\quad ,\quad \chi \in (0,2\pi )%
\end{array}
\label{c62}
\end{equation}%
It has the form \ 
\begin{equation}
\begin{array}{l}
x_{+}^{0}=\frac{\sin \tau }{\cos \tau +\cos \rho } \\ 
x_{+}^{1}+ix_{+}^{2}=\frac{\sin \rho }{\cos \tau +\cos \rho }\sin \sigma \
e^{i\chi } \\ 
x_{+}^{3}=\frac{\sin \rho }{\cos \tau +\cos \rho }\cos \sigma \\ 
\tau \in (-\pi ,\pi )\ ,\ \rho \in \lbrack 0,\pi )\ ,\ \sigma \in \lbrack
0,\pi )\ ,\ \chi \in (0,2\pi ) \\ 
\\ 
s:=\frac{\sin \rho }{\cos \tau \ +\cos \rho }>0\quad \leftrightarrow \quad
\cos \tau \ +\ \cos \rho >0%
\end{array}
\label{c63}
\end{equation}%
which describes the one $%
\mathbb{R}
^{4}$-chart around the point $w=I$.

One may view the rest of $U(2)$ as the Cayley transformation centered at $%
w=-I$. But I find more convenient the extension of the above parameter $s$
to negative values to correspond to the rest of $U(2)$, the region $\frac{%
\sin \rho }{\cos \tau \ +\cos \rho }<0$ to be the other sheet of $%
\mathbb{R}
^{4}$ with \ 
\begin{equation}
\begin{array}{l}
x_{-}^{0}=\frac{\sin \tau }{\cos \tau +\cos \rho } \\ 
x_{-}^{1}+ix_{-}^{2}=-\frac{\sin \rho }{\cos \tau +\cos \rho }\sin \sigma \
e^{i\chi } \\ 
x_{-}^{3}=-\frac{\sin \rho }{\cos \tau +\cos \rho }\cos \sigma \\ 
\tau \in (-\pi ,\pi )\ ,\ \rho \in \lbrack 0,\pi )\ ,\ \sigma \in \lbrack
0,\pi )\ ,\ \chi \in (0,2\pi ) \\ 
\\ 
-s:=\frac{\sin \rho }{\cos \tau +\cos \rho }<0\quad \leftrightarrow \quad
\cos \tau +\cos \rho <0%
\end{array}
\label{c64}
\end{equation}%
Hence we conclude that the unbounded realization shows only the one sheet of
the universe. These two cartesian sheets do not overlap, therefore they do
not constitute an atlas of the universe $U(2)$. Recall that for the $SU(1,1)$
classical domain, the correspondence between the circle of the disc and the
real line is bijective%
\begin{equation}
\begin{array}{l}
w=e^{i\varphi },\ \varphi \in (-\pi ,\pi )\quad \longleftrightarrow \quad
x=\tan \frac{\varphi }{2}\in 
\mathbb{R}
\\ 
\end{array}
\label{c65}
\end{equation}%
All the circle (boundary) is covered by one real line $%
\mathbb{R}
$.

The Shilov boundary of the $SU(2,2)$ symmetric classical domain is

\begin{equation}
\begin{array}{l}
\rho _{ij}(\overline{X^{mi}},X^{nj})=\overline{X^{mi}}E_{mn}^{U}X^{nj}=0 \\ 
\\ 
K(X^{mj})=0\quad ,\quad E_{mn}^{U}:=%
\begin{pmatrix}
0 & I \\ 
I & 0%
\end{pmatrix}%
\end{array}
\label{c66}
\end{equation}%
in the unbounded realization. Hence, for these LCR-structures, which we call
"algebraically flat", we have real projective coordinates $\widehat{r}=%
\widehat{x}=\widehat{x}^{\dag }$, and the LCR-structure is determined only
by the homogeneous holomorphic function $K(X^{m})$. That is, the points of
the Shilov boundary take the representation 
\begin{equation}
\begin{array}{l}
X^{mj}=%
\begin{pmatrix}
\lambda \\ 
-i\widehat{x}\lambda%
\end{pmatrix}%
=%
\begin{pmatrix}
\lambda ^{Aj} \\ 
-ix_{A^{\prime }B}\lambda ^{Bj}%
\end{pmatrix}
\\ 
\\ 
\widehat{x}=\eta _{\mu \nu }x^{\mu }\sigma ^{\nu }=%
\begin{pmatrix}
x^{0}-x^{3} & -(x^{1}-ix^{2}) \\ 
-(x^{1}+ix^{2}) & x^{0}+x^{3}%
\end{pmatrix}%
=x_{A^{\prime }B}%
\end{array}
\label{c67}
\end{equation}%
in homogeneous coordinates. The Poincar\'{e}$\times $dilation
transformations preserve the above infinity 
\begin{equation}
\begin{array}{l}
\begin{pmatrix}
\lambda ^{\prime } \\ 
-i\widehat{x}^{\prime }\lambda ^{\prime }%
\end{pmatrix}%
=%
\begin{pmatrix}
B & 0 \\ 
-iTB & (B^{\dag })^{-1}%
\end{pmatrix}%
\begin{pmatrix}
\lambda \\ 
-i\widehat{x}\lambda%
\end{pmatrix}
\\ 
\\ 
\lambda ^{\prime }=B\lambda \quad ,\quad \widehat{x}^{\prime
}=(B^{-1})^{\dagger }\widehat{x}B^{-1}+T \\ 
\det B=\overline{\det B}\quad ,\quad T^{\dagger }=T%
\end{array}
\label{c68}
\end{equation}%
It is the spinorial Poincar\'{e}$\times $dilation group. Its Poincar\'{e}
subgroup is identified with the Poincar\'{e} symmetry observed in nature,
but there is an essential difference between the present symmetry and that
implied by the metric of general relativity. The present group is the proper
orthochronous group of special relativity (Minkowski space). The spatial and
temporal reflections are \emph{external automorphisms} of its Lie algebra,
i.e. they do not exponentiate into elements of the group (\ref{c68}).
Therefore they may be broken in PCFT as it has been observed in nature.

In order to osculate the general LCR-structure relations with the
"algebraically flat" LCR-structure conditions, I write

\begin{equation}
\begin{array}{l}
\rho _{ij}(\overline{X^{mi}},X^{nj})=\overline{X^{mi}}X^{nj}E_{mn}-G_{ij}(%
\overline{X^{mi}},X^{nj})=0 \\ 
\\ 
K(X^{mj})=0%
\end{array}
\label{c69}
\end{equation}%
where $G_{ij}(\overline{X^{mi}},X^{nj})$ is viewed as the non-flat part of
the LCR-structure. Notice that the Kerr homogeneous function persists even\
in the deformed boundary.

\subsection{LCR-structures determined with complex trajectories}

Newman introduced\cite{Newm1973} complex trajectories to describe geodetic
and shear free null congruences in Minkowski space. Besides, it is well
known that the Lienard-Wiechert radiating electromagnetic potentials are
related to accelerating trajectories. In the present formalism and in the
case of embeddable LCR-structures (\ref{c66}) the ruled surfaces of $CP(3)$
are characterized by complex trajectories $\xi ^{b}(\tau )$ in the
corresponding grassmannian $G(4,2)$. The ruled surfaces have a special
explicit parameterization, which reveals their internal property to be made
up of lines, which determine a generally complex trajectory in the
grassmannian space. That is, they have the form 
\begin{equation}
\begin{array}{l}
Z^{m}(\tau ,s)=(1-s)Z^{m1}(\tau )+sZ^{m2}(\tau )= \\ 
\qquad =Z^{m1}(\tau )+sT^{m}(\tau ) \\ 
\\ 
T^{m}(\tau ):=Z^{m2}(\tau )-Z^{m1}(\tau )%
\end{array}
\label{c70}
\end{equation}%
where $T^{m}(\tau )$ indicates the direction\ of the generating line which
meets $Z^{m1}(\tau )$ (the generatrix) at $\tau $. In this definition the
Kerr function is replaced with its proper parametrization.

The generating lines correspond to complex points of the grassmannian
manifold $G(4,2)$, with projective coordinates 
\begin{equation}
\begin{array}{l}
\widehat{\xi }(\tau )=:iX_{2}X_{1}^{-1}=:%
\begin{pmatrix}
\xi ^{0}-\xi ^{3} & -(\xi ^{1}-i\xi ^{2}) \\ 
-(\xi ^{1}+i\xi ^{2}) & \xi ^{0}+\xi ^{3}%
\end{pmatrix}
\\ 
\\ 
X_{1}=:%
\begin{pmatrix}
X^{01} & X^{02} \\ 
X^{11} & X^{12}%
\end{pmatrix}%
=%
\begin{pmatrix}
Z^{0}(\tau ,0) & Z^{0}(\tau ,1) \\ 
Z^{0}(\tau ,0) & Z^{1}(\tau ,1)%
\end{pmatrix}
\\ 
X_{2}=:%
\begin{pmatrix}
X^{21} & X^{22} \\ 
X^{31} & X^{32}%
\end{pmatrix}%
=%
\begin{pmatrix}
Z^{2}(\tau ,0) & Z^{2}(\tau ,1) \\ 
Z^{3}(\tau ,0) & Z^{3}(\tau ,1)%
\end{pmatrix}%
\end{array}
\label{c71}
\end{equation}%
Using homogeneous coordinates, this curve of $G(4,2)$ is spanned by the two
points $Z^{ni}(\tau );\ i=1,2$ of $%
\mathbb{C}
^{4}$. The curve is called non-degenerate if the following determinant\cite%
{GrHa2} does not identically vanish 
\begin{equation}
\begin{array}{l}
\det [Z^{n1},Z^{n2},\frac{dZ^{n1}}{d\tau },\frac{dZ^{n2}}{d\tau }]=\det 
\begin{pmatrix}
X_{1} & \overset{.}{X_{1}} \\ 
-i\widehat{\xi }X_{1} & -i(\overset{.}{\widehat{\xi }}X_{1}+\widehat{\xi }%
\overset{.}{X_{1}})%
\end{pmatrix}%
= \\ 
=\det [%
\begin{pmatrix}
1 & 0 \\ 
-i\widehat{\xi } & 1%
\end{pmatrix}%
\begin{pmatrix}
X_{1} & \overset{.}{X_{1}} \\ 
0 & -i\overset{.}{\widehat{\xi }}X_{1}%
\end{pmatrix}%
]=-\det (\overset{.}{\widehat{\xi }})(\det X_{1})^{2}%
\end{array}
\label{c72}
\end{equation}%
This happens if and only if $\overset{.}{\xi }^{a}\overset{.}{\xi }^{b}\eta
_{ab}\neq 0$, because $\det X_{1}\neq 0$ in this projective coordinate
chart. This condition will differentiate the massive from the massless
partner (neutrino) of a leptonic generation. The complex trajectory is
related to the ordinary classical trajectory of the particle viewed as a
soliton. If they are real, they are identified with the well known
trajectories of the Lienard-Wiechert potential. If the curve is degenerate ($%
\overset{.}{\xi }^{a}\overset{.}{\xi }^{b}\eta _{ab}=0$), the gaussian
curvature of the ruled surface vanishes and the ruled surface is called
developable. The developable surfaces of $CP(3)$ are cones, cylinders and
tangent developables with $T^{n}(\tau )=\frac{dZ_{1}^{m}(\tau )}{d\tau }$.

In the context of special relativity, Newman\cite{Newm1973} showed that a
complex trajectory in complex Minkowski spacetime defines a geodetic and
shear free null congruence. A quite general Poincar\'{e} covariant explicit
parameterization of a ruled hypersurface of $CP(3)$ and its corresponding
grassmannian patch, is\ 
\begin{equation}
\begin{array}{l}
X^{ni}=%
\begin{pmatrix}
Z^{0}(\tau ,s) & Z^{0}(\widetilde{\tau },\widetilde{s}) \\ 
Z^{1}(\tau ,s) & Z^{1}(\widetilde{\tau },\widetilde{s}) \\ 
Z^{2}(\tau ,s) & Z^{2}(\widetilde{\tau },\widetilde{s}) \\ 
Z^{3}(\tau ,s) & Z^{3}(\widetilde{\tau },\widetilde{s})%
\end{pmatrix}%
=%
\begin{pmatrix}
\lambda ^{Ai} \\ 
-ir_{B^{\prime }B}\lambda ^{Bi}%
\end{pmatrix}
\\ 
\end{array}
\label{c73}
\end{equation}%
where the $r_{B^{\prime }B}=r_{b}\sigma _{B^{\prime }B}^{b}$ are the
projective coordinates, generally \emph{outside} the $\xi _{b}(\tau )$
trajectory of the ruled surface. Because simply not all the \emph{pairs} of
points of a ruled surface belong to rulings. If $r_{b}\in \xi _{b}(\tau )$
the projective line of $CP(3)$ coincides with a ruling line of the ruled
surface.

The reparametrization ($\tau $) ambiguity may be fixed with either the
condition $\xi ^{0}(\tau )=\tau $ or the more restrictive one $\overset{.}{%
\xi }^{a}\overset{.}{\xi }^{b}\eta _{ab}=1$. In the coordinate chart $%
Z^{0}=1 $ of $CP(3)$, a \ general point of the ruled surface determined by a
trajectory $\xi ^{b}(\tau )$ has the form 
\begin{equation}
\begin{array}{l}
Z^{n}(\tau ,\lambda )=%
\begin{pmatrix}
1 \\ 
0 \\ 
-i(\xi ^{0}-\xi ^{3}) \\ 
i(\xi ^{1}+i\xi ^{2})%
\end{pmatrix}%
+\lambda 
\begin{pmatrix}
0 \\ 
1 \\ 
i(\xi ^{1}-i\xi ^{2}) \\ 
-i(\xi ^{0}+\xi ^{3})%
\end{pmatrix}
\\ 
\\ 
\lambda =:\frac{(1-s)\lambda ^{11}(\tau )+s\lambda ^{12}(\tau )}{%
(1-s)\lambda ^{01}(\tau )+s\lambda ^{02}(\tau )}%
\end{array}
\label{c74}
\end{equation}%
The first term is the directrix curve of the ruled surface and the second is
the generating line (ruling) of the surface. We will see that the linear
trajectory $\xi ^{b}(\tau )=v^{b}\tau +d^{b}$ with $v^{a}v^{b}\eta _{ab}=1$
corresponds to the "free" electron and with $v^{a}v^{b}\eta _{ab}=0$
corresponds to its neutrino.

A general line ($\widehat{r}\in G(4,2)$) of $CP(3)$, generally intersects $d$
times the ruled surface and $d$ coincides with the algebraic degree of the
ruled surface. Two of these intersection points ($X^{n1}(\tau _{1},s_{1})\
,\ X^{n2}(\tau _{2},s_{2})$) determine the line and subsequently\ 
\begin{equation}
\begin{array}{l}
X^{ni}=%
\begin{pmatrix}
\lambda ^{A1}(\tau _{1},s_{1}) & \lambda ^{A2}(\tau _{2},s_{2}) \\ 
-i\xi _{B^{\prime }B}(\tau _{1})\lambda ^{B1} & -i\xi _{B^{\prime }B}(\tau
_{2})\lambda ^{B2}%
\end{pmatrix}%
=%
\begin{pmatrix}
\lambda ^{Ai} \\ 
-ir_{B^{\prime }B}\lambda ^{Bi}%
\end{pmatrix}
\\ 
\\ 
\lambda ^{Ai}(\tau _{i},s_{i})=%
\begin{pmatrix}
\lambda ^{0i}(\tau _{i},s_{i}) \\ 
\lambda ^{1i}(\tau _{i},s_{i})%
\end{pmatrix}%
\quad ,\quad i=1,2%
\end{array}
\label{c75}
\end{equation}

Using spinorial coordinates, the above relation takes the form\ 
\begin{equation}
\begin{array}{l}
\begin{pmatrix}
\lambda ^{Aj} \\ 
-ir_{A^{\prime }B}\lambda ^{Bj}%
\end{pmatrix}%
=%
\begin{pmatrix}
\lambda ^{Aj} \\ 
-i\xi _{A^{\prime }B}(\tau _{j})\lambda ^{Bj}%
\end{pmatrix}
\\ 
\\ 
(r_{A^{\prime }B}-\xi _{A^{\prime }B}(\tau _{j}))\lambda ^{Bj}=0%
\end{array}
\label{c76}
\end{equation}%
which are two homogeneous linear equations for every $j=1,2$. They admit a
(projectively) non-vanishing solution $\lambda ^{Bj}$ for every $j$, if\ 
\begin{equation}
\begin{array}{l}
\det (r_{A^{\prime }B}-\xi _{A^{\prime }B}(\tau ))=\det (\widehat{r}-%
\widehat{\xi }(\tau ))= \\ 
\qquad =(r^{a}-\xi ^{a})(r^{b}-\xi ^{b})\eta _{ab}=0 \\ 
\end{array}
\label{c77}
\end{equation}%
Every generally complex solution $\tau (r_{A^{\prime }A})$ of this equation
is replaced back into (\ref{c76}) and find the corresponding spinor $\lambda
^{A}$. For every column of $X^{ni}$ (point of $CP(3)$) we get a pair of
generally complex functions \ 
\begin{equation}
\begin{array}{l}
z^{0}(r)=\tau _{1}(r)\quad ,\quad z^{1}(r)=\frac{\lambda ^{11}(\tau _{1}(r))%
}{\lambda ^{01}(\tau _{1}(r))} \\ 
z^{\widetilde{0}}(r)=\tau _{2}(r)\quad ,\quad z^{\widetilde{1}}(r)=-\frac{%
\lambda ^{01}(\tau _{2}(r))}{\lambda ^{11}(\tau _{2}(r))} \\ 
\end{array}
\label{c78}
\end{equation}%
which may be assumed as the structure coordinates in the ambient complex
manifold of the LCR-structure. Notice that $z^{b}(r^{a})$ are generally
holomorphicfunctions. After the projection to the real LCR-submanifold, they
become proper structure coordinates.

The reader must not confuse the set of "straight" lines of $CP(3)$, which
are all the points $r_{A^{\prime }A}$ of the grassmannian manifold $G(4,2)$,
with the rulings of the ruled surface, i.e. the "straight" lines which
belong (as sets of points) to the ruled hypersurface of $CP(3)$, and which
are just the points of the complex trajectory $\xi _{A^{\prime }A}(\tau )$
in $G(4,2)$.

A geometric visualization of the above mathematical procedure is the
following. A point of the grassmannian manifold with projective coordinates $%
\widehat{r}$ determines a line of $CP(3)$. This line intersects the
hypersurface of $CP(3)$ at a number of points (equal to the polynomial
degree of the surface), which belong to different sheets of the surface of $%
CP(3)$. Every pair of intersection points with homogeneous coordinates $%
X^{ni}$ may be taken as the corresponding homogeneous coordinates of the
grassmannian point $\widehat{r}$. Hence every point $\widehat{r}\in G(4,2)$,
determines (and is determined by) two points $\xi ^{b}(\tau _{1})$ and $\xi
^{b}(\tau _{2})$ of the complex trajectory $\xi ^{b}(\tau )$ with two
corresponding spinors $\lambda ^{A1}(\tau _{1},s_{1})$ and $\lambda
^{A2}(\tau _{2},s_{2})$. Notice that the trajectory $\xi ^{b}(\tau )$
determines the algebraic subcurve of $CP(3)$, where the two sheets
intersect, $\lambda ^{A}(\tau _{1})=\lambda ^{A}(\tau _{2})$. This is valid
for any point $\widehat{r}$ of the ambient complex manifold. If the point
belongs in the real LCR-submanifold, i.e. if $r^{a}=x^{a}+iy^{a}(x)$, the
"observer" at the point $x^{a}$, has the local null system

\begin{equation}
\begin{array}{l}
L^{a}=\frac{1}{\sqrt{2}}\sigma _{A^{\prime }A}^{a}\overline{\lambda }%
^{A^{\prime }1}\lambda ^{A1},\quad N^{a}=\frac{1}{\sqrt{2}}\sigma
_{A^{\prime }A}^{a}\,\overline{\lambda }^{A^{\prime }2}\lambda ^{A2},\quad
M^{a}=\frac{1}{\sqrt{2}}\sigma _{A^{\prime }A}^{a}\,\overline{\lambda }%
^{A^{\prime }2}\lambda ^{A1} \\ 
\\ 
\epsilon _{AB}\lambda ^{A1}\lambda ^{B2}=1%
\end{array}
\label{c79}
\end{equation}%
and conditions (\ref{c69}) are formally "solved" by \ 
\begin{equation}
\begin{array}{l}
y^{a}=\frac{1}{2\sqrt{2}}[G_{22}N^{a}+G_{11}L^{a}-G_{12}M^{a}-\overline{%
G_{12}}\overline{M}^{a}] \\ 
\end{array}
\label{c79a}
\end{equation}%
The computation of $\lambda ^{Ai}$\ as functions of $r^{a}$ using the Kerr
condition $K(X^{mi})$, permits us to perturbatively compute $y^{a}$ as
functions of the real part $x^{a}$ of $r^{a}$. This procedure gives the
canonical form $y^{a}=$ $h^{a}(x)$ of the (totally real) lorentzian CR
submanifold expressed in the projective coordinates of $G(4,2)$. The
explicit form of $h^{a}(x)$\ is perturbatively implied by the precise
dependence of $G_{ij}(\overline{X^{mi}},X^{mj})$ from $X^{mj}$.

If we choose the $\xi ^{0}(\tau )=\tau $ normalization, we find

\begin{equation}
\begin{array}{l}
z^{0}(x)=r^{0}(x)-\sqrt{(r^{i}(x)-\xi ^{i}(z^{0}))^{2}}\quad ,\quad z^{1}(x)=%
\frac{r^{1}+ir^{2}-\xi ^{1}(z^{0})-i\xi ^{2}(z^{0})}{r^{0}+r^{3}-\xi
^{0}(z^{0})-\xi ^{3}(z^{0})} \\ 
z^{\widetilde{0}}(x)=r^{0}(x)+\sqrt{(r^{i}(x)-\xi ^{i}(z^{\widetilde{0}%
}))^{2}}\quad ,\quad z^{\widetilde{1}}(x)=\frac{r^{1}-ir^{2}-\xi ^{1}(z^{%
\widetilde{0}})+i\xi ^{2}(z^{\widetilde{0}})}{r^{0}-r^{3}-\xi ^{0}(z^{%
\widetilde{0}})+\xi ^{3}(z^{\widetilde{0}})} \\ 
\end{array}
\label{c80}
\end{equation}%
Hence, the left column of $X^{n1}$ provides the retarded coordinates $%
z^{\alpha }(x)$ and the right column $X^{n2}$ provides the advanced
coordinates $z^{\widetilde{\alpha }}(x)$. A (curved) LCR-tetrad is found as
usual by simply taking the differential forms of the structure coordinates
and using their reality conditions.

One can easily see that in the zero gravity approximation $y^{a}(x)=0$, the
structure coordinates (and the null tetrad) are completely determined by the
generally complex trajectory as we should expect from Kerr's theorem (in
Minkowski spacetime). In the first $\frac{1}{c}$ approximation, the
LCR-structure coordinates take the form

\begin{equation}
\begin{array}{l}
z^{0}(x)\simeq x^{0}-\frac{1}{c}\sqrt{(x^{i}-\xi ^{i}(x^{0}))^{2}}\quad
,\quad z^{1}(x)\simeq \frac{x^{1}+ix^{2}-\xi ^{1}(x^{0})-i\xi ^{2}(x^{0})}{%
x^{0}+x^{3}-\xi ^{0}(x^{0})-\xi ^{3}(x^{0})} \\ 
z^{\widetilde{0}}(x)\simeq x^{0}+\frac{1}{c}\sqrt{(x^{i}-\xi ^{i}(x^{0}))^{2}%
}\quad ,\quad z^{\widetilde{1}}(x)\simeq \frac{x^{1}-ix^{2}-\xi
^{1}(x^{0})+i\xi ^{2}(x^{0})}{x^{0}-x^{3}-\xi ^{0}(x^{0})+\xi ^{3}(x^{0})}
\\ 
\end{array}
\label{c81}
\end{equation}%
where the (dimensional) light velocity factor is made apparent in order to
reveal the newtonian approximation.

The points of the trajectory $\xi ^{i}(x^{0})$ are the singularities of the
structure coordinates. If the trajectory is real, the singularity is just a
curve in LCR-manifold. But if the trajectory is complex $\xi ^{j}(x^{0})=\xi
_{R}^{j}(x^{0})+i\xi _{I}^{j}(x^{0})$, the singularity is the surface

\begin{equation}
\begin{array}{l}
\tsum\limits_{j}(x^{i}-\xi ^{i}(x^{0}))^{2}=\tsum\limits_{j}[(x^{j}-\xi
_{R}^{j}(x^{0}))^{2}-(\xi _{I}^{j}(x^{0}))^{2}-2i(x^{j}-\xi
_{R}^{j}(x^{0}))\xi _{I}^{j}(x^{0})]=0 \\ 
\\ 
\tsum\limits_{j}[(x^{j}-\xi _{R}^{j}(x^{0}))^{2}-(\xi
_{I}^{j}(x^{0}))^{2}]=0\quad ,\quad \tsum\limits_{j}(x^{j}-\xi
_{R}^{j}(x^{0}))\xi _{I}^{j}(x^{0})=0%
\end{array}
\label{c82}
\end{equation}%
This is the well-known ring-like singularity of Kerr-type metrics in general
relativity. The imaginary part of the trajectory is related to the spin of
the LCR-structure. It is exactly this imaginary part that generates the
fermionic gyromagnetic ratio of the Kerr-Newman spacetime. Hence the complex
trajectory $\xi ^{a}(\tau )$ is the singular curve where two sheets of the
ruled surface of $CP(3)$ intersect.

\section{EMERGENCE\ OF\ EINSTEIN'S METRICS}

\setcounter{equation}{0}

The LCR-tetrad ($\ell ,m;n,\overline{m}$) may be identified with a geodetic
and shear-free null tetrad\cite{Flahe} of general relativity, which is
defined with the condition $\kappa =\sigma =0=\lambda =\nu $ in the
Newman-Penrose formalism. But the LCR-tetrad is not uniquely determined by
the LCR-structure. The tetrad-Weyl symmetry (\ref{c33}) implies that we may
only define a class of metrics and the corresponding self-dual 2-forms

\begin{equation}
\begin{array}{l}
\lbrack g_{\mu \nu }]=\ell _{\mu }n_{\nu }+\ell _{\nu }n_{\mu }-m_{\mu }%
\overline{m}_{\nu }-m_{\nu }\overline{m}_{\mu } \\ 
\\ 
\lbrack V_{1}]=\ell \wedge m\quad ,\quad \lbrack V_{2}]=n\wedge \overline{m}
\\ 
\lbrack V_{3}]=\ell \wedge n-m\wedge \overline{m}%
\end{array}
\label{e1}
\end{equation}%
Notice that \emph{only} metrics compatible with two geodetic and shear-free
null congruences can be defined in PCFT. The other metrics cannot have any
physical meaning. The observed in nature Kerr type metrics admit two
geodetic and shear-free null congruences and are defined through (\ref{e1}).
I want to point out that the additional Einstein's equations will be derived
by the BEGS procedure, which constitutes the QFT emergence in the context of
PCFT and will be described in section VII.

Recall that the Newman-Penrose formalism is essentially the Cartan moving
frame formalism adapted to a null tetrad, not necessarily geodetic and
shear-free. Its fundamental connection symbols are%
\begin{equation}
\begin{tabular}{|l|}
\hline
$\alpha =\frac{1}{4}[(\ell n\partial \overline{m})+(\ell \overline{m}%
\partial n)-(n\overline{m}\partial \ell )-2(m\overline{m}\partial \overline{m%
})]$ \\ \hline
$\beta =\frac{1}{4}[(\ell n\partial m)+(\ell m\partial n)-(nm\partial \ell
)-2(m\overline{m}\partial m)]$ \\ \hline
$\gamma =\frac{1}{4}[(nm\partial \overline{m})-(n\overline{m}\partial m)-(m%
\overline{m}\partial n)+2(\ell n\partial n)]$ \\ \hline
$\varepsilon =\frac{1}{4}[(\ell m\partial \overline{m})-(\ell \overline{m}%
\partial m)-(m\overline{m}\partial \ell )+2(\ell n\partial \ell )]$ \\ \hline
$\mu =-\frac{1}{2}[(m\overline{m}\partial n)+(nm\partial \overline{m})+(n%
\overline{m}\partial m)]$ \\ \hline
$\pi =\frac{1}{2}[(\ell n\partial \overline{m})-(n\overline{m}\partial \ell
)-(\ell \overline{m}\partial n)]$ \\ \hline
$\rho =\frac{1}{2}[(\ell \overline{m}\partial m)+(\ell m\partial \overline{m}%
)-(m\overline{m}\partial \ell )]$ \\ \hline
$\tau =\frac{1}{2}[(nm\partial \ell )+(\ell m\partial n)+(\ell n\partial m)]$
\\ \hline
$\kappa =(\ell m\partial \ell )\quad ,\quad \sigma =(\ell m\partial m)$ \\ 
\hline
$\nu =-(n\overline{m}\partial n)\quad ,\quad \lambda =-(n\overline{m}%
\partial \overline{m})$ \\ \hline
\end{tabular}
\label{e2}
\end{equation}%
where the symbols $(...)$ are constructed according to the rule of the
following example $(\ell m\partial n)=(\ell ^{\mu }m^{\nu }-\ell ^{\nu
}m^{\mu })(\partial _{\mu }n_{\nu })$, without using a metric or connection.
In general relativity, where the metric is the fundamental quantity, the
symmetry is not the tetrad-Weyl transformations but the well known internal
tetrad-Lorentz transformations\cite{Chand}, which generally \emph{do not
conserve} the LCR-structure conditions $\kappa =\sigma =0=\lambda =\nu $.

Hence the Newman-Penrose formalism is easily incorporated in PCFT, by simply
imposing the conditions $\kappa =\sigma =0=\lambda =\nu $. The general
Cartan relation for the independent covectors of a basis in the cotangent
vector basis are%
\begin{equation}
\begin{array}{l}
d\ell =-(\varepsilon +\overline{\varepsilon })\ell \wedge n+(\alpha +%
\overline{\beta }-\overline{\tau })\ell \wedge m+(\overline{\alpha }+\beta
-\tau )\ell \wedge \overline{m}- \\ 
\qquad -\overline{\kappa }n\wedge m-\kappa n\wedge \overline{m}+(\rho -%
\overline{\rho })m\wedge \overline{m} \\ 
dn=-(\gamma +\overline{\gamma })\ell \wedge n+\nu \ell \wedge m+\overline{%
\nu }\ell \wedge \overline{m}+(\pi -\alpha -\overline{\beta })n\wedge m+ \\ 
\qquad +(\overline{\pi }-\overline{\alpha }-\beta )n\wedge \overline{m}+(\mu
-\overline{\mu })m\wedge \overline{m} \\ 
dm=-(\tau +\overline{\pi })\ell \wedge n+(\gamma -\overline{\gamma }+%
\overline{\mu })\ell \wedge m+\overline{\lambda }\ell \wedge \overline{m}+
\\ 
\qquad \quad +(\varepsilon -\overline{\varepsilon }-\rho )n\wedge m-\sigma
n\wedge \overline{m}+(\beta -\overline{\alpha })m\wedge \overline{m}%
\end{array}
\label{e3}
\end{equation}%
without any reference to the metric. If we apply the $\kappa =\sigma
=0=\lambda =\nu $ conditions it takes the form of the LCR-structure
condition (\ref{c31}).\ That is $\ell _{\mu }dx^{\mu }$ and $n_{\mu }dx^{\mu
}$\ are not qualified as null vectors. They are just real vectors, which
combine with the complex vector $m_{\mu }dx^{\mu }$\ to form a basis. The
corresponding relations in the tangent space is

\begin{equation}
\begin{array}{l}
\lbrack \ell ^{\mu }\partial _{\mu }\ ,\ m^{\nu }\partial _{\nu }]=(%
\overline{\pi }-\overline{\alpha }-\beta )\ell ^{\rho }\partial _{\rho
}-\kappa n^{\rho }\partial _{\rho }+(\overline{\rho }+\varepsilon -\overline{%
\varepsilon })m^{\rho }\partial _{\rho }+\sigma \overline{m}^{\rho }\partial
_{\rho } \\ 
\lbrack n^{\mu }\partial _{\mu }\ ,\ \overline{m}^{\nu }\partial _{\nu
}]=\nu \ell ^{\rho }\partial _{\rho }+(\alpha +\overline{\beta }-\overline{%
\tau })n^{\rho }\partial _{\rho }+(\overline{\gamma }-\gamma -\overline{\mu }%
)\overline{m}^{\rho }\partial _{\rho }+\lambda m^{\rho }\partial _{\rho } \\ 
\end{array}
\label{e4}
\end{equation}%
which becomes the LCR-condition (\ref{c32}) if $\kappa =\sigma =0=\lambda
=\nu $ is imposed. Hence we may cast the LCR-tetrad in the form of
Newman-Penrose formalism and use it freely, but we can apply the tetrad-Weyl
transformation only for the particular LCR-integrable tetrad.

Notice that the real quantity \ 
\begin{equation}
\begin{array}{l}
i(\rho -\overline{\rho })(\mu -\overline{\mu })(\tau +\overline{\pi })(%
\overline{\tau }+\pi )\ell \wedge m\wedge n\wedge \overline{m} \\ 
\end{array}
\label{e5}
\end{equation}%
is an LCR-structure invariant, but it is not invariant under a
tetrad-Lorentz transformation. Another interesting property is that the
complex scalars of the conformal tensor ($\Psi _{0},\Psi _{1},\Psi _{2},\Psi
_{3},\Psi _{4}$) which correspond to a LCR-tetrad satisfy the condition $%
\Psi _{0}=0=\Psi _{4}$. Hence the LCR-tetrad is determined\cite{Chand} by
two solutions of the quartic polynomial \ 
\begin{equation}
\begin{array}{l}
\Psi _{0}^{\prime }+4b\Psi _{1}^{\prime }+6b^{2}\Psi _{2}^{\prime
}+4b^{3}\Psi _{3}^{\prime }+b^{4}\Psi _{4}^{\prime }=0 \\ 
\end{array}
\label{e6}
\end{equation}%
where $\Psi _{i}^{\prime }$ are the conformal tensor components relative to
any other null tetrad of the metric. That is the class of LCR-tetrads is
determined from the principal null directions of the conformal tensor. On
the other hand the above relation poses a maximum number of LCR-tetrads of
the solitonic gravitating LCR-structures which will be identified below with
the leptonic generations (it will be described in section IV).

The LCR-structure condition (\ref{c25}) determines a class of Kaehler
metrics and symplectic forms \ 
\begin{equation}
\begin{array}{l}
ds^{2}=2\frac{\partial ^{2}\det (\rho _{ij})}{\partial z^{a}\partial 
\overline{z^{b}}}dz^{a}d\overline{z^{b}}\quad ,\quad \omega =2i\frac{%
\partial ^{2}\det (\rho _{ij})}{\partial z^{a}\partial \overline{z^{b}}}%
dz^{a}\wedge d\overline{z^{b}} \\ 
\end{array}
\label{e7}
\end{equation}%
in the ambient complex manifold. Like in the 3-dimensional CR-structure it
is not unique nor the induced Einstein metric which is analogous to (\ref{e1}%
). Because a given LCR-structure may be implied by any $\rho _{ij}^{\prime
}=f_{ij}\rho _{ij}$ without summation but the same dependence on the $%
z^{\alpha }$and $z^{\widetilde{\beta }}$ structure variables. In fact the
symplectic form vanishes implying that the LCR-manifold is a lagrangian
submanifold. We will consider below the two symmetric "algebraically flat"
LCR-structures and their corresponding symmetric metrics.

Recall that the "algebraically flat" LCR-structures (\ref{c66}) are
essentially determined by the Kerr function. In the unbounded realization we
consider the reducible quadric polynomial $X^{0}X^{1}=0$. Then the
homogeneous coordinates are 
\begin{equation}
\begin{array}{l}
X^{ni}=%
\begin{pmatrix}
1 & 0 \\ 
0 & 1 \\ 
-iz^{0} & iz^{\widetilde{1}} \\ 
iz^{1} & -iz^{\widetilde{0}}%
\end{pmatrix}%
=:%
\begin{pmatrix}
I \\ 
-i\widehat{x}%
\end{pmatrix}
\\ 
\end{array}
\label{e8}
\end{equation}%
where we have introduced the convenient structure coordinates so that the
structure conditions are 
\begin{equation}
\begin{array}{l}
\rho _{ij}=\overline{X^{ni}}E_{nm}^{U}X^{mj}=0 \\ 
\\ 
\rho _{11}=-i(z^{0}-\overline{z^{0}})=0 \\ 
\rho _{12}=i(z^{\widetilde{1}}-\overline{z^{1}})=0 \\ 
\rho _{22}=i(z^{\widetilde{0}}-\overline{z^{\widetilde{0}}})=0%
\end{array}
\label{e9}
\end{equation}%
and the dependence of the structure coordinates with the cartesian
coordinates is 
\begin{equation}
\begin{array}{l}
z^{0}:=i\frac{X^{21}}{X^{01}}=x^{0}-x^{3}\quad ,\quad z^{1}:=-i\frac{X^{31}}{%
X^{01}}=x^{1}+ix^{2} \\ 
\\ 
z^{\widetilde{0}}:=i\frac{X^{32}}{X^{12}}=x^{0}+x^{3}\quad ,\quad z^{%
\widetilde{1}}:=-i\frac{X^{22}}{X^{12}}=x^{1}-ix^{2}%
\end{array}
\label{e10}
\end{equation}%
which we call "light-cone" LCR-structure. The general LCR-tetrad is 
\begin{equation}
\begin{array}{l}
\ell _{\mu }dx^{\mu }=\Lambda dz^{0}|_{M}\quad ,\quad m_{\mu }dx^{\mu
}=Mdz^{1}|_{M} \\ 
\\ 
n_{\mu }dx^{\mu }=Ndz^{\widetilde{0}}|_{M}\quad ,\quad \overline{m}_{\mu
}dx^{\mu }=\overline{M}dz^{\widetilde{1}}|_{M}%
\end{array}
\label{e11}
\end{equation}%
with $\Lambda $, $N$ and $M$ the arbitrary non vanishing factors of the
tetrad-Weyl transformation. Notice that this LCR-structure is degenerate
with vanishing all its relative invariants, $\Phi _{j}=0$.

After a straight forward computation we find for the ambient Kaehler
manifold 
\begin{equation}
\begin{array}{l}
ds^{2}=2(dz^{0}d\overline{z^{\widetilde{0}}}+dz^{\widetilde{0}}d\overline{%
z^{0}}-dz^{1}d\overline{z^{1}}-dz^{\widetilde{1}}d\overline{z^{\widetilde{1}}%
}) \\ 
\\ 
\omega =2i(dz^{0}\wedge d\overline{z^{\widetilde{0}}}+dz^{\widetilde{0}%
}\wedge d\overline{z^{0}}-dz^{1}\wedge d\overline{z^{1}}-dz^{\widetilde{1}%
}\wedge d\overline{z^{\widetilde{1}}})%
\end{array}
\label{e12}
\end{equation}%
and the induced manifold has $\omega |_{M}=0$ as expected and the induced
metric 
\begin{equation}
\begin{array}{l}
ds^{2}|_{M}=(dx^{0})^{2}-(dx^{1})^{2}-(dx^{2})^{2}-(dx^{3})^{2} \\ 
\end{array}
\label{e13}
\end{equation}%
properly normalized. It is the symmetric Minkowski spacetime with vanishing
curvature.

Apparently the present "light-cone" LCR-structure is regular $\forall x^{\mu
}\in 
\mathbb{R}
^{4}$. But singularities may be hidden at the "infinities". Therefore this
problem should be treated in the largest LCR-manifold $%
\mathbb{R}
\times SU(2)$ being the boundary of the bounded realization of the $SU(2,2)$
classical domain. This is formally done by transcribing the Kerr polynomial
condition $X^{0}X^{1}=0$ into the bounded homogeneous coordinates 
\begin{equation}
\begin{array}{l}
X^{11}=\frac{1}{\sqrt{2}}(Y^{11}+Y^{31})=0 \\ 
\\ 
X^{02}=\frac{1}{\sqrt{2}}(Y^{02}+Y^{22})=0%
\end{array}
\label{e14}
\end{equation}%
In the normalization $Y^{01}=1=Y^{12}$ we have 
\begin{equation}
\begin{array}{l}
\widehat{w}=%
\begin{pmatrix}
Y^{\prime 21} & -Y^{\prime 02} \\ 
-Y^{\prime 11} & Y^{\prime 02}%
\end{pmatrix}%
\begin{pmatrix}
1 & Y^{\prime 02} \\ 
Y^{\prime 11} & 1%
\end{pmatrix}%
^{-1}= \\ 
\qquad =e^{i\tau }\left( 
\begin{array}{cc}
\cos \rho +i\sin \rho \cos \sigma & -i\sin \rho \sin \sigma \ e^{-i\chi } \\ 
-i\sin \rho \sin \sigma \ e^{i\chi } & \cos \rho -i\sin \rho \cos \sigma%
\end{array}%
\right) \\ 
\end{array}
\label{e15}
\end{equation}%
which implies that the singularities occur when 
\begin{equation}
\begin{array}{l}
\det 
\begin{pmatrix}
w_{11}+1 & w_{12} \\ 
w_{21} & w_{22}+1%
\end{pmatrix}%
=0 \\ 
\multicolumn{1}{c}{\Downarrow} \\ 
\cos \tau +\cos \rho =0%
\end{array}
\label{e16}
\end{equation}%
which are the future and past celestial spheres. In fact the singularities
happen when the two roots of the quadratic Kerr polynomial $%
(Y^{0}+Y^{2})(Y^{1}+Y^{3})$ vanish. Apparently the "light-cone"
LCR-structure has singularities (hidden) at the conformal infinities of $%
\mathbb{R}
^{4}$.

In the unbounded realization we consider the reducible quadric polynomial $%
Y^{0}Y^{1}=0$. Then the homogeneous coordinates are 
\begin{equation}
\begin{array}{l}
Y^{nj}=%
\begin{pmatrix}
1 & 0 \\ 
0 & 1 \\ 
w^{0} & -iw^{\widetilde{1}} \\ 
-iw^{1} & w^{\widetilde{0}}%
\end{pmatrix}%
=:\left( 
\begin{array}{c}
I \\ 
\widehat{w}%
\end{array}%
\right) \\ 
\end{array}
\label{e17}
\end{equation}%
where the proper structure coordinates are fixed. Then the LCR-structure
embedding conditions are 
\begin{equation}
\begin{array}{l}
\rho _{ij}=\overline{Y^{ni}}E_{nm}^{B}Y^{mj}=0 \\ 
\\ 
\rho _{11}=w^{0}\overline{w^{0}}+w^{1}\overline{w^{1}}-1=0 \\ 
\rho _{12}=\overline{w^{0}}w^{\widetilde{1}}-w^{\widetilde{0}}\overline{w^{1}%
}=0 \\ 
\rho _{22}=w^{\widetilde{0}}\overline{w^{\widetilde{0}}}+w^{\widetilde{1}}%
\overline{w^{\widetilde{1}}}-1=0%
\end{array}
\label{e18}
\end{equation}%
This LCR-structure will be called "natural $U(2)$" structure. One may check
that the Taub-NUT spacetime\cite{GrPo}\cite{Flahe} admits a LCR-structure
equivalent to the present one. From (\ref{e17}) we find the LCR-structure
coordinates 
\begin{equation}
\begin{array}{l}
w^{0}=(\cos \rho +i\sin \rho \cos \sigma )e^{i\tau }\quad ,\quad w^{1}=\sin
\rho \sin \sigma \ e^{i\chi }e^{i\tau } \\ 
\\ 
w^{\widetilde{0}}=(\cos \rho -i\sin \rho \cos \sigma )e^{i\tau }\quad ,\quad
w^{\widetilde{1}}=\sin \rho \sin \sigma \ e^{-i\chi }e^{i\tau } \\ 
\end{array}
\label{e19}
\end{equation}%
and the LCR-tetrad \ 
\begin{equation}
\begin{array}{l}
\ell =d\tau +\cos \sigma d\rho -\sin \rho \cos \rho \sin \sigma d\sigma
+\sin ^{2}\rho \sin ^{2}\sigma d\chi \\ 
n=d\tau -\cos \sigma d\rho +\sin \rho \cos \rho \sin \sigma d\theta -\sin
^{2}\rho \sin ^{2}\sigma d\chi \\ 
m=-e^{i\chi }[\sin \sigma d\rho +\sin \rho (\cos \rho \cos \sigma +i\sin
\rho )d\sigma + \\ 
\qquad +(\cos \rho +i\sin \rho \cos \sigma )\sin \rho \sin \sigma d\chi%
\end{array}
\label{e20}
\end{equation}%
which satisfies the conditions \ 
\begin{equation}
\begin{array}{l}
\begin{pmatrix}
\ell & \overline{m} \\ 
m & n%
\end{pmatrix}%
:=\widehat{e}\quad ,\quad d\widehat{e}-i\widehat{e}\wedge \widehat{e}=0 \\ 
\multicolumn{1}{c}{\Downarrow} \\ 
d\ell =im\wedge \overline{m}\quad ,\quad dn=-im\wedge \overline{m}\quad
,\quad dm=i(\ell -n)\wedge m%
\end{array}
\label{e21}
\end{equation}%
Notice that this LCR-structure has relative invariants $\Phi _{1}=i$, $\Phi
_{2}=-i$ and $\Phi _{3}=0$. Hence the "natural $U(2)$" structure is not
equivalent to the degenerate "light-cone" LCR-structure. Besides, the
possibility to cast the LCR-tetrad (with $\Phi _{1}\neq 0$, $\Phi _{2}\neq 0$%
) into the above hermitian matrix form with vanishing Cartan $U(2)$
curvature provides the possibility to interpret it as the electroweak $U(2)$
potential with vanishing field strength. This will be verified in section V
by simply computing this field strength (Cartan curvature) in the electron
LCR-structure solitonic sector, where its electromagnetic potential is
derived.

We will now see how the "natural $U(2)$" looks like in a projective chart of
the unbounded realization. The $Y^{0}Y^{1}=0$ Kerr polynomial takes the form 
$(X^{0}+X^{2})(X^{1}+X^{3})=0$ and the LCR-conditions 
\begin{equation}
\begin{array}{l}
X^{\dag }E^{(U)}X=0 \\ 
\\ 
X^{nj}:=%
\begin{pmatrix}
1 & -z^{\widetilde{1}} \\ 
z^{1} & 1 \\ 
-i(z^{0}+i) & z^{\widetilde{1}} \\ 
-z^{1} & -i(z^{\widetilde{0}}+i)%
\end{pmatrix}%
=:%
\begin{pmatrix}
\lambda \\ 
-i\widehat{x}\lambda%
\end{pmatrix}%
\end{array}
\label{e22}
\end{equation}%
are satisfied with $\widehat{x}^{\dag }=\widehat{x}$ and the explicit forms
of the structure coordinates are 
\begin{equation}
\begin{array}{l}
z^{1}=\frac{x^{1}+ix^{2}}{i+x^{0}+x^{3}}\quad ,\quad z^{0}=x^{0}-x^{3}-i-%
\frac{(x^{1})^{2}+(x^{2})^{2}}{i+x^{0}+x^{3}} \\ 
z^{\widetilde{1}}=-\frac{x^{1}-ix^{2}}{i+x^{0}-x^{3}}\quad ,\quad z^{%
\widetilde{0}}=x^{0}+x^{3}-i-\frac{(x^{1})^{2}+(x^{2})^{2}}{i+x^{0}-x^{3}}
\\ 
\\ 
z^{0}-\overline{z^{0}}+2i(1-z^{1}\overline{z^{1}})=0\ ,\quad z^{\widetilde{0}%
}-\overline{z^{\widetilde{0}}}+2i(1-z^{\widetilde{1}}\overline{z^{\widetilde{%
1}}})=0\ ,\quad z^{\widetilde{1}}\overline{z^{0}}+z^{\widetilde{0}}\overline{%
z^{1}}=0%
\end{array}
\label{e23}
\end{equation}%
There is no singularity in $%
\mathbb{R}
^{4}$, because $\det \lambda \neq 0$.

After a straight forward computation we find the metric and the symplectic
form for the ambient Kaehler manifold. The induced metric is \ 
\begin{equation}
\begin{array}{l}
ds^{2}|_{M}=(d\tau )^{2}-(d\rho )^{2}-\sin ^{2}\rho (d\theta )^{2}-\sin
^{2}\rho \sin ^{2}\theta (d\varphi )^{2} \\ 
\end{array}
\label{e24}
\end{equation}%
properly normalized. Recall that $\rho _{ij}$ is determined up to conformal
factors. Hence the above metric is equivalent to the symmetric de Sitter
metric%
\begin{equation}
\begin{array}{l}
ds_{S}^{2}=(dt)^{2}-T_{0}^{2}\cosh ^{2}\frac{t}{T_{0}}[(d\rho )^{2}+\sin
^{2}\rho (d\sigma )^{2}+\sin ^{2}\rho \sin ^{2}\sigma (d\chi )^{2}]= \\ 
\qquad =T_{0}^{2}\cosh ^{2}\frac{t}{T_{0}}[ds^{2}|_{M}] \\ 
\\ 
\tau =2\arctan (e^{\frac{t}{T_{0}}})\quad ,\quad T_{0}:=\sqrt{\frac{3}{%
\Lambda }}%
\end{array}
\label{e24a}
\end{equation}%
with $\rho \in \lbrack 0,2\pi )$. It is fixed being the general symmetric
space which covers the entire covering spacetime $R\times SU(2)$. We will
use this result (or rather the inverse argument) to show the existence of
dark energy. The existence of dark energy suggests us to assume that the
empty (symmetric) universe is the de Sitter metric and not its conformally
equivalent Minkowski metric.

\section{LEPTONIC\ LCR-STRUCTURES}

\setcounter{equation}{0}

Identifying the Poincar\'{e} subgroup of the $SU(2,2)$ group of the
classical domain with the observed Poincar\'{e} symmetry in nature opens up
the possibility to look for automorphic LCR-structures relative to time
translation and z-rotation, i.e. static and axially symmetric
LCR-structures. Trying to impose the additional dilation we find a singular
LCR-structure. We will essentially work in two steps. First we will look for
automorphic algebraically flat LCR-structures and after we will find a
curved deformation using the Kerr-Schild ansatz adapted to the LCR-tetrad.

\subsection{Free electron (and positron) LCR-structure}

Recall that an algebraically flat LCR-structure is completely determined by
the roots of the Kerr function. We need two roots to fix the flat null
tetrad (\ref{c79}). On the other hand higher than 4-degree polynomials are
excluded by the condition (\ref{e6}) on the non-vanishing conformal tensor.
Therefore we will study quadric, cubic and quartic polynomials.

The infinitesimal z-rotation, time translation and dilation are 
\begin{equation}
\begin{array}{l}
\delta _{z}X^{0i}=-i\frac{\varepsilon ^{12}}{2}X^{0i}\ ,\ \delta _{z}X^{1i}=i%
\frac{\varepsilon ^{12}}{2}X^{1i}\ ,\ \delta _{z}X^{2i}=-i\frac{\varepsilon
^{12}}{2}X^{2i}\ ,\ \delta _{z}X^{3i}=i\frac{\varepsilon ^{12}}{2}X^{3i} \\ 
\\ 
\delta _{t}X^{0i}=0\ ,\ \delta _{t}X^{1i}=0\ ,\ \delta
_{t}X^{2i}=-i\varepsilon ^{0}X^{0i}\ ,\ \delta _{t}X^{3i}=-i\varepsilon
^{0}X^{1i} \\ 
\\ 
\delta _{d}X^{0i}=-\frac{\varepsilon }{2}X^{0i}\ ,\ \delta _{d}X^{1i}=-\frac{%
\varepsilon }{2}X^{1i}\ ,\ \delta _{d}X^{2i}=\frac{\varepsilon }{2}X^{2i}\
,\ \delta _{d}X^{3i}=\frac{\varepsilon }{2}X^{3i}%
\end{array}
\label{l1}
\end{equation}%
The automorphic quadratic Kerr polynomial $K(X^{m})=\tsum%
\limits_{m,n}A_{mn}X^{m}X^{n}$ relative to the z-rotation has the form%
\begin{equation}
\begin{array}{l}
K_{z}=A_{01}X^{0}X^{1}+A_{03}X^{0}X^{3}+A_{12}X^{1}X^{2}+A_{23}X^{2}X^{3} \\ 
\end{array}
\label{l2}
\end{equation}%
The two solutions of this quadratic polynomial are generally time dependent.
The time translation automorphism restricts it into the form%
\begin{equation}
\begin{array}{l}
K_{zt}=A_{01}X^{0}X^{1}+A_{12}(X^{1}X^{2}-X^{0}X^{3}) \\ 
\end{array}
\label{l3}
\end{equation}%
A singular point $X^{n}\in CP(3)$ is solution of%
\begin{equation}
\begin{array}{l}
\frac{\partial K_{zt}}{\partial X^{n}}=(A_{01}X^{1}-A_{12}X^{3},\
A_{01}X^{0}+A_{12}X^{2},\ A_{12}X^{1},\ -A_{12}X^{0})=\overrightarrow{0} \\ 
\end{array}
\label{l4}
\end{equation}%
If $A_{01}\neq 0\neq A_{12}$ we find $X^{n}=0$, $\forall n$. But this
solution of the homogeneous coordinates does not represent a point of $CP(3)$%
. Hence this is a regular surface of $CP(3)$ and gives the algebraically
regular "free electron" LCR-structure. If $A_{01}\neq 0=A_{12}$ we find the
"light-cone" LCR-structure. If $A_{01}=0\neq A_{12}$ 
\begin{equation}
\begin{array}{l}
K_{ztd}=(X^{1}X^{2}-X^{0}X^{3}) \\ 
\end{array}
\label{l5}
\end{equation}%
we find an algebraically regular LCR-structure, which is also automorphic
relative to the dilation transformation implying that the parameter $A_{01}$
(which will become the rotating parameter) essentially breaks the dilation
transformation. But the corresponding dilation invariant LCR-manifold is
singular, because the $\ell ^{\mu }$ and $n^{\mu }$ congruences have \
infinite concentrations into finite points. We will investigate this effect
below in comparison with the "free electron" LCR-structure. The
non-existence of the automorphic LCR-structure relative to all the three (%
\ref{l1}) automorphisms is directly related to the "spontaneous" breaking of
the $SU(2,2)$ group to its Poincar\'{e} subgroup as we will see in the
section VII of the SM derivation.

If we make a general translation and after a boost transformation, the
automorphic Kerr polynomial (\ref{l3}) gives the general flatprint form of
the "free electron" LCR-structure. But I find more impressive to start from
the linear trajectory $\xi ^{a}=v^{a}\tau +c^{a}$ with $v^{a}$ a general
real velocity and $c^{a}$ generally complex. After eliminating the complex
parameter $\tau $ using (\ref{c74}), the Kerr polynomial takes the form \ 
\begin{equation}
\begin{array}{l}
K(X^{n})=iX^{0}X^{0}[(v^{0}-v^{3})(c^{1}+ic^{2})-(v^{1}+iv^{2})(c^{0}-c^{3})]+
\\ 
\qquad +iX^{0}X^{1}[(v^{0}+v^{3})(c^{0}-c^{3})-(v^{0}-v^{3})(c^{0}+c^{3})+
\\ 
\qquad +(v^{1}+iv^{2})(c^{1}-ic^{2})-(v^{1}-iv^{2})(c^{1}+ic^{2})]- \\ 
\qquad -X^{0}X^{2}(v^{1}+iv^{2})-X^{0}X^{3}(v^{0}-v^{3})+ \\ 
\qquad +iX^{1}X^{1}[(v^{1}-iv^{2})(c^{0}+c^{3})-(v^{0}+v^{3})(c^{1}-ic^{2})]+
\\ 
\qquad +X^{1}X^{2}(v^{0}+v^{3})+X^{1}X^{3}(v^{1}-iv^{2}) \\ 
\end{array}
\label{l6}
\end{equation}%
This is the most general quadratic Kerr polynomial which incorporates all
the parameters of the Poincar\'{e} representation. The singular points of
this quadratic surface satisfy the relations $\partial _{n}K(X^{m})=0$ and $%
X^{n}\neq \overrightarrow{0}$. We finally find that there are the following
two cases \ 
\begin{equation}
\begin{array}{l}
1st:\quad If\ \ v^{a}v^{b}\eta _{ab}\neq 0\quad the\ surface\ is\ irreducible
\\ 
\\ 
2nd:\quad If\ \ v^{a}v^{b}\eta _{ab}=0\quad the\ surface\ is\ reducible \\ 
\end{array}
\label{l7}
\end{equation}%
The first case gives the electron and positron LCR-solitons and the second
reducible surface gives the left-handed chiral part of the neutrino. The
electron LCR-structure is determined with an irreducible quadratic
polynomial and the neutrino LCR-structure is determined with the
corresponding reducible quadratic polynomial.

Returning back to the automorphic Kerr polynomial $K_{zt}$, we will first
find the algebraically flat LCR-structure with 
\begin{equation}
\begin{array}{l}
X^{mi}=%
\begin{pmatrix}
1 & -z^{\widetilde{1}} \\ 
z^{1} & 1 \\ 
-i(z^{0}-ia) & i(z^{\widetilde{0}}-ia)z^{\widetilde{1}} \\ 
-i(z^{0}+ia)z^{1} & -i(z^{\widetilde{0}}+ia)%
\end{pmatrix}
\\ 
\end{array}
\label{l8}
\end{equation}%
where ($z^{\alpha };z^{\widetilde{\beta }}$) are now the structure
coordinates, which imply simple formula. The LCR-structure coordinates are%
\begin{equation}
\begin{array}{l}
z^{0}=t-r+ia\cos \theta \quad ,\quad z^{1}=e^{i\varphi }\tan \frac{\theta }{2%
} \\ 
z^{\widetilde{0}}=t+r-ia\cos \theta \quad ,\quad z^{\widetilde{1}}=\frac{r+ia%
}{r-ia}e^{-i\varphi }\tan \frac{\theta }{2} \\ 
\end{array}
\label{l9}
\end{equation}%
from which we find the LCR-tetrad%
\begin{equation}
\begin{array}{l}
\ell _{\mu }dx^{\mu }=\Lambda \lbrack dt-dr-a\sin ^{2}\theta d\varphi ] \\ 
n_{\mu }dx^{\mu }=N[dt+\frac{r^{2}+2a^{2}\cos ^{2}\theta -a^{2}}{r^{2}+a^{2}}%
dr-a\sin ^{2}\theta \ d\varphi ] \\ 
m_{\mu }dx^{\mu }=M[-ia\sin \theta \ (dt-dr)+(r^{2}+a^{2}\cos ^{2}\theta
)d\theta + \\ 
\qquad \qquad +i\sin \theta (r^{2}+a^{2})d\varphi ] \\ 
\end{array}
\label{l10}
\end{equation}%
where the tetrad-Weyl factors are not determined, as expected. They are
determined by simply imposing that the tetrad gives the symmetric Minkowski
metric. But for that, we have to find first the relation of the cartesian
coordinates $x^{\mu }$ with the present convenient coordinates ($t,r,\theta
,\varphi $).

The general relation between the projective coordinates and the homogeneous
coordinates of $G(4,2)$ is found by simply inverting their definition
formula. We finally find%
\begin{equation}
\begin{array}{l}
r^{0}=i\frac{(X^{01}X^{32}-X^{31}X^{02})+(X^{21}X^{12}-X^{11}X^{22})}{%
2(X^{01}X^{12}-X^{11}X^{02})} \\ 
r^{1}=i\frac{(X^{11}X^{32}-X^{31}X^{12})+(X^{21}X^{02}-X^{01}X^{22})}{%
2(X^{01}X^{12}-X^{11}X^{02})} \\ 
r^{2}=\frac{(X^{11}X^{32}-X^{31}X^{12})-(X^{21}X^{02}-X^{01}X^{22})}{%
2(X^{01}X^{12}-X^{11}X^{02})} \\ 
r^{3}=i\frac{(X^{01}X^{32}-X^{31}X^{02})-(X^{21}X^{12}-X^{11}X^{22})}{%
2(X^{01}X^{12}-X^{11}X^{02})}%
\end{array}
\label{l11}
\end{equation}%
We already know that the imaginary part of $r^{b}=x^{b}+iy^{b}$ determines
the gravitational "dressing", because the "flatness" condition implies $%
y^{b}=0$. The Minkowski coordinates $x^{\mu }$ are related with ($t,r,\theta
,\varphi $) via the relation%
\begin{equation}
\begin{array}{l}
x^{0}=t \\ 
x^{1}+ix^{2}=(r-ia)\sin \theta e^{i\varphi }=\sqrt{r^{2}+a^{2}}e^{-i\arctan 
\frac{a}{r}}\sin \theta e^{i\varphi } \\ 
x^{3}=r\cos \theta \\ 
\\ 
r^{4}-[(x^{1})^{2}+(x^{2})^{2}+(x^{3})^{2}-a^{2}]r^{2}-a^{2}(x^{3})^{2}=0 \\ 
\cos \theta =\frac{x^{3}}{r},\quad \sin \theta =\sqrt{\frac{%
(x^{1})^{2}+(x^{2})^{2}}{r^{2}+a^{2}}},\quad \frac{(x^{1})^{2}+(x^{2})^{2}}{%
r^{2}+a^{2}}+\frac{(x^{3})^{2}}{r^{2}}=1%
\end{array}
\label{l12}
\end{equation}%
We finally find the geodetic and shear free null tetrad 
\begin{equation}
\begin{array}{l}
L_{\mu }dx^{\mu }=[dt-dr-a\sin ^{2}\theta d\varphi ] \\ 
N_{\mu }dx^{\mu }=\frac{r^{2}+a^{2}}{2(r^{2}+a^{2}\cos ^{2}\theta )}[dt+%
\frac{r^{2}+2a^{2}\cos ^{2}\theta -a^{2}}{r^{2}+a^{2}}dr-a\sin ^{2}\theta \
d\varphi ] \\ 
M_{\mu }dx^{\mu }=\frac{-1}{\sqrt{2}(r+ia\cos \theta )}[-ia\sin \theta \
(dt-dr)+(r^{2}+a^{2}\cos ^{2}\theta )d\theta + \\ 
\qquad \qquad +i\sin \theta (r^{2}+a^{2})d\varphi ] \\ 
\end{array}
\label{l13}
\end{equation}%
of the Minkowski metric.

A static axisymmetric gravitating tetrad is found with the "Kerr-Schild"
ansatz adapted to the LCR-structure formalism%
\begin{equation}
\begin{array}{l}
\ell _{\mu }=L_{\mu }\quad ,\quad m_{\mu }=M_{\mu }\quad ,\quad n_{\mu
}=N_{\mu }+\frac{h(r)}{2(r^{2}+a^{2}\cos ^{2}\theta )}\ L_{\mu } \\ 
\end{array}
\label{l14}
\end{equation}%
I want to point out that we find the same static LCR-structure looking for a
general LCR-structure admitting time translation and axisymmetric
automorphisms applied on the regular coordinates.

With the above definition of the coordinates ($t,r,\theta ,\varphi $), the
structure coordinates have the form

\begin{equation}
\begin{array}{l}
z^{0}=t-r+ia\cos \theta \quad ,\quad z^{1}=e^{i\varphi }\tan \frac{\theta }{2%
} \\ 
\\ 
z^{\widetilde{0}}=t+r-ia\cos \theta -2f_{1}\quad ,\quad z^{\widetilde{1}}=%
\frac{r+ia}{r-ia}\ e^{2iaf_{2}}\ e^{-i\varphi }\tan \frac{\theta }{2}%
\end{array}
\label{l15}
\end{equation}%
where the two new functions are%
\begin{equation}
\begin{array}{l}
f_{1}(r)=\int \frac{h}{r^{2}+a^{2}+h}\ dr\quad ,\quad f_{2}(r)=\int \frac{h}{%
(r^{2}+a^{2}+h)(r^{2}+a^{2})}\ dr \\ 
\end{array}
\label{l16}
\end{equation}

The Newman-Penrose spin coefficients are found to be%
\begin{equation}
\begin{tabular}{|l|}
\hline
$\alpha =\frac{ia(1+\sin ^{2}\theta )-r\cos \theta }{2\sqrt{2}\sin \theta \
(r-ia\cos \theta )^{2}}\quad ,\quad \beta =\frac{\cos \theta }{2\sqrt{2}\sin
\theta \ (r+ia\cos \theta )}$ \\ \hline
$\gamma =-\frac{a^{2}+iar\cos \theta +h}{2\rho ^{2}\ (r-ia\cos \theta )}+%
\frac{h^{\prime }}{4\rho ^{2}}\quad ,\quad \varepsilon =0$ \\ \hline
$\mu =-\frac{r^{2}+a^{2}+h}{2\rho ^{2}\ (r-ia\cos \theta )}\quad ,\quad \pi =%
\frac{ia\sin \theta }{\sqrt{2}(r-ia\cos \theta )^{2}}$ \\ \hline
$\rho =-\frac{1}{r-ia\cos \theta }\quad ,\quad \tau =-\frac{ia\sin \theta }{%
\sqrt{2}\rho ^{2}}$ \\ \hline
$\kappa =0\quad ,\quad \sigma =0\quad ,\quad \nu =0\quad ,\quad \lambda =0$
\\ \hline
\end{tabular}
\label{l17}
\end{equation}%
which will be useful for our computations. Recall that the Kerr-Newman
spacetime has $h(r)=-2Mr+e^{2}$. In this case the integrals are%
\begin{equation}
\begin{array}{l}
f_{1}(r)=\int \frac{-2Mr+e^{2}}{r^{2}+a^{2}-2Mr+e^{2}}\ dr=-M\ln \frac{%
|\Delta |}{r1}+\frac{2M^{2}-e^{2}}{\Theta }\arctan \frac{\Theta }{r-M} \\ 
\\ 
f_{2}(r)=\int \frac{-2Mr+e^{2}}{(r^{2}+a^{2}-2Mr+e^{2})(r^{2}+a^{2})}\ dr=%
\frac{1}{2ia}\ln [r_{2}\frac{r-ia}{r+ia}(\frac{r-M+i\Theta }{r-M-i\Theta })^{%
\frac{a}{\Theta }}] \\ 
\\ 
\Delta :=r^{2}+a^{2}-2Mr+e^{2}\quad ,\quad \Theta :=\sqrt{a^{2}+e^{2}-M^{2}}%
\end{array}
\label{l18}
\end{equation}%
and the structure coordinates of the "free electron" LCR-manifold are

\begin{equation}
\begin{array}{l}
z^{0}=t-r+ia\cos \theta \quad ,\quad z^{1}=e^{i\varphi }\tan \frac{\theta }{2%
} \\ 
z^{\widetilde{0}}=t+r-ia\cos \theta +2M\ln \frac{|\Delta |}{r_{1}}+\frac{%
2(e^{2}-2M^{2})}{\Theta }\arctan \frac{\Theta }{r-M} \\ 
z^{\widetilde{1}}=r_{2}(\frac{r-M+i\Theta }{r-M-i\Theta })^{\frac{a}{\Theta }%
}\ e^{-i\varphi }\tan \frac{\theta }{2}%
\end{array}
\label{l19}
\end{equation}%
in the coordinates ($t,r,\theta ,\varphi $). The constants $r_{1}$ and $%
r_{2} $ are normalization constants. Notice the singularities in the ambient
complex manifold occur at the two complex values of $r=M\pm i\Theta $. It is
well known to general relativists that this choice of tetrad-Weyl factors
preserve the electromagnetic current, and the energy-momentum and angular
momentum currents. Identifying the "electron" LCR-manifold with the above
Kerr-Newman manifold we fix the electromagnetic and gravitational
"dressings" of the electron to be%
\begin{equation}
\begin{array}{l}
A=\frac{qr}{4\pi (r^{2}+a^{2}\cos ^{2}\theta )}(dt-dr-a\sin ^{2}\theta
d\varphi ) \\ 
\\ 
h_{\mu \nu }=\frac{-2Mr+e^{2}}{(r^{2}+a^{2}\cos ^{2}\theta )}\ L_{\mu
}N_{\nu }%
\end{array}
\label{l19a}
\end{equation}

The general form (\ref{c25}) of the embedding of the LCR-manifold in the
ambient complex manifold may be viewed as a deformation of the 3-dimensional
CR-manifold $\rho _{11}(\overline{z^{\alpha }},z^{\beta })=0$ through a
formal \textbf{anti-}meromorphic transformation

\begin{equation}
\begin{array}{l}
z^{\widetilde{\beta }}=f^{\widetilde{\beta }}(\overline{z^{\alpha }};s) \\ 
\end{array}
\label{l20}
\end{equation}%
which generalizes the trivial transformation of the degenerate
LCR-structure. In the present electron LCR-structure this deformation takes
the form

\begin{equation}
\begin{array}{l}
z^{\widetilde{0}}=\overline{z^{0}}+2(r-f_{1}) \\ 
\\ 
z^{\widetilde{1}}=r_{2}\overline{z^{1}}(\frac{r-M+i\Theta }{r-M-i\Theta })^{%
\frac{a}{\Theta }}%
\end{array}
\label{l21}
\end{equation}%
where the deformation parameter is the real variable $r$.

The static axially symmetric LCR-structure (identified with the electron) is
stable, because all its relative invariants 
\begin{equation}
\begin{array}{l}
\Phi _{1}=\frac{\rho -\overline{\rho }}{i}=\frac{-2a\cos \theta }{%
r^{2}+a^{2}\cos ^{2}\theta } \\ 
\\ 
\Phi _{2}=\frac{\mu -\overline{\mu }}{i}=-\frac{(r^{2}+a^{2}+h)a\cos \theta 
}{(r^{2}+a^{2}\cos ^{2}\theta )^{2}} \\ 
\\ 
\Phi _{3}=-(\tau +\overline{\pi })=\frac{2iar\sin \theta }{\sqrt{2}(r+ia\cos
\theta )^{2}(r-ia\cos \theta )}%
\end{array}
\label{l22}
\end{equation}%
do not vanish.

The positron LCR-structure is the conjugate of the electron one 
\begin{equation}
\begin{array}{l}
z^{\prime 0}=\overline{z^{0}}=t-r-ia\cos \theta \quad ,\quad z^{\prime 1}=%
\overline{z^{1}}=e^{-i\varphi }\tan \frac{\theta }{2} \\ 
\\ 
z^{\prime \widetilde{0}}=\overline{z^{\widetilde{0}}}=t+r+ia\cos \theta
+2M\ln \frac{|\Delta |}{r_{1}}+\frac{2(e^{2}-2M^{2})}{\Theta }\arctan \frac{%
\Theta }{r-M} \\ 
\\ 
z^{\prime \widetilde{1}}=\overline{z^{\widetilde{1}}}=r_{2}(\frac{%
r-M-i\Theta }{r-M+i\Theta })^{\frac{a}{\Theta }}\ e^{i\varphi }\tan \frac{%
\theta }{2}%
\end{array}
\label{l22a}
\end{equation}%
which has the LCR-tetrad 
\begin{equation}
\begin{array}{l}
\ell _{\mu }^{\prime }=\ell _{\mu }\quad ,\quad m_{\mu }^{\prime }=\overline{%
m}_{\mu }\quad ,\quad n_{\mu }^{\prime }=n_{\mu }\quad ,\quad \overline{m}%
_{\mu }^{\prime }=m_{\mu } \\ 
\end{array}
\label{l22b}
\end{equation}%
which has the same gravitational dressing but opposite charge
electromagnetic dressing.

\subsection{Solving the electron naked singularity "problem"}

We saw that the LCR-structure implies Einstein's general relativity and the
"free electron" LCR-structure is related to the Kerr-Newman metric. This
metric admits two geodetic and shear free null congruences, which are
related with the LCR-tetrad. It also admits two commuting killing vectors,
which are identified with the time-translation and $z$-rotation generators
of the Poincar\'{e} group. Carter's\cite{Cart} discovery that the
gyromagnetic ratio of the Kerr-Newman manifold is fermionic, which implies
that it has gyromagnetic ratio $g=2$ \cite{NW1974} shocked the community of
general relativists. Many tried to identify the Kerr-Newman spacetime with
the electron without success, because the electron constants imply the
existence of a naked singularity in the Kerr-Newman spacetime.

The electron mass $M_{e}$, charge $e^{2}$ and spin parameter $a$ have the
values%
\begin{equation}
\begin{array}{l}
M=\frac{M_{e}}{M_{P}}=4.18\ast 10^{-23} \\ 
e^{2}=\frac{q^{2}}{4\pi \varepsilon _{0}\hbar c}=\frac{1}{137} \\ 
a=\frac{\hbar }{2M_{e}}=2.09\ast 10^{23} \\ 
\\ 
a^{2}>>e^{2}>>M^{2}%
\end{array}
\label{l23}
\end{equation}%
Hence $a^{2}+e^{2}-M^{2}>0$, and the electron metric has an essential naked
singularity, which is not permitted in riemannian geometry. This is a
problem for general relativity, because its fundamental quantity, the
metric, does not "see" the algebraic structure. It is known (and well
described in many books of general relativity) that its analytic extension
has two sheets $x^{b}$ and $x^{\prime b}$ which are determined by the two
roots%
\begin{equation}
\begin{array}{l}
r=\pm \left\{ \frac{(x^{1})^{2}+(x^{2})^{2}+(x^{3})^{2}-a^{2}}{2}+\sqrt{[%
\frac{(x^{1})^{2}+(x^{2})^{2}+(x^{3})^{2}-a^{2}}{2}]^{2}+a^{2}(x^{3})^{2}}%
\right\} ^{\frac{1}{2}} \\ 
\end{array}
\label{l24}
\end{equation}%
These two surfaces constitute the boundary $U(2)$ of the bounded realization
of the $SU(2,2)$ classical domain and their correspondence is the well known
Cayley transformation, which has been described through the formula (\ref%
{c60}-\ref{c64}). The spinorial electron naked singularity in $U(2)$
universe can be properly incorporated in PCFT, while it is rejected as
"unphysical" by the riemannian formalism. In the context of the unbounded
realization this may be studied using the ray tracing in the flatprint "free
electron" LCR-structure (\ref{l9}) and the cartesian coordinates (\ref{l12}%
). Then $L^{\mu }\partial _{\mu }z^{\alpha }=0$ implies that the outgoing $%
L^{\mu }$ null integral curves (rays) are determined by the surfaces%
\begin{equation}
\begin{array}{l}
s_{1}:=t-r\quad ,\quad s_{2}:=\theta \quad ,\quad s_{3}:=\varphi \\ 
\end{array}
\label{l25}
\end{equation}%
Assuming the coordinates ($r,s_{1},s_{2},s_{3}$), which have the property ($%
0,s_{1},\frac{\pi }{2},s_{3}$) to be on the caustic. In this caustic
coordinate system the LCR-rays are traced by the relation%
\begin{equation}
\begin{array}{l}
x_{L}^{0}(r)=s_{1}+r \\ 
x_{L}^{1}(r)=(r\cos \varphi +a\sin \varphi )\sin \theta \\ 
x_{L}^{2}(r)=(r\sin \varphi -a\cos \varphi )\sin \theta \\ 
x_{L}^{3}(r)=r\cos \theta \\ 
\\ 
Jacobian=[r^{2}+a^{2}\cos ^{2}\theta ]\sin \theta%
\end{array}
\label{l26}
\end{equation}%
The source of the LCR-rays are at $r=0$, i.e.%
\begin{equation}
\begin{array}{l}
x_{L}^{0}(0)=s_{1} \\ 
x_{L}^{1}(0)=a\sin \varphi \sin \theta \\ 
x_{L}^{2}(0)=-a\cos \varphi \sin \theta \\ 
x_{L}^{3}(0)=0%
\end{array}
\label{l27}
\end{equation}%
which is the disc found above. Notice that the rays with $s_{2}:=\theta \neq 
\frac{\pi }{2}$ pass through the disc.

The $N^{\mu }\partial _{\mu }z^{\widetilde{\alpha }}=0$ implies that its
incoming $N^{\mu }$ rays are determined by the surfaces%
\begin{equation}
\begin{array}{l}
s_{1}^{\prime }:=t+r\quad ,\quad s_{2}^{\prime }:=\theta \quad ,\quad
s_{3}^{\prime }:=\varphi +\arctan \frac{2ar}{a^{2}-r^{2}} \\ 
\end{array}
\label{l28}
\end{equation}%
Then we find the congruence%
\begin{equation}
\begin{array}{l}
x_{N}^{0}(r)=s_{1}^{\prime }-r \\ 
x_{N}^{1}(r)=(r\cos s_{3}^{\prime }-a\sin s_{3}^{\prime })\sin \theta \\ 
x_{N}^{2}(r)=(r\sin s_{3}^{\prime }+a\cos s_{3}^{\prime })\sin \theta \\ 
x_{N}^{3}(r)=r\cos \theta \\ 
\\ 
Jacobian=[r^{2}+a^{2}\cos ^{2}\theta ]\sin \theta%
\end{array}
\label{l29}
\end{equation}

We will now show that the origin of the essential singularity of the Kerr
manifold is the intersection of the two sheets of the static electron
regular quadric (in the unbounded Siegel realization) (\ref{l3}) of $CP^{3}$%
. In the flatprint case and after the projection to the real surface $%
\mathbb{R}
^{4}$ we have 
\begin{equation}
\begin{array}{l}
X^{0}=1\quad ,\quad X^{1}=\lambda \quad ,\quad
X^{2}=-i[(x^{0}-x^{3})-(x^{1}-ix^{2})\lambda ] \\ 
X^{3}=-i[-(x^{1}+ix^{2})+(x^{0}+x^{3})\lambda ] \\ 
\\ 
K_{zt}(X^{n})=X^{1}X^{2}-X^{0}X^{3}+2aX^{0}X^{1}=0%
\end{array}
\label{l30}
\end{equation}%
The two solutions (sheets) of the above quadric are 
\begin{equation}
\begin{array}{l}
\lambda _{1(2)}=\frac{-x^{3}+ia\ \pm \sqrt{\Delta }}{x^{1}-ix^{2}}\quad
,\quad \Delta =(x^{1})^{2}+(x^{2})^{2}+(x^{3})^{2}-a^{2}-2iax^{3} \\ 
\end{array}
\label{l30a}
\end{equation}%
The intersection of the two sheets (which occur for $\lambda _{1}=\lambda
_{2}$) is the singularity ring of the "free electron" structure. Notice that
the quadratic surface is algebraically regular and the intersection of the
two branches is implied by the projection. The points of the algebraic
intersection curve (the branch curve) of the (regular) quadric of $CP^{3}$
are regular points like any other point of the quadric.

The bounded realization of a flat LCR-manifold is $U(2)$, which is covered
by two $%
\mathbb{R}
^{4}$ sheets through the Cayley $2\rightarrow 1$ transformation with the $%
x_{+}^{\mu }$ (\ref{c63}) written in the present context%
\begin{equation}
\begin{array}{l}
For\ s:=R_{0}\frac{\sin \rho }{\cos \tau +\cos \rho }>0 \\ 
\\ 
x^{0}=T_{0}\frac{\sin \tau }{\cos \tau +\cos \rho } \\ 
x^{1}+ix^{2}=R_{0}\frac{\sin \rho }{\cos \tau +\cos \rho }\sin \sigma \
e^{i\chi } \\ 
x^{3}=R_{0}\frac{\sin \rho }{\cos \tau +\cos \rho }\cos \sigma%
\end{array}
\label{l31}
\end{equation}%
and $x_{-}^{\mu }$ (\ref{c64}) the second $%
\mathbb{R}
^{4}$ is identified with $s<0$,%
\begin{equation}
\begin{array}{l}
For\ s:=R_{0}\frac{\sin \rho }{\cos \tau +\cos \rho }<0 \\ 
\\ 
x^{\prime 0}=T_{0}\frac{\sin \tau }{\cos \tau +\cos \rho } \\ 
x^{\prime 1}+ix^{\prime 2}=-R_{0}\frac{\sin \rho }{\cos \tau +\cos \rho }%
\sin \sigma \ e^{i\chi } \\ 
x^{\prime 3}=-R_{0}\frac{\sin \rho }{\cos \tau +\cos \rho }\cos \sigma%
\end{array}
\label{l32}
\end{equation}%
The constants $T_{0}$ and $R_{0}$\ are related to the time and space sizes.
Notice that this is the Penrose artificial compactification of the Minkowski
spacetime, but in the context of PCFT, this is implied by the formalism
itself. In the case of the Penrose artificial compactification these two
sheets $s\gtrless 0$ communicate through the scri+ and scri- infinities. In
the case of the electron flatprint LCR-structure, these two sheets
communicate through the glued two discs $(x^{1})^{2}+(x^{2})^{2}<a^{2}$ too,
because we may assume%
\begin{equation}
\begin{array}{l}
r=+\left\{ \frac{s^{2}-a^{2}}{2}+\sqrt{[\frac{s^{2}-a^{2}}{2}%
]^{2}+a^{2}(x^{3})^{2}}\right\} ^{\frac{1}{2}}\ for\ s>0 \\ 
r=-\left\{ \frac{s^{2}-a^{2}}{2}+\sqrt{[\frac{s^{2}-a^{2}}{2}%
]^{2}+a^{2}(x^{3})^{2}}\right\} ^{\frac{1}{2}}\ for\ s<0%
\end{array}
\label{l33}
\end{equation}%
Notice that in the identified region (the disc for both sheets) $r=0$ in
both sheets. That is, $r=0$ occurs at $x^{3}=0$ and $s^{2}\leq a^{2}$ for
both sheets $s\gtrless 0$.

The two LCR-congruences $L^{\mu }=\frac{dx_{L}^{\mu }}{dr}$ and $N^{\mu }=%
\frac{dx_{N}^{\mu }}{dr}$\ of the flatprint electron LCR-manifold can be
easily implied from the calculations of the previous section. The starting
idea is that the structure coordinates $z^{\alpha }(x)$ provide the three
invariants ($s_{1},s_{2},s_{3}$) along the ray, which\ label the $L$-ray $%
x_{L}^{\mu }(r)$, and the structure coordinates $z^{\widetilde{\alpha }}(x)$
provide the invariants ($s_{1}^{\prime },s_{2}^{\prime },s_{3}^{\prime }$),
which\ label the $N$-ray $x_{n}^{\mu }(r)$. Hence we simply have the same
forms, but we let $r\in (-\infty ,+\infty )$ and at $r=0$ we pass from the
one $%
\mathbb{R}
^{4}$ sheet to the other.

A complete visualization of the rays $w_{L,N}(r;s_{1},s_{2},s_{3})\in U(2)$
taking $r\in (-\infty ,+\infty )$ can be done in the bounded realization of
the flatprint electron (as the $U(2)$ boundary of the $SU(2,2)$\ classical
domain). From (\ref{c58}) we find the relation 
\begin{equation}
\begin{array}{l}
Y^{0}=\frac{1}{\sqrt{2}}(X^{0}+X^{2})\quad ,\quad Y^{1}=\frac{1}{\sqrt{2}}%
(X^{1}+X^{3}) \\ 
\\ 
Y^{2}=\frac{1}{\sqrt{2}}(X^{0}-X^{2})\quad ,\quad Y^{3}=\frac{1}{\sqrt{2}}%
(X^{1}-X^{3})%
\end{array}
\label{l34}
\end{equation}%
between the bounded $Y^{ni}$ and unbounded $X^{ni}$ homogeneous coordinates
and from (\ref{l8}) we find%
\begin{equation}
\begin{array}{l}
Y^{mi}=\frac{1}{\sqrt{2}}%
\begin{pmatrix}
1-i(z^{0}-ia) & (-1+i(z^{\widetilde{0}}-ia))z^{\widetilde{1}} \\ 
(1-i(z^{0}+ia))z^{1} & 1-i(z^{\widetilde{0}}+ia) \\ 
1+i(z^{0}-ia) & -(1+i(z^{\widetilde{0}}-ia))z^{\widetilde{1}} \\ 
(1+i(z^{0}+ia))z^{1} & 1+i(z^{\widetilde{0}}+ia)%
\end{pmatrix}
\\ 
\end{array}
\label{l35}
\end{equation}%
The relation between the projective coordinates and the homogeneous
coordinates of $G(4,2)$ is found by simply inverting their definition
formula in the bounded realization. We finally find the relation 
\begin{equation}
\begin{array}{l}
w_{11}=\frac{Y^{21}Y^{12}-Y^{11}Y^{22}}{Y^{01}Y^{12}-Y^{11}Y^{02}}\quad
,\quad w_{12}=\frac{Y^{01}Y^{22}-Y^{21}Y^{02}}{Y^{01}Y^{12}-Y^{11}Y^{02}} \\ 
\\ 
w_{21}=\frac{Y^{31}Y^{12}-Y^{11}Y^{32}}{Y^{01}Y^{12}-Y^{11}Y^{02}}\quad
,\quad w_{12}=\frac{Y^{01}Y^{32}-Y^{31}Y^{02}}{Y^{01}Y^{12}-Y^{11}Y^{02}}%
\end{array}
\label{l36}
\end{equation}%
between the bounded projective $\widehat{w}\in U(2)$ and homogeneous $Y^{ni}$
coordinates. In principle we may compute the rays $\widehat{w}%
_{L,N}(r;s_{1},s_{2},s_{3})\in U(2)$ in the complete bounded universe $U(2)$%
, but it seems to be complicated. But the intersection (touching) of the two 
$%
\mathbb{R}
^{4}$ sheets in $U(2)$ coordinates can be computed by simply making the
Cayley transformation of the cartesian form of the ring singularity. Then we
find that in ($\tau ,\rho ,\sigma ,\chi $) coordinates the ring singularity
(the caustic of the congruence) is 
\begin{equation}
\begin{array}{l}
\sigma =\frac{\pi }{2}\quad ,\quad R_{0}^{2}\frac{\sin ^{2}\rho }{(\cos \tau
+\cos \rho )^{2}}\leq a^{2} \\ 
-\pi <\rho <\pi \quad ,\quad -\pi <\tau <\pi \\ 
\end{array}
\label{l37}
\end{equation}

We have already identified (\ref{l22a}-\ref{l22b}) the positron with the
conjugate LCR-structure of the electron. In order to geometrically
distinguish the positron from the electron we have to consider the
(unbounded) two $%
\mathbb{R}
^{4}$-sheets realization, where the $\ell ^{\mu }$ and $n^{\mu }$
congruences have opposite directions passing through the ring singularity.
In a given $%
\mathbb{R}
^{4}$-sheet, assuming that the electron is the LCR-manifold with the $\ell
^{\mu }$ congruence outgoing and the $n^{\mu }$ congruence ingoing, the
positron is the LCR-manifold with the $\ell ^{\prime \mu }$ congruence
outgoing too but with opposite twist $a$ and the $n^{\prime \mu }$
congruence ingoing too but with opposite twist too. Hence the outgoing
(retarded) and ingoing (advanced) character of their LCR-rays is preserved.

In order to make clear the importance of the spin parameter $a$ to the
regularity of "free electron" LCR-structure, we will describe in details the
simple quadratic surface (\ref{l5}) of $CP(3)$. It is the quadratic Kerr
polynomial which is symmetric relative to z-rotations, time translations and
dilations. Apparently the quadric of $CP(3)$ is algebraically regular as we
have already showed in (\ref{l4}). But, we will see that its reduction to
the real LCR-manifold is going to generate a non-permitting singularity. The
(\ref{l30}) quadratic polynomial now becomes 
\begin{equation}
\begin{array}{l}
K_{ztd}(X^{n})=X^{1}X^{2}-X^{0}X^{3}=0 \\ 
\\ 
X^{0}=1\quad ,\quad X^{1}=\lambda \quad ,\quad
X^{2}=-i[(x^{0}-x^{3})-(x^{1}-ix^{2})\lambda ] \\ 
X^{3}=-i[-(x^{1}+ix^{2})+(x^{0}+x^{3})\lambda ]%
\end{array}
\label{l38}
\end{equation}%
The Kerr polynomial and its two solutions are 
\begin{equation}
\begin{array}{l}
(x^{1}-ix^{2})\lambda ^{2}+2x^{3}\lambda -(x^{1}+ix^{2})=0 \\ 
\\ 
\lambda _{1(2)}=\frac{-x^{3}\pm \sqrt{\Delta }}{x^{1}-ix^{2}}\quad ,\quad
\Delta =(x^{1})^{2}+(x^{2})^{2}+(x^{3})^{2} \\ 
\end{array}
\label{l39}
\end{equation}%
where $\lambda _{1(2)}$\ are the two values of $\lambda $ on the two sheets
of the quadric. The intersection of the two sheets of $CP(3)$ becomes 
\begin{equation}
\begin{array}{l}
\Delta =(x^{1})^{2}+(x^{2})^{2}+(x^{3})^{2}=0 \\ 
\end{array}
\label{l40}
\end{equation}%
a point.

The preceding calculations are described as follows in the algebraic
picture. The two points 
\begin{equation}
\begin{array}{l}
X^{n1}=%
\begin{pmatrix}
1 \\ 
\lambda _{1}(x) \\ 
-i[x^{0}-x^{3}-(x^{1}-ix^{2})\lambda _{1}] \\ 
-i[-(x^{1}-ix^{2})+(x^{0}+x^{3})\lambda _{1}]%
\end{pmatrix}
\\ 
\\ 
X^{n2}=%
\begin{pmatrix}
1 \\ 
\lambda _{2}(x) \\ 
-i[x^{0}-x^{3}-(x^{1}-ix^{2})\lambda _{2}] \\ 
-i[-(x^{1}-ix^{2})+(x^{0}+x^{3})\lambda _{2}]%
\end{pmatrix}%
\end{array}
\label{l41}
\end{equation}%
of the above quadric belong to different sheets created by the considered
projection and they correspond to a point $x^{a}$ of the characteristic
boundary $%
\mathbb{R}
^{4}$ of the "upper half-plane" domain of $G(4,2)$. If $\det (\lambda
^{Ai})=\lambda _{2}-\lambda _{1}=0$, the two points coincide, that is, the
projection line is tangent to the quadric.

In the present case the branch curve is reduced to a point $\overrightarrow{x%
}=\overrightarrow{0}$. Therefore the branch cut should be reduced to a line
joining $\overrightarrow{0}$ and $\infty $. The structure coordinates are 
\begin{equation}
\begin{array}{l}
z^{0}=iX^{21}=x^{0}-|\overrightarrow{x}|\quad ,\quad z^{1}=\lambda _{1}=%
\frac{|\overrightarrow{x}|-x^{3}}{x^{1}-ix^{2}}=\frac{x^{1}+ix^{2}}{|%
\overrightarrow{x}|+x^{3}} \\ 
z^{\widetilde{0}}=iX^{22}=x^{0}+|\overrightarrow{x}|\quad ,\quad z^{%
\widetilde{1}}=\frac{-1}{\lambda _{2}}=\frac{x^{1}-ix^{2}}{|\overrightarrow{x%
}|+x^{3}}=\overline{z^{1}} \\ 
\end{array}
\label{l42}
\end{equation}%
and the derived tetrad is 
\begin{equation}
\begin{array}{l}
\ell _{\mu }dx^{\mu }=\Lambda \lbrack |\overrightarrow{x}|dx^{0}-%
\overrightarrow{x}\cdot d\overrightarrow{x}] \\ 
m_{\mu }dx^{\mu }=M[(|\overrightarrow{x}|(x^{3}+|\overrightarrow{x}%
|)-(x^{1}+ix^{2})x^{1})dx^{1}+ \\ 
\qquad +(i|\overrightarrow{x}|(x^{3}+|\overrightarrow{x}%
|)-(x^{1}+ix^{2})x^{2})dx^{2}-(x^{1}+ix^{2})(x^{3}+|\overrightarrow{x}%
|)dx^{3}] \\ 
n_{\mu }dx^{\mu }=N[|\overrightarrow{x}|dx^{0}+\overrightarrow{x}\cdot d%
\overrightarrow{x}] \\ 
\\ 
\ell \wedge m\wedge n\wedge \overline{m}=-4i|\overrightarrow{x}|^{4}(x^{3}+|%
\overrightarrow{x}|)^{2}dx^{0}\wedge dx^{1}\wedge dx^{2}\wedge dx^{3}\neq
0,\quad \forall x^{\mu }\in 
\mathbb{R}
^{4}-\{%
\mathbb{R}
_{-}\}%
\end{array}
\label{l43}
\end{equation}%
where the tetrad-Weyl factors are arbitrary as expected. The tetrad is
singular (because it cannot be a basis of the tangent space) in the negative
z-axis, where the branch cut in the algebraic quadric is reduced.

We can make the same calculations in the compact realization of complete
spacetime. In this coordinate patch, $Y^{n}$ is given by the linear
transformation (\ref{l34}). Then the Kerr polynomial has the same form \ 
\begin{equation}
\begin{array}{l}
K_{ztd}^{B}(Y^{n})=Y^{1}Y^{2}-Y^{0}Y^{3} \\ 
\end{array}
\label{l44}
\end{equation}%
as in the unbounded realization. This quadratic LCR-structure has the same
form in the bounded and unbounded realizations. In the bounded realization
the homogeneous coordinates of $G(4,2)$ have the form 
\begin{equation}
\begin{array}{l}
Y^{ni}=%
\begin{pmatrix}
Y^{01} & Y^{02} \\ 
Y^{11} & Y^{12} \\ 
Y^{21} & Y^{22} \\ 
Y^{31} & Y^{32}%
\end{pmatrix}%
=\left( 
\begin{array}{c}
k \\ 
\widehat{w}k%
\end{array}%
\right) \\ 
\\ 
\det k\neq 0\quad ,\quad \widehat{w}\in U(2) \\ 
\end{array}
\label{l45}
\end{equation}%
where the elements of the 2$\times $2 matrix $\widehat{w}$ are the
projective coordinates. Hence we will substitute 
\begin{equation}
\begin{array}{l}
Y^{n}=%
\begin{pmatrix}
1 \\ 
k \\ 
w_{00}+w_{01}k \\ 
w_{10}+w_{11}k%
\end{pmatrix}
\\ 
\end{array}
\label{l46}
\end{equation}%
in the new (bounded) form of the Kerr quadric. Then it takes the form

\begin{equation}
\begin{array}{l}
k^{2}w_{01}+k(w_{00}-w_{11})-w_{10}=0 \\ 
\end{array}
\label{l47}
\end{equation}%
with singularities at the points 
\begin{equation}
\begin{array}{l}
w=e^{i\tau }I\quad and\quad w=-e^{i\tau }I \\ 
\multicolumn{1}{c}{\Downarrow} \\ 
\rho =0\quad and\quad \rho =\pi%
\end{array}
\label{l48}
\end{equation}

The $\ell ^{\mu }$ and $n^{\mu }$ rays, which pass from these two points are
determined by $s_{1}:=\frac{\sin \tau \mp \sin \rho }{\cos \tau +\cos \rho }=%
\frac{\sin \tau }{\cos \tau \ \pm 1}$ , $s_{2}:=\sigma $ and $s_{3}:=\chi $
respectively. Notice that the concentration of rays at the above points of $%
SU(2)$ stop at two points of the regular manifold $U(2)$. Hence this
LCR-structure should be rejected, because it is not defined in the entire $%
U(2)$\ universe. I want to point out that all the spherically symmetric
metrics of general relativity are compatible with this singular
LCR-structure. Hence the fact, that the $K_{ztd}$ symmetric soliton does not
exist for PCFT, its physical space is going to restrict the $SU(2,2)$ rigged
Hilbert space down to the rigged Hilbert space of the Poincar\'{e}
representations as it will be described in section VII.

\subsection{Free neutrino LCR-structure}

We know that the neutrino is a massless particle therefore we will search
for the corresponding automorphisms. The form of the infinitesimal
z-rotation is the same. But the infinitesimal translation has to be taken
for $\delta (x^{0}-x^{3})$, which is

\begin{equation}
\begin{array}{l}
\delta X^{0}=0\quad ,\quad \delta X^{1}=0 \\ 
\delta X^{2}=-i\varepsilon ^{03}X^{0}\quad ,\quad \delta X^{3}=0 \\ 
\multicolumn{1}{c}{\Downarrow} \\ 
K_{zt^{\prime }}=(A_{01}X^{1}+A_{03}X^{3})X^{0}%
\end{array}
\label{l48a}
\end{equation}%
A linear trajectory $\xi ^{a}=v^{a}\tau +c^{a}$ of a free massless particle
implies a reducible quadratic Kerr polynomial (\ref{i6}), which has an
analogous form. The two linear polynomials are arranged such that

\begin{equation}
\begin{array}{l}
X^{11}+aX^{31}=0\quad ,\quad X^{02}=0 \\ 
\end{array}
\label{l49}
\end{equation}%
for the left and right homogeneous coordinates. Then in $CP(3)$, the
intersections of the line $\widehat{x}$ with the first and second planes are

\begin{equation}
\begin{array}{l}
\widehat{x}:=iX_{2}X_{1}^{-1} \\ 
\\ 
-a(x^{1}+ix^{2})X^{01}+[a(x^{0}+x^{3})+i]X^{11}=0 \\ 
X^{02}=0 \\ 
\end{array}
\label{l50}
\end{equation}%
Notice that this LCR-structure is regular in the present affine chart ($\det
(X_{1})\neq 0$), because $[a(x^{0}+x^{3})+i]\neq 0$.

The convenient structure coordinates are

\begin{equation}
\begin{array}{l}
X^{ni}=%
\begin{pmatrix}
1 & 0 \\ 
az^{1} & 1 \\ 
-iz^{0} & z^{\widetilde{1}}(1+iaz^{\widetilde{0}}) \\ 
-z^{1} & -iz^{\widetilde{0}}%
\end{pmatrix}
\\ 
\end{array}
\label{l51}
\end{equation}%
The LCR-structure conditions ($X^{\dag }E_{U}X=0$) are

\begin{equation}
\begin{array}{l}
\frac{z^{0}-\overline{z^{0}}}{2i}-az^{1}\overline{z^{1}}=0 \\ 
z^{\widetilde{1}}-\overline{z^{1}}=0\quad ,\quad \frac{z^{\widetilde{0}}-%
\overline{z^{\widetilde{0}}}}{2i}=0 \\ 
\end{array}
\label{l52}
\end{equation}%
and the structure coordinates are

\begin{equation}
\begin{array}{l}
z^{0}=x^{0}-x^{3}-a\frac{(x^{1})^{2}+(x^{2})^{2}}{a(x^{0}+x^{3})+i}\quad
,\quad z^{1}=-\frac{X^{31}}{X^{01}}=\frac{x^{1}+ix^{2}}{a(x^{0}+x^{3})+i} \\ 
z^{\widetilde{0}}=x^{0}+x^{3}\quad ,\quad z^{\widetilde{1}}=\frac{%
x^{1}-ix^{2}}{a(x^{0}+x^{3})-i}=\overline{z^{1}} \\ 
\end{array}
\label{l53}
\end{equation}%
The flat LCR-tetrad and its differential conditions are found following the
general computational rules

\begin{equation}
\begin{array}{l}
L=du-ia\overline{z^{1}}dz^{1}+iaz^{1}d\overline{z^{1}}\quad ,\quad
M=dz^{1}\quad ,\quad N=dv \\ 
\\ 
u:=\frac{z^{0}+\overline{z^{0}}}{2}\quad ,\quad v:=\frac{z^{\widetilde{0}}+%
\overline{z^{\widetilde{0}}}}{2} \\ 
\\ 
dL=2iaM\wedge \overline{M}\quad ,\quad dM=0\quad ,\quad dN=0%
\end{array}
\label{l54}
\end{equation}

The integral curves $x_{L}^{\mu }(s)$ of the causal vector $L^{\mu }\partial
_{\mu }$ are found using the definition of the projective coordinates and
the fact that $z^{0}$ and $z^{1}$ are constant along the curves, because $%
L^{\mu }\partial _{\mu }z^{0}=0=L^{\mu }\partial _{\mu }z^{1}$. We assume
the ray (affine) parameter $s:=z^{\widetilde{0}}=x^{0}+x^{3}$ and the ray
labels $s_{1}=\func{Re}(z^{0})$, $s_{2}=\func{Re}(z^{1})$, and $s_{3}=\func{%
Im}(z^{1})$. Then the jacobian is

\begin{equation}
\begin{array}{l}
ds\wedge ds_{1}\wedge ds_{2}\wedge ds_{3}=\frac{8}{a^{2}(x^{0}+x^{3})^{2}+1}%
dx^{0}\wedge dx^{1}\wedge dx^{2}\wedge dx^{3} \\ 
\end{array}
\label{l55}
\end{equation}%
Hence the caustic of the causal rays is at infinity as we algebraically
found it above.

The coordinate singularity is hidden at infinity, like in the "light-cone"
LCR-structure. In order to "see" it, we have to work in the patch $\det
X_{2}\neq 0$, where

\begin{equation}
\begin{array}{l}
\widehat{x}^{\prime }:=-iX_{1}(X_{2})^{-1}=\widehat{x}^{-1} \\ 
\\ 
-(x^{\prime 1}+ix^{\prime 2})X^{21}+[(x^{\prime 0}+x^{\prime 3})-ia]X^{31}=0
\\ 
(x^{\prime 0}-x^{\prime 3})X^{22}-(x^{\prime 1}-ix^{\prime 2})X^{32}=0 \\ 
\end{array}
\label{l56}
\end{equation}%
The singularity occurs at

\begin{equation}
\begin{array}{l}
\det X_{2}=0 \\ 
\multicolumn{1}{c}{\Downarrow} \\ 
x^{\prime 0}-x^{\prime 3}=0\quad ,\quad x^{\prime 1}=0=x^{\prime 2} \\ 
\end{array}
\label{l57}
\end{equation}%
which describes an object moving with the velocity of light.

Using the transformation (\ref{l34}), we may have a global view of the
LCR-manifold in the "bounded realization" coordinate chart $Y^{ni}$ where
the two planes take the form

\begin{equation}
\begin{array}{l}
(a+1)Y^{11}+(1-a)Y^{31}=0\quad ,\quad Y^{02}+Y^{22}=0 \\ 
\multicolumn{1}{c}{\Downarrow} \\ 
(1-a)w_{21}Y^{01}+[(a+1)+(1-a)w_{22}]Y^{11}=0 \\ 
(1+w_{11})Y^{02}+w_{12}Y^{12}=0%
\end{array}
\label{l58}
\end{equation}%
and the singular surface ($\det Y_{1}=0$) is

\begin{equation}
\begin{array}{l}
\lbrack
w_{11}w_{22}-w_{12}w_{21}+1+w_{11}+w_{22}]+a[-w_{11}w_{22}+w_{12}w_{21}+1+w_{11}-w_{22}]=0
\\ 
\multicolumn{1}{c}{\Downarrow} \\ 
\cos \tau +\cos \rho =0\quad ,\quad \sin \tau +\sin \rho \cos \sigma =0 \\ 
\tau \in (0,2\pi )\quad ,\quad \rho \in \lbrack 0,2\pi )\quad ,\quad \sigma
\in \lbrack 0,\pi ]%
\end{array}
\label{l59}
\end{equation}%
Notice the difference between the "light-cone" LCR-manifold and the present
"neutrino" LCR-manifold. The former is singular in the entire null infinity
(scri+ and scri-)(\ref{e16}), while the present is singular at a part of it.
This subset must satisfy

\begin{equation}
\begin{array}{l}
(\sin \rho )(\sin \sigma )=0\quad ;\quad \rho \in \lbrack 0,2\pi ),\quad
\sigma \in \lbrack 0,\pi ] \\ 
\end{array}
\label{l60}
\end{equation}%
For $\sigma =0$, we find the $\tau =2\pi -\rho $ curve and for $\sigma =\pi $%
, we find the $\tau =\rho $ curve.

The (\ref{l54}) integrability conditions of the flat "neutrino" LCR-tetrad
implies that its relative invariants are $\Phi _{1}=2a$ and $\Phi
_{2}=0=\Phi _{3}$. But the LCR-tetrad implied by the Kerr-Schild ansatz is

\begin{equation}
\begin{array}{l}
\ell =L\quad ,\quad m=M\quad ,\quad n=N+f(v)L \\ 
\end{array}
\label{l61}
\end{equation}%
where $f(v)$\ is an arbitrary function of $v$. The LCR-structure relations
are

\begin{equation}
\begin{array}{l}
d\ell =2iam\wedge \overline{m}\quad ,\quad dm=0\quad ,\quad dn=-f^{\prime
}\ell \wedge n+2iafm\wedge \overline{m} \\ 
\end{array}
\label{l62}
\end{equation}%
Notice that the curved LCR-structure changes category with the relative
invariants being now $\Phi _{1}=2a$, $\Phi _{2}=2af$ and $\Phi _{3}=0$. This
is the category of the "natural $U(2)$" LCR-structure (\ref{e21}). We see
that the right part of the massless soliton is no longer trivial, like in
the flatprint case. That is the curved neutrino is not a smooth deformation
of the flat one. Does it mean that the neutrino small mass is due to its
gravitational dressing?

\subsection{The muon and tau generations}

In the spinorial formulation\cite{PenRin} of general relativity the
conformal tensor is a spinorial tensor $\Psi _{ABCD}$. A geodetic and shear
free null congruence $\ell ^{\mu }=\overline{\lambda }^{A^{\prime }}\lambda
^{B}\sigma _{A^{\prime }B}^{b}e_{b}^{\ \mu }$ satisfies the condition $\Psi
_{ABCD}\lambda ^{A}\lambda ^{B}\lambda ^{C}\lambda ^{D}=0$. This is the
spinorial form of the relation (\ref{e6}). Formally expanding this relation
relative to the gravitational constant at a point, we find a general quartic
polynomial plus higher order terms. We may generally arrange the tetrad $%
e_{b}^{\ \mu }$ such that the first term of the expansion is identified with
the Kerr function of the formal flatprint of $\ell _{.}^{\mu }$, and
subsequently $\lambda ^{A}$ is projectively related with ($1,b$). But the
two real vectors $\ell ^{\mu }$and $n^{\mu }$ of the LCR-tetrad are geodetic
and shear-free null congruences relative to the defined class of metrics (%
\ref{e1}). Hence locally the number of gravitationally permitted solutions
are restricted to be roots of quadratic, cubic and quartic Kerr polynomials.
We claim that the limited number of three particle generations observed in
nature is related to the above limitation on the degree of the Kerr
polynomial.

In the previous subsections we have found that an irreducible and a
reducible quadratic Kerr polynomial provides the static "electron" and the
stationary "neutrino" LCR-structures respectively. Proceeding into the case
of cubic and quartic polynomials I have not found \emph{regular} irreducible
LCR-structures. We should expect it, because we know that muon and tau
leptons are not static particles. But reducible symmetric cubic and quartic
polynomials may be found. The z-rotation invariant cubic polynomial is%
\begin{equation}
\begin{array}{l}
K_{z}=A_{001}(X^{0})^{2}X^{1}+A_{003}(X^{0})^{2}X^{3}+A_{012}X^{0}X^{1}X^{2}+
\\ 
\quad +A_{023}X^{0}X^{2}X^{3}+A_{122}X^{1}(X^{2})^{2}+A_{223}(X^{2})^{2}X^{3}
\\ 
\end{array}
\label{l63}
\end{equation}%
The static axially symmetric cubic polynomial is reducible%
\begin{equation}
\begin{array}{l}
K_{zt}=[A_{001}X^{0}X^{1}+A_{003}(X^{0}X^{3}-X^{1}X^{2})]X^{0} \\ 
\end{array}
\label{l64}
\end{equation}%
which has the form of quadratic electron up to a factor $X^{0}$. On the
other hand the massless automorphism restricts (\ref{l63}) to

\begin{equation}
\begin{array}{l}
\delta X^{0}=0\quad ,\quad \delta X^{1}=0 \\ 
\delta X^{2}=-i\varepsilon ^{03}X^{0}\quad ,\quad \delta X^{3}=0 \\ 
\multicolumn{1}{c}{\Downarrow} \\ 
K_{zt^{\prime }}=[A_{001}X^{1}+A_{003}X^{3}](X^{0})^{2} \\ 
\end{array}
\label{l65}
\end{equation}%
which is analogous to electron neutrino two planes of $CP(3)$, up to a
factor $(X^{0})^{2}$.

The z-rotation quartic invariant polynomial is

\begin{equation}
\begin{array}{l}
K_{z}=A_{0011}(X^{0})^{2}(X^{1})^{2}+A_{0013}(X^{0})^{2}(X^{1})(X^{3})+A_{0033}(X^{0})^{2}(X^{3})^{2}+
\\ 
\quad
+A_{0112}(X^{0})(X^{1})^{2}(X^{2})+A_{0123}(X^{0})(X^{1})(X^{2})(X^{3})+ \\ 
\quad +A_{0233}(X^{0})(X^{2})(X^{3})^{2}+A_{1122}(X^{1})^{2}(X^{2})^{2}+ \\ 
\quad +A_{1223}(X^{1})(X^{2})^{2}(X^{3})+A_{2233}(X^{2})^{2}(X^{3})^{2} \\ 
\end{array}
\label{l66}
\end{equation}%
The static axially symmetric quartic polynomial has the form%
\begin{equation}
\begin{array}{l}
K_{zt}=A(X^{1}X^{2}-X^{0}X^{3})^{2}+BX^{0}X^{1}(X^{1}X^{2}-X^{0}X^{3})+C(X^{0}X^{1})^{2}
\\ 
\end{array}
\label{l67}
\end{equation}%
which is essentially the product of two independent quadrics (\ref{l3}). The
massless automorphism restricts (\ref{l66}) to%
\begin{equation}
\begin{array}{l}
K_{zt^{\prime
}}=[A_{0011}(X^{1})^{2}+A_{0013}X^{1}X^{3}+A_{0033}(X^{3})^{2}](X^{0})^{2}
\\ 
\end{array}
\label{l68}
\end{equation}%
which is essentially the product of two independent (\ref{l48a}) reducible
quadrics of $CP(3)$.

The general meaning of the LCR-structures locally described with reducible
surfaces of $CP(3)$ may be realized considering the case of the product of
ruled surfaces with different trajectories $\xi _{a}(\tau )$ and $\widetilde{%
\xi }_{a}(\tau )$. Then we may have the left and right columns from
different trajectories

\ 
\begin{equation}
\begin{array}{l}
X^{nj}=\left( 
\begin{array}{cc}
\lambda & \widetilde{\lambda } \\ 
-i\xi _{a}(z^{0})\sigma ^{a}\lambda & -i\widetilde{\xi }_{a}(z^{\widetilde{0}%
})\sigma ^{a}\widetilde{\lambda }%
\end{array}%
\right) \\ 
\lambda =%
\begin{pmatrix}
1 \\ 
z^{1}%
\end{pmatrix}%
\quad ,\quad \widetilde{\lambda }=%
\begin{pmatrix}
-z^{\widetilde{1}} \\ 
1%
\end{pmatrix}%
\end{array}
\label{l69}
\end{equation}%
In the zero algebraic gravity approximation we have the structure conditions

\ 
\begin{equation}
\begin{array}{l}
\rho _{ij}=X^{\dag }\left( 
\begin{array}{cc}
0 & I \\ 
I & 0%
\end{array}%
\right) X=\left( 
\begin{array}{cc}
i(\overline{\xi _{a}}-\xi _{a})\lambda ^{\dag }\sigma ^{a}\lambda & i(%
\overline{\xi _{a}}-\widetilde{\xi }_{a})\lambda ^{\dag }\sigma ^{a}%
\widetilde{\lambda } \\ 
-i(\xi _{a}-\overline{\widetilde{\xi }_{a}})\widetilde{\lambda }^{\dag
}\sigma ^{a}\lambda & i(\overline{\widetilde{\xi }_{a}}-\widetilde{\xi }_{a})%
\widetilde{\lambda }^{\dag }\sigma ^{a}\widetilde{\lambda }%
\end{array}%
\right) =0 \\ 
\end{array}
\label{l70}
\end{equation}%
where $z^{0}$, $z^{1}$\ and $z^{\widetilde{0}}$, $z^{\widetilde{0}}$\ are
the structure coordinates of the two independent trajectories. The null
integral curves (LCR-rays) of the retarded $\ell ^{\mu }$ congruence are
labeled by the three independent functions $\func{Re}(z^{0})$, $\func{Re}%
(z^{1})$ and $\func{Im}(z^{1})$ of the structure coordinates $z^{\beta }$.
The first is the wavefront surface. Hence, the algebraic trajectory $\xi
^{a}(z^{0})$ is the movement of the source of retarded rays in the
grassmannian space $G(4,2)$. In the case of the advanced $n^{\mu }$ rays the 
$z^{\widetilde{\beta }}$ structure coordinates are related to the second
trajectory.

\section{EMERGENCE\ OF\ ELECTROWEAK\ FIELD}

\setcounter{equation}{0}

In conventional field theory the electroweak interactions are imposed as $%
U(2)$ connections. In the Cartan formalism the group $U(2)$ gauge field $%
B=B_{I\mu }dx^{\mu }t_{I}$ with $t_{I}$ being the generators of the Lie
algebra of $U(2)$ have generally non-vanishing curvature $F=dB-iB\wedge B$.
We precisely have \ 
\begin{equation}
\begin{array}{l}
B=B_{I\mu }dx^{\mu }t_{I}=%
\begin{pmatrix}
A & \overline{W} \\ 
W & Z%
\end{pmatrix}%
\quad ,\quad t_{0}=I\ ,\ t_{j}=\frac{1}{2}\sigma ^{j} \\ 
\\ 
F=dB-iB\wedge B\ \longrightarrow DF:=\ dF+iB\wedge F-iF\wedge B=0 \\ 
\end{array}
\label{ew1}
\end{equation}%
where $A_{\mu }$, $Z_{\mu }$ and $W_{\mu }$ are the electromagnetic, neutral
and charged fields respectively. In the computation of the electron and its
neutrino distributional LCR-structures, these fields appear as gravitational
and electromagnetic dressing distributions with precise compact singular
support. The purpose of the present section is to provide the algorithmic
derivation of the electroweak connection observed in the SM formulation. The
surprising result is that the electroweak connection is directly related
with a precise LCR-tetrad. This essentially generalizes the surprising
identification of the electron electromagnetic dressing $A_{\mu }(x)$ with
the vector $\ell _{\mu }$ of the LCR-tetrad (and the induced gravitational
dressing).

The first observation is that the LCR-structure conditions $\rho _{ij}$ (\ref%
{c25}) form a 2$\times $2 hermitian matrix, implying that the LCR-tetrad $%
i(\partial -\overline{\partial })\rho _{ij}$ may be cast as a hermitian
1-form. But recall that the LCR-tetrad is not uniquely determined because of
the tetrad-Weyl symmetry (\ref{c33}-\ref{c33a}). Hence we may make a
tetrad-Weyl transformation (\ref{c33a}) fixing the relative invariants $\Phi
_{1}=1=-\Phi _{2}$. Then this special hermitian LCR-tetrad%
\begin{equation}
\begin{array}{l}
\widehat{e}^{\prime }:=%
\begin{pmatrix}
\ell ^{\prime } & \overline{m^{\prime }} \\ 
m^{\prime } & n^{\prime }%
\end{pmatrix}%
=i(\partial -\overline{\partial })%
\begin{pmatrix}
\rho _{11}^{\prime } & \rho _{12}^{\prime } \\ 
\overline{\rho _{12}^{\prime }} & \rho _{22}^{\prime }%
\end{pmatrix}
\\ 
\end{array}
\label{ew2}
\end{equation}%
considered as a $U(2)$ connection has the generally non-vanishing Cartan
curvature (\ref{e21}) \ 
\begin{equation}
\begin{array}{l}
d\widehat{e}^{\prime }-i\widehat{e}^{\prime }\wedge \widehat{e}^{\prime
}=\Omega \neq 0 \\ 
\multicolumn{1}{c}{\Downarrow} \\ 
d\ell ^{\prime }-im^{\prime }\wedge \overline{m}^{\prime }=Z_{1}^{\prime
}\wedge \ell ^{\prime }\quad ,\quad dn^{\prime }+im^{\prime }\wedge 
\overline{m}^{\prime }=Z_{2}^{\prime }\wedge n^{\prime } \\ 
dm^{\prime }-i(\ell ^{\prime }-n^{\prime })\wedge m^{\prime }=(Z_{3}^{\prime
}-i(\ell ^{\prime }-n^{\prime }))\wedge m^{\prime }+\Phi _{3}^{\prime }\ell
^{\prime }\wedge n^{\prime }%
\end{array}
\label{ew3}
\end{equation}%
Then identifying the Cartan $U(2)$ connection $\widehat{e}^{\prime }$ (\ref%
{ew2}) with the (\ref{ew1}) gauge potential \ 
\begin{equation}
\begin{array}{l}
\widehat{e}^{\prime }:=%
\begin{pmatrix}
\ell ^{\prime } & \overline{m^{\prime }} \\ 
m^{\prime } & n^{\prime }%
\end{pmatrix}%
=:%
\begin{pmatrix}
A & \overline{W} \\ 
W & Z%
\end{pmatrix}
\\ 
\end{array}
\label{ew4}
\end{equation}%
we find the direct relation between the electroweak potentials with a member
of the LCR-tetrad class of the solitonic LCR-structure. We precisely have\ 
\begin{equation}
\begin{array}{l}
B_{0\mu }+\frac{1}{2}B_{3\mu }=\ell _{\mu }^{\prime }\quad ,\quad B_{0\mu }-%
\frac{1}{2}B_{3\mu }=n_{\mu }^{\prime }\quad ,\quad \frac{1}{2}(B_{1\mu
}+iB_{2\mu })=m_{\mu }^{\prime } \\ 
\\ 
F_{0\mu \nu }=\partial _{\mu }B_{0\nu }-\partial _{\nu }B_{0\mu } \\ 
F_{i\mu \nu }=\partial _{\mu }B_{i\nu }-\partial _{\nu }B_{i\mu }-\epsilon
_{ijk}B_{j\mu }B_{k\nu }%
\end{array}
\label{ew5}
\end{equation}

In order to clarify the procedure we will give two precise examples. The
first is the simple observation that The "natural U(2)" LCR-structure (\ref%
{e18}) has vanishing electroweak field strength. The second example is the
"electron" LCR-structure. The relative invariant (\ref{l22}) are tetrad-Weyl
transformed to reach the conditions $\Phi _{1}^{\prime }=1=-\Phi
_{2}^{\prime }$. We find \ \ 
\begin{equation}
\begin{array}{l}
N=-\frac{2(r^{2}+a^{2}\cos ^{2}\theta )}{r^{2}+a^{2}+h}\Lambda \\ 
\\ 
M\overline{M}=-\frac{2a\cos \theta }{r^{2}+a^{2}\cos ^{2}\theta }\Lambda%
\end{array}
\label{ew6}
\end{equation}%
The electromagnetic dressing (\ref{l19a}) is found with $\Lambda =\frac{qr}{%
4\pi (r^{2}+a^{2}\cos ^{2}\theta )}$. Then the electroweak connection $B$ (%
\ref{ew4}) is found with\ \ 
\begin{equation}
\begin{array}{l}
\Lambda =\frac{qr}{4\pi (r^{2}+a^{2}\cos ^{2}\theta )} \\ 
\\ 
N=-\frac{qr}{2\pi (r^{2}+a^{2}+h)} \\ 
\\ 
M\overline{M}=-\frac{qra\cos \theta }{2\pi (r^{2}+a^{2}\cos ^{2}\theta )^{2}}%
\end{array}
\label{ew7}
\end{equation}%
up to an $M$ phase tetrad-Weyl transformation. That is, we find the
following electroweak potentials (dressings) of the electron%
\begin{equation}
\begin{array}{l}
A=\frac{qr}{4\pi (r^{2}+a^{2}\cos ^{2}\theta )}(dt-dr-a\sin ^{2}\theta
d\varphi ) \\ 
\\ 
Z=\frac{-qr}{4\pi (r^{2}+a^{2}\cos ^{2}\theta )}(dt+\frac{r^{2}+2a^{2}\cos
^{2}\theta -a^{2}-h}{r^{2}+a^{2}+h}dr-a\sin ^{2}\theta d\varphi ) \\ 
\\ 
W=\frac{-M}{\sqrt{2}(r+ia\cos \theta )}[-ia\sin \theta \
(dt-dr)+(r^{2}+a^{2}\cos ^{2}\theta )d\theta + \\ 
\qquad \qquad +i\sin \theta (r^{2}+a^{2})d\varphi ]%
\end{array}
\label{ew8}
\end{equation}%
where the $h=-2mr+q^{2}$ and the tetrad-Weyl factor $M$ will be computed
below through the third scalar (relative invariant) dressing.

The third (complex) relative invariant $\Phi _{3}^{\prime }$ is not
completely fixed. Its phase is absorbed by $W$ and the remaining scalar real
field \ \ 
\begin{equation}
\begin{array}{l}
|\Phi _{3}^{\prime }|^{2}=\frac{(4\pi a)^{3}\sin ^{2}\theta \cos \theta
(r^{2}+a^{2}+h)^{2}}{q^{3}r(r^{2}+a^{2}\cos ^{2}\theta )^{2}} \\ 
\end{array}
\label{ew9}
\end{equation}%
will be finally related with the Higgs field of the SM. The $U(2)$ gauge
field is directly related to the LCR-tetrad and the breaking of the
tetrad-Weyl symmetry through $\Phi _{1}^{\prime }=1=-\Phi _{2}^{\prime }$,
while the value of the the third relative invariant $|\Phi _{3}^{\prime }|$
indicates the difference of the compact deSitter vacuum (with dark Energy)
and the matter sector. The electroweak field strength and $\Phi _{3}^{\prime
}$ exist even in the case of vanishing \emph{algebraic} gravity, that is
electroweak field strength (de Sitter space curvature) and $\Phi
_{3}^{\prime }$ may not vanish even without algebraic gravity. The above
form (\ref{ew9}) of $|\Phi _{3}^{\prime }|$ indicates that at the large
universe scale we have the regions of matter concentrations with
non-vanishing algbraic gravity, and the asymptotic empty space being a de
Sitter space with vanishing algbraic gravity, electroweak field strength and 
$|\Phi _{3}^{\prime }|\simeq 0$. But there may be an intermediate region
where algebraic gravity vanishes while the third scalar relative invariant
does not vanish.

The tetrad-Weyl transformation is the local symmetry of the fundamental
geometric LCR-structure. This symmetry is broken by the local Lorentz $%
SO(1,3)$ transformation of the tetrad, which preserves the Einstein metric.
It is also broken by the electroweak $U(2)$ transformation. That is the $%
SO(1,3)$ and $U(2)$ transformations are transversal to the tetrad-Weyl
transformations.

\section{GLUONIC\ FIELD AND\ QUARKS}

\setcounter{equation}{0}

The hadronic sector of the elementary particles is (about) a copy of the
leptonic sector relative to the electroweak interactions. Quarks simply have
the additional gluonic interaction, which should provide a confining
mechanism. The SM does not explain the lepton-quark correspondence, while
the artificial add-on of the $SU(3)$ gauge group gives some answers to some
phenomena and the annihilation of the anomalies, but it fails to imply (in
the continuum) confinement, which is the characteristic property of strong
interactions.

The tetrad-Weyl symmetry (\ref{c33}) of PCFT is highly constrained. The only
permitted field equations have the following two dual forms 
\begin{equation}
\begin{array}{l}
\frac{1}{\sqrt{-g}}(D_{\mu })_{ij}[\sqrt{-g}(\Gamma ^{\mu \nu \rho \sigma
}\mp \overline{\Gamma ^{\mu \nu \rho \sigma }})F_{j\rho \sigma }]=0 \\ 
\\ 
\Gamma ^{\mu \nu \rho \sigma }:=\frac{1}{2}[(\ell ^{\mu }m^{\nu }-\ell ^{\nu
}m^{\mu })(n^{\rho }\overline{m}^{\sigma }-n^{\sigma }\overline{m}^{\rho
})+(n^{\mu }\overline{m}^{\nu }-n^{\nu }\overline{m}^{\mu })(\ell ^{\rho
}m^{\sigma }-\ell ^{\sigma }m^{\rho })] \\ 
F_{j\mu \nu }:=\partial _{\mu }A_{j\nu }-\partial _{\nu }A_{j\mu }-\gamma
\,f_{jik}A_{i\mu }A_{k\nu } \\ 
(D_{\mu })_{ij}:=\delta _{ij}\partial _{\mu }-\gamma f_{ikj}A_{k\mu } \\ 
\sqrt{-g}=\frac{i}{4!}\epsilon ^{\mu \nu \rho \sigma }\ell _{\mu }n_{\nu
}m_{\rho }\overline{m}_{\sigma }%
\end{array}
\label{g1}
\end{equation}%
No scalar or spinorial sources are permitted because simply they would break
tetrad-Weyl symmetry. Besides only the tetrad-Weyl covariant self-dual forms
[$V_{1}$] and [$V_{2}$] (\ref{e1}) appear in the differential operator.
Essentially the two $\mp $ differential equations are the imaginary and real
part of the sum of [$V_{1}$] and [$V_{2}$] self-dual forms. Notice that the
one equation is the dual of the other equation.

In the first subsection we define the quarks as the permitted real source of
the sum of [$V_{1}$] and [$V_{2}$] self-dual forms. The immediate
consequence is the observed in nature lepton-quark correspondence, based on
the same LCR-structure between the leptons and the corresponding quarks. I
explicitly solve the differential equations in the "electron" LCR-structure
and find the gluonic dressing of the corresponding quark which apparently
continues to have the electroweak dressings. The magneto-gluonic field has
line singularities not permitted for free field configurations. In the BEGS
procedure this must be described with colorless distributions of compact
support. In the second subsection I find convex bag-like picture for the
hadronic structures. In the third subsection I explicitly find that the
"natural U(2)" LCR-structure does not have gluonic sources. Recall that this
LCR-structure is the vacuum of the electroweak interactions and the origin
of the de Sitter metric (\ref{e24}), which describes dark energy.

\subsection{Emergence of quarks}

The quarks emerge in PCFT as the permitted distributional sources 
\begin{equation}
\begin{array}{l}
(-)\quad \rightarrow \quad \frac{1}{\sqrt{-g}}(D_{\mu })_{ij}\{\sqrt{-g}%
[(\ell ^{\mu }m^{\nu }-\ell ^{\nu }m^{\mu })(n^{\rho }\overline{m}^{\sigma
}F_{j\rho \sigma })+ \\ 
\qquad \qquad +(n^{\mu }\overline{m}^{\nu }-n^{\nu }\overline{m}^{\mu
})(\ell ^{\rho }m^{\sigma }F_{j\rho \sigma })]\}=-k_{i}^{\nu } \\ 
\\ 
(+)\quad \rightarrow \quad \frac{1}{\sqrt{-g}}(D_{\mu })_{ij}\{\sqrt{-g}%
[(\ell ^{\mu }m^{\nu }-\ell ^{\nu }m^{\mu })(n^{\rho }\overline{m}^{\sigma
}F_{j\rho \sigma })+ \\ 
\qquad \qquad +(n^{\mu }\overline{m}^{\nu }-n^{\nu }\overline{m}^{\mu
})(\ell ^{\rho }m^{\sigma }F_{j\rho \sigma })]\}=-ik_{i}^{\nu }%
\end{array}
\label{g2}
\end{equation}%
where $k_{i}^{\nu }(x)$ is a real vector field. That is, taking into account
that the sum of the two terms is self-dual we find 
\begin{equation}
\begin{array}{l}
\frac{1}{2}(G_{j}^{\mu \nu }-i\ast G_{j}^{\mu \nu }):=\Gamma ^{\mu \nu \rho
\sigma }F_{j\rho \sigma }=\frac{1}{2}\Gamma ^{\mu \nu \rho \sigma }(F_{j\rho
\sigma }-i\ast F_{j\rho \sigma }) \\ 
\\ 
\Gamma ^{\mu \nu \rho \sigma }=\frac{1}{2}[(\ell ^{\mu }m^{\nu }-\ell ^{\nu
}m^{\mu })(n^{\rho }\overline{m}^{\sigma }-n^{\sigma }\overline{m}^{\rho
})+(n^{\mu }\overline{m}^{\nu }-n^{\nu }\overline{m}^{\mu })(\ell ^{\rho
}m^{\sigma }-\ell ^{\sigma }m^{\rho })] \\ 
\end{array}
\label{g3}
\end{equation}%
In (-) PDE $k_{i}^{\nu }$\ is a source of the real part $G_{j}^{\mu \nu }$,\
while in (+) $k_{i}^{\nu }$\ is a source of its dual part imaginary part.
The PDEs look like the equations of a gauge field with a color-electric and
color-magnetic source respectively. Notice the essential difference of the
present equations (\ref{g1}) and the conventional gluonic equations. The
covariant gauge field derivative $(D_{\mu })_{ij}$ applies only on the
tetrad-Weyl invariant part of the gauge field and \emph{not} over the entire
gauge field. Notice that the LCR-structure defining equations completely
decuple from the gauge field equations. The LCR-structure is first fixed and
after we proceed to the solution of the field equations, which involve the
gluonic gauge field. This property of PCFT is essentially behind the
physical observation of the lepton-quark correspondence! That is, a quark
has the same LCR-structure with the corresponding lepton and the implied
electroweak gauge field (\ref{ew2}). But the quark has in addition a stable
non-vanishing distributional gauge field configuration "dressing" (from
which it gets its color), while the lepton has vanishing gluonic "dressing".

Like in the case of the static axisymmetric LCR-structure (static electron)
we looked for a solution in the abelian Cartan subgroup of the Poincar\'{e}
group, we now look for a solution with only $A_{3\mu }$ and $A_{8\mu }$
non-vanishing, the abelian subgroup of $SU(3)$. That is we will look for
solutions in the abelian PDE (\ref{g2}), which admit \emph{compact}
distributional sources. Then Stoke's theorem will be applied on their
distributional singularities. We will first look for a solution of the
electro-gluonic form (-) and finally we will show that $\epsilon ^{\mu \nu
\rho \sigma }\partial _{\nu }F_{\rho \sigma }=0$ does not permit
magneto-gluonic solutions of (+).

It is more convenient to make calculations using the differential forms. In
the vanishing gravity case (\ref{l13}),\ the (-) PDE takes the form

\begin{equation}
\begin{array}{l}
d\{(N^{\rho }\overline{M}^{\sigma }F_{j\rho \sigma }^{+})L\wedge M+(L^{\rho
}M^{\sigma }F_{j\rho \sigma }^{+})N\wedge \overline{M}\}=i\ast k_{j} \\ 
\end{array}
\label{g4}
\end{equation}%
We find the non-vanishing closed 2-forms (with real compact sources) in the
case of the flatprint massive LCR-tetrad

\begin{equation}
\begin{array}{l}
d\{\frac{C_{j}^{\prime }}{\sin \theta (r-ia\cos \theta )}L\wedge M+\frac{%
C_{j}^{\prime \prime }(r-ia\cos \theta )}{(r^{2}+a^{2})\sin \theta }N\wedge 
\overline{M}\}=i\ast k_{j} \\ 
\\ 
G_{j}^{+}:=\frac{1}{2}(G_{j}-i\ast G_{j})=\frac{C_{j}^{\prime }}{\sin \theta
(r-ia\cos \theta )}L\wedge M+\frac{C_{j}^{\prime \prime }(r-ia\cos \theta )}{%
(r^{2}+a^{2})\sin \theta }N\wedge \overline{M}%
\end{array}
\label{g5}
\end{equation}%
where $C_{j}^{\prime }$ and $C_{j}^{\prime \prime }$ with $j=3,8$ are
arbitrary complex constants, which are fixed using Stokes' theorem and the
reality conditions for gluonic sources. That is, we have

\begin{equation}
\begin{array}{l}
(L^{\rho }M^{\sigma }F_{j\rho \sigma })=\frac{C_{j}^{\prime \prime
}(r-ia\cos \theta )}{(r^{2}+a^{2})\sin \theta }\quad ,\quad (N^{\rho }%
\overline{M}^{\sigma }F_{j\rho \sigma })=\frac{C_{j}^{\prime }}{\sin \theta
(r-ia\cos \theta )} \\ 
\\ 
(L^{\rho }M^{\sigma }F_{j\rho \sigma }^{+})=\frac{C_{j}^{\prime \prime
}(r-ia\cos \theta )}{(r^{2}+a^{2})\sin \theta }\quad ,\quad (N^{\rho }%
\overline{M}^{\sigma }F_{j\rho \sigma }^{+})=\frac{C_{j}^{\prime }}{\sin
\theta (r-ia\cos \theta )} \\ 
\\ 
F_{j\rho \sigma }^{+}:=\frac{1}{2}(F_{j\rho \sigma }-i\ast F_{j\rho \sigma })
\\ 
F_{j\rho \sigma }:=\partial _{\rho }A_{j\sigma }-\partial _{\sigma }A_{j\rho
}\quad ,\quad j=3,8%
\end{array}
\label{g6}
\end{equation}%
We apparently have 
\begin{equation}
\begin{array}{l}
\lbrack (L^{\mu }N^{\nu }-M^{\mu }\overline{M}^{\nu })G_{j\mu \nu }^{+}]=0
\\ 
\\ 
F^{+}=-G^{+}-f(L\wedge N-M\wedge \overline{M})%
\end{array}
\label{g7}
\end{equation}%
In order to avoid any ambiguity, we assume $f=0$. Hence

\begin{equation}
\begin{array}{l}
F_{j\rho \sigma }^{+}=-\frac{C_{j}^{\prime }}{\sin \theta (r-ia\cos \theta )}%
(L_{\rho }M_{\sigma }-L_{\sigma }M_{\sigma })-\frac{C_{j}^{\prime \prime
}(r-ia\cos \theta )}{(r^{2}+a^{2})\sin \theta }(N_{\rho }\overline{M}%
_{\sigma }-N_{\sigma }\overline{M}_{\sigma }) \\ 
\end{array}
\label{g8}
\end{equation}

For the static flatprint LCR-tetrad the solutions have the explicit forms

\begin{equation}
\begin{array}{l}
F_{j}-i\ast F_{j}:=-\frac{2C_{j}^{\prime }}{\sin \theta (r-ia\cos \theta )}%
L\wedge M-\frac{2C_{j}^{\prime \prime }(r-ia\cos \theta )}{(r^{2}+a^{2})\sin
\theta }N\wedge \overline{M}= \\ 
\quad =\frac{2C_{j}^{\prime }+C_{j}^{\prime \prime }}{\sqrt{2}}[\frac{-ia}{%
r^{2}+a^{2}}dt\wedge dr+\frac{1}{\sin \theta }dt\wedge d\theta +\frac{%
a^{2}\sin \theta }{r^{2}+a^{2}}dr\wedge d\theta \\ 
\quad -idr\wedge d\varphi +a\sin \theta d\theta \wedge d\varphi ]+ \\ 
\quad +\frac{2C_{j}^{\prime }-C_{j}^{\prime \prime }}{\sqrt{2}}[\frac{ia}{%
(r^{2}+a^{2})}dt\wedge dr+idt\wedge d\varphi -\frac{r^{2}+a^{2}\cos
^{2}\theta }{(r^{2}+a^{2})\sin \theta }dr\wedge d\theta ]%
\end{array}
\label{g9}
\end{equation}%
After a straightforward calculation I find

\begin{equation}
\begin{array}{l}
\tint\limits_{t,r=const}[\frac{-2C_{j}^{\prime }}{\sin \theta (r-ia\cos
\theta )}L\wedge M+\frac{-2C_{j}^{\prime \prime }(r-ia\cos \theta )}{%
(r^{2}+a^{2})\sin \theta }N\wedge \overline{M}]=\frac{(2C_{j}^{\prime
}+C_{j}^{\prime \prime })4\pi a}{\sqrt{2}}=:-i\gamma _{j} \\ 
\end{array}
\label{g10}
\end{equation}%
which implies that the (dimensionless) constants $\gamma _{3}$ and $\gamma
_{8}$ must be real for the sources to be real and the original field
equations to be satisfied.

From (\ref{g9}) we see that we actually have the sum of a retarded and an
advanced null 2-form. Assuming $2C_{j}^{\prime }-C_{j}^{\prime \prime }=0$
(because it is not determined by the gluonic charge), the arbitrary
constants are completely fixed and the solutions are

\begin{equation}
\begin{array}{l}
F_{j}=\frac{-\gamma _{j}}{4\pi a}[\frac{a}{r^{2}+a^{2}}dt\wedge dr+dr\wedge
d\varphi ]= \\ 
\qquad =d[\frac{\gamma _{j}}{4\pi a}(\tan ^{-1}\frac{r}{a}dt-rd\varphi )] \\ 
\\ 
\ast F_{j}=\frac{\gamma _{j}}{4\pi }[\frac{1}{a\sin \theta }dt\wedge d\theta
+\frac{a\sin \theta }{r^{2}+a^{2}}dr\wedge d\theta +\sin \theta d\theta
\wedge d\varphi ]%
\end{array}
\label{g11}
\end{equation}%
where $j=3,8$ for the $su(3)$ color algebra and the corresponding potentials
being apparent.

It is interesting to compare the electromagnetic dressing potential of the
"electron" LCR-structure (\ref{l19a})%
\begin{equation}
\begin{array}{l}
F^{EM}=\frac{q}{4\pi (r^{2}+a^{2}\cos ^{2}\theta )^{2}}[(r^{2}-a^{2}\cos
^{2}\theta )dt\wedge dr-2a^{2}r\cos \theta \sin \theta dt\wedge d\theta + \\ 
\qquad +2a^{2}r\cos \theta \sin \theta dr\wedge d\theta +a(r^{2}-a^{2}\cos
^{2}\theta )\sin ^{2}\theta dr\wedge d\varphi - \\ 
\qquad -2ar(r^{2}+a^{2})\cos \theta \sin \theta d\theta \wedge d\varphi = \\ 
\\ 
\qquad =d[\frac{qr}{4\pi (r^{2}+a^{2}\cos ^{2}\theta )}(dt-dr-a\sin
^{2}\theta d\varphi )] \\ 
\\ 
r^{4}-[(x^{1})^{2}+(x^{2})^{2}+(x^{3})^{2}-a^{2}]r^{2}-a^{2}(x^{3})^{2}=0%
\end{array}
\label{g12}
\end{equation}%
with the electromagnetic potential in cartesian coordinates%
\begin{equation}
\begin{array}{l}
A^{EM}=\frac{qr}{4\pi (r^{2}+a^{2}\cos ^{2}\theta )}(dt-dr-a\sin ^{2}\theta
d\varphi )= \\ 
\\ 
\quad =\frac{qr^{3}}{4\pi (r^{4}+a^{2}(x^{3})^{2})}(dx^{0}-\frac{%
rx^{1}-ax^{2}}{r^{2}+a^{2}}dx^{1}-\frac{rx^{2}+ax^{1}}{r^{2}+a^{2}}dx^{2}-%
\frac{x^{3}}{r}dx^{3}) \\ 
\end{array}
\label{g13}
\end{equation}%
The most important differences are the singularities relative to the
rotation parameter $a$, and the cylindrical variable $\rho $ of the
magneto-gluonic field (and potential) as we will explicitly show below.

The gluonic dressing potential of the static quark LCR-structure in
cartesian coordinates%
\begin{equation}
\begin{array}{l}
x^{0}=t \\ 
x^{1}=(r\cos \varphi +a\sin \varphi )\sin \theta \\ 
x^{2}=(r\sin \varphi -a\cos \varphi )\sin \theta \\ 
x^{3}=r\cos \theta \\ 
\\ 
dx^{1}\wedge dx^{2}\wedge dx^{3}=(r^{2}+a^{2}\cos ^{2}\theta )\sin \theta
dr\wedge d\theta \wedge d\varphi%
\end{array}
\label{g14}
\end{equation}%
inverted into%
\begin{equation}
\begin{array}{l}
dt=dx^{0} \\ 
dr=\frac{r}{r^{4}+a^{2}(x^{3})^{2}}%
[r^{2}x^{1}dx^{1}+r^{2}x^{2}dx^{2}+(r^{2}+a^{2})x^{3}dx^{3}] \\ 
d\theta =\frac{r}{\rho (r^{4}+a^{2}(x^{3})^{2})\sqrt{(r^{2}+a^{2})}}%
[(r^{2}+a^{2})x^{3}(x^{1}dx^{1}+x^{2}dx^{2})-r^{2}\rho ^{2}dx^{3}] \\ 
d\varphi =-\frac{x^{2}}{\rho ^{2}}dx^{1}+\frac{x^{1}}{\rho ^{2}}dx^{2}-\frac{%
a}{r^{2}+a^{2}}dr \\ 
\\ 
\rho ^{2}:=(x^{1})^{2}+(x^{2})^{2}\quad ,\quad \frac{\rho ^{2}}{r^{2}+a^{2}}+%
\frac{(x^{3})^{2}}{r^{2}}=1%
\end{array}
\label{g15}
\end{equation}%
I find the following gauge potential (up to a gauge transformation)%
\begin{equation}
\begin{array}{l}
A_{j}^{(g)}=\frac{\gamma _{j}}{4\pi a}(\tan ^{-1}\frac{r}{a}dt-rd\varphi )=
\\ 
\qquad =\frac{\gamma _{j}}{4\pi a}[\tan ^{-1}\frac{r}{a}dx^{0}+\frac{rx^{2}}{%
(x^{1})^{2}+(x^{2})^{2}}dx^{1}-\frac{rx^{1}}{(x^{1})^{2}+(x^{2})^{2}}dx^{2}]=
\\ 
\qquad =\frac{\gamma _{j}}{4\pi a}[\tan ^{-1}\frac{r}{a}dx^{0}-rd\arctan 
\frac{x^{2}}{x^{1}}] \\ 
\\ 
d\arctan \frac{x^{2}}{x^{1}}=\frac{-x^{2}}{(x^{1})^{2}+(x^{2})^{2}}dx^{1}+%
\frac{x^{1}}{(x^{1})^{2}+(x^{2})^{2}}dx^{2}%
\end{array}
\label{g16}
\end{equation}%
where I considered the gauge transformation $A_{j}^{(g)}\rightarrow
A_{j}^{(g)}-\frac{\gamma _{j}}{4\pi a}\frac{a}{r^{2}+a^{2}}dr$. The gauge
field strength is%
\begin{equation}
\begin{array}{l}
F_{j}^{(g)}=\frac{-\gamma _{j}}{4\pi a}(\frac{a}{r^{2}+a^{2}}dt\wedge
dr+dr\wedge d\varphi )= \\ 
\qquad =\frac{-\gamma _{j}r}{4\pi (r^{2}+a^{2})(r^{4}+a^{2}(x^{3})^{2})}%
[r^{2}x^{1}dx^{0}\wedge dx^{1}+r^{2}x^{2}dx^{0}\wedge
dx^{2}+(r^{2}+a^{2})x^{3}dx^{0}\wedge dx^{3}]+ \\ 
\qquad +\frac{-\gamma _{j}r}{4\pi a(r^{4}+a^{2}(x^{3})^{2})}%
[r^{2}dx^{1}\wedge dx^{2}+\frac{(r^{2}+a^{2})x^{3}}{\rho ^{2}}%
(x^{2}dx^{1}\wedge dx^{3}-x^{1}dx^{2}\wedge dx^{3})] \\ 
\end{array}
\label{g17}
\end{equation}%
Notice that the electromagnetic potential is real analytic relative to the
rotation parameter $a$, while the magneto-gluonic potential and field
strength are singular. Both also has a line singularity along the z-axis.
Recall that the Dirac magnetic monopole field strength has the ordinary
Coulomb singularity while line singularity appears in its magnetic
potential. The above field strength line singularity \{$x^{1}=0=x^{2},$ $%
x^{3}\neq 0$\} cannot be removed. But the components of the gauge potential $%
A_{j}^{(g)}$ are locally integrable functions while the components of the
field $F_{j}^{(g)}$ are not, as it usually happens in a proper Schwartz
distribution.

Using the notation 
\begin{equation}
\begin{array}{l}
F_{\mu \nu }^{(g)}\equiv 
\begin{pmatrix}
0 & -E_{x} & -E_{y} & -E_{z} \\ 
E_{x} & 0 & B_{z} & -B_{y} \\ 
E_{y} & -B_{z} & 0 & B_{x} \\ 
E_{z} & B_{y} & -B_{x} & 0%
\end{pmatrix}
\\ 
F^{(g)}=-E^{1}dt\wedge dx-E^{2}dt\wedge dy-E^{3}dt\wedge dz+ \\ 
\quad +B^{3}dx\wedge dy-B^{2}dx\wedge dz+B^{1}dy\wedge dz%
\end{array}
\label{g18}
\end{equation}%
we find 
\begin{equation}
\begin{array}{l}
\overrightarrow{E_{j}^{(g)}}=\frac{-\gamma _{j}r}{4\pi
(r^{2}+a^{2})(r^{4}+a^{2}(x^{3})^{2})}[r^{2}x^{1},\ r^{2}x^{2},\
(r^{2}+a^{2})x^{3}]=-\overrightarrow{\nabla }A_{j0}^{(g)} \\ 
\\ 
\overrightarrow{B_{j}^{(g)}}=\frac{\gamma _{j}r}{4\pi a\rho
^{2}(r^{4}+a^{2}(x^{3})^{2})}[(r^{2}+a^{2})x^{1}x^{3},\
(r^{2}+a^{2})x^{2}x^{3},\ -\rho ^{2}r^{2}]=\overrightarrow{\nabla }\times 
\overrightarrow{A^{(g)}}_{j} \\ 
\\ 
r^{4}-[(x^{1})^{2}+(x^{2})^{2}+(x^{3})^{2}-a^{2}]r^{2}-a^{2}(x^{3})^{2}=0%
\end{array}
\label{g19}
\end{equation}%
where the electro-gluonic $\overrightarrow{E_{j}^{(g)}}$ and magneto-gluonic 
$\overrightarrow{B_{j}^{(g)}}$ fields are singular at the disc \{$x^{3}=0,\
\rho \leq a$\} and besides $\overrightarrow{A^{(g)}}_{j}$ is singular at the
line $\rho =0$. The corresponding potentials are%
\begin{equation}
\begin{array}{l}
A_{i0}^{(g)}=\frac{\gamma _{i}}{4\pi a}(\arctan \frac{r}{a}) \\ 
\\ 
\overrightarrow{A^{(g)}}_{j}=\frac{\gamma _{i}r}{4\pi a}[-\frac{x^{2}}{%
(x^{1})^{2}+(x^{2})^{2}}\ ,\quad \frac{x^{1}}{(x^{1})^{2}+(x^{2})^{2}}\
,\quad 0] \\ 
\quad =\frac{\gamma _{i}r}{4\pi a}\overrightarrow{\nabla }\arctan \frac{x^{2}%
}{x^{1}} \\ 
\\ 
\overrightarrow{\nabla }\cdot \overrightarrow{A^{(g)}}_{j}=0%
\end{array}
\label{g20}
\end{equation}%
where the vector potential $\overrightarrow{A^{(g)}}_{j}$ has a line
singularity at the z-axis and a singularity along the negative part of the $%
x^{1}$ axis, which is the characteristic singularity implied by the gap $%
\varphi +2\pi $ chosen to be the negative $x^{1}$ axis. Notice that both the
gluonic potential and the field strength are singular while they do not have
magnetic charge. They are not Dirac gluonic monopoles. These singularities
are compatible with the Schwartz distribution, because%
\begin{equation}
\begin{array}{l}
\underset{\varepsilon \rightarrow 0}{\lim }\tint\limits_{\varepsilon
}^{\Lambda }A_{j0}^{(g)}dx^{1}dx^{2}dx^{3}=\underset{\varepsilon \rightarrow
0}{\lim }\tint\limits_{\varepsilon }^{\Lambda }\frac{-\gamma _{j}}{4\pi a}%
(\arctan \frac{r}{a})\rho d\rho d\psi dz=finite \\ 
\\ 
\underset{\varepsilon \rightarrow 0}{\lim }\tint\limits_{\varepsilon
}^{\Lambda }\frac{rx^{1}}{(x^{1})^{2}+(x^{2})^{2}}dx^{1}dx^{2}dx^{3}\leq 
\underset{\varepsilon \rightarrow 0}{\lim }\tint\limits_{\varepsilon
}^{\Lambda }\frac{r|x^{1}|}{\rho }d\rho d\psi dz=finite%
\end{array}
\label{g21}
\end{equation}%
where $\Lambda $ is an arbitrary upper bound of the integration region.
Hence the gauge potential is a distribution, but analogous calculations
imply that the field strength components are not locally integrable. As
usual, the potential is the base locally integrable function, on which the
"ladder" of the corresponding Schwartz distributions is built.

The energy density is 
\begin{equation}
\begin{array}{l}
u=\frac{1}{2}\tsum\limits_{3+8}(\overrightarrow{E}_{j}\cdot \overrightarrow{%
E_{j}}+\overrightarrow{B_{j}}\cdot \overrightarrow{B_{j}})=\tsum\limits_{3+8}%
\frac{\gamma _{j}^{2}r^{2}}{32\pi ^{2}(r^{4}+a^{2}(x^{3})^{2})}[\frac{1}{%
r^{2}+a^{2}}+\frac{r^{2}+a^{2}}{a^{2}\rho ^{2}}] \\ 
\\ 
\overrightarrow{E_{j}}\cdot \overrightarrow{E_{j}}=\frac{\gamma _{j}^{2}r^{2}%
}{16\pi ^{2}(r^{2}+a^{2})(r^{4}+a^{2}(x^{3})^{2})}\quad ,\quad 
\overrightarrow{B_{j}}\cdot \overrightarrow{B_{j}}=\frac{\gamma
_{j}^{2}r^{2}(r^{2}+a^{2})}{16\pi ^{2}a^{2}\rho ^{2}(r^{4}+a^{2}(x^{3})^{2})}
\\ 
\\ 
\rho ^{2}:=(x^{1})^{2}+(x^{2})^{2}\quad ,\quad \frac{\rho ^{2}}{r^{2}+a^{2}}+%
\frac{(x^{3})^{2}}{r^{2}}=1%
\end{array}
\label{g22}
\end{equation}%
and the momentum density is 
\begin{equation}
\begin{array}{l}
\overrightarrow{s}=\tsum\limits_{3+8}\overrightarrow{E_{j}}\times 
\overrightarrow{B_{j}}=\tsum\limits_{3+8}\frac{\gamma _{j}^{2}r^{2}}{16\pi
^{2}a\rho ^{2}(r^{4}+a^{2}(x^{3})^{2})}[x^{2}\overrightarrow{e_{1}}-x^{1}%
\overrightarrow{e_{2}}] \\ 
\end{array}
\label{g23}
\end{equation}%
The singularities of the magneto-gluonic dressing of quark is apparent in
the corresponding energy densities and the circularly rotating Poynting
vector.

\subsection{Confinement problem}

The well-known confinement mechanism\cite{Fels} based on the surface
singularity of a locally defined potential of the Dirac magnetic formalism
does not apply here, because the quark defining PDE (\ref{g2}) (case (-))
assures that the magneto-gluonic charge vanishes.

From (\ref{g11}) we see that in the oblate coordinates the electro-gluonic
part $E_{j}^{(g)}=\frac{\gamma _{j}}{4\pi (r^{2}+a^{2})}dr$ of the gluonic
field is rather conventional but its magneto-gluonic part $B_{j}^{(g)}=\frac{%
-\gamma _{j}}{4\pi a}dr\wedge d\varphi $ is a constant 2-form in the
3-dimensional space. The oblate coordinates are singular, therefore it is
convenient to use the vector fields (\ref{g16}) and the corresponding
electro-gluonic and magneto-gluonic potentials (\ref{g18}) where $%
\overrightarrow{A^{(g)}}_{j}$ is defined up to a gauge transformation.
Notice that this line singularity is present in the magneto-gluonic field
too. It is not a singularity only on a non-globally defined magnetic
potential like in the Dirac magnetic monopole.

Such a singularity appears in the magneto-gluonic flux through a surface $S$
bounded by the closed loop $\Gamma $ which has the form 
\begin{equation}
\begin{array}{l}
\phi _{j}=\tint\limits_{S}\overrightarrow{B^{(g)}}_{j}\cdot \overrightarrow{n%
}dS=\tint\limits_{S}(\overrightarrow{\nabla }\times \overrightarrow{A^{(g)}}%
_{j})\cdot \overrightarrow{n}dS=\toint\limits_{\Gamma }\overrightarrow{%
A^{(g)}}_{j}\cdot d\overrightarrow{r} \\ 
\end{array}
\label{g24}
\end{equation}%
In the cylindrical coordinates the above magneto-gluonic flux can be
computed at a circle of radius $\rho =\sqrt{((x^{1})^{2}+(x^{2})^{2})}$at a
point $x^{3}=z$. We find 
\begin{equation}
\begin{array}{l}
x^{1}=\rho \cos \psi \quad ,\quad x^{2}=\rho \sin \psi \\ 
dx^{1}=-\rho \sin \psi d\psi \quad ,\quad dx^{2}=\rho \cos \psi d\psi \\ 
\\ 
\phi _{j}=\frac{\gamma _{j}r_{0}}{2a} \\ 
r_{0}:=\pm \left\{ \frac{\rho ^{2}+z^{2}-a^{2}}{2}+\sqrt{[\frac{\rho
^{2}+z^{2}-a^{2}}{2}]^{2}+a^{2}z^{2}}\right\} ^{\frac{1}{2}}%
\end{array}
\label{g25}
\end{equation}%
Hence we precisely find 
\begin{equation}
\begin{array}{l}
at\quad \{z=0,\ \rho \leq a\}\quad \rightarrow \quad \phi _{j}=0 \\ 
\\ 
at\quad \{z=0,\ \rho >a\}\quad \rightarrow \quad \phi _{j}=\frac{\gamma _{j}%
}{2a}\sqrt{(\rho ^{2}-a^{2})} \\ 
\\ 
at\quad \{z\neq 0\}\quad \rightarrow \quad \phi _{j}=\frac{\gamma _{j}r_{0}}{%
2a}%
\end{array}
\label{g26}
\end{equation}%
But at a closed surface the magneto-gluonic flux must vanish, because $%
\overrightarrow{\nabla }\cdot \overrightarrow{B^{(g)}}_{j}=0$ i.e. no
compact (distributional) magneto-gluonic source exists. We interpret it that
the line singularity enters and leaves the closed surface without internal
source or sink. There is an additional singularity implied by the jump of
the angle $\psi =\arctan \frac{x^{2}}{x^{1}}$ at the ends of the $[-\pi ,\pi
)$ interval. This means that the singularity is the entire $x-z$ plane,
without magneto-gluonic charge.

In (\ref{g21}) we checked that the potentials define proper Schwartz
distributions. Hence the line and surface singularities are "compatible"
with the Schwartz distributions in $%
\mathbb{R}
^{3}$. They do not affect it because the singular surfaces in $%
\mathbb{R}
^{3}$, have dimension less than three. But they cannot be physically
acceptable. Infinite singular straight lines and surfaces cannot traverse
the universe, but curved line singularities (vortices) in a bounded domain
may exist\cite{Fels}.

In the present case the extended singularities are removed by simply
considering that the hadronic Schwartz distributions have compact support.
Apparently this implies confinement (a MIT bag-like picture), because the
gluonic fields and their sources are restricted into convex 3-dimensional
bounded subsets of $%
\mathbb{R}
^{3}$. Because of (\ref{i5}), it is a subset of tempered distributions,
opening up the possibility to study hadrons in the context of BEGS procedure.

In the case of the "quantum" approach (the BEGS procedure) the computations
seem to be feasible. We start by including to the asymptotic rigged Hilbert
space the MIT bag bound states in dynamical spheroids. The relation (Lemma
1.4.3 in Hormander's book\cite{Horm}) 
\begin{equation}
\begin{array}{l}
B\subset K\subset B\sqrt{3} \\ 
\end{array}
\label{g27}
\end{equation}%
of the compact support $K$\ between two spheroids provides a good
approximation for the hadron. On the other hand from a Payley-Wienner
theorem (7.3.1 in Hormander's book\cite{Horm}) we know that the
Fourier-Laplace transform of a distribution with compact support is an
entire holomorphic function of the exponential type, which permits the
application of the dispersion relations. The study of hydrogen-like atoms in
the rigged Hilbert space of the tempered distributions is a very effective
method\cite{BoShir}. Hence the above two ways of the incorporation of the
compactness of the hadronic Schwartz distributions into the BEGS procedure
seem to provide the feasible computations in the multihadron interactions.

\subsection{The gluonic dressing of the "natural U(2)" LCR-structure}

We consider the "natural $U(2)$" LCR-manifold as the compact "vacuum"
universe, because it is compatible with the Minkowski metric and the
corresponding U(2)-Cartan curvature (the electroweak field strength)
vanishes. In this subsection I will show that the gluonic field on the
"natural $U(2)$" LCR-manifold does not have any source, which I interpret
that it does not correspond to any hadronic particle. These calculations are
more convenient to be done in the following local coordinates ($z^{\alpha
},z^{\widetilde{\beta }}$) of "natural $U(2)$" LCR-structure which has the
structure coordinates ($w^{\alpha },w^{\widetilde{\beta }}$) satisfying the
structure conditions (\ref{e18}). The precise relations are 
\begin{equation}
\begin{array}{l}
w^{0}=:e^{i\frac{z^{0}}{4l}}=\cos \frac{\theta }{2}e^{i\frac{t-r^{\prime }}{%
4l}}\quad ,\quad \frac{w^{1}}{w^{0}}=:z^{1}=e^{i\varphi }\tan \frac{\theta }{%
2} \\ 
\\ 
w^{\widetilde{0}}=:e^{i\frac{z^{\widetilde{0}}}{4l}}=\cos \frac{\theta }{2}%
e^{i\frac{t+r^{\prime }}{4l}}\quad ,\quad \frac{w^{\widetilde{1}}}{w^{%
\widetilde{0}}}=:z^{\widetilde{1}}=e^{-i\varphi }\tan \frac{\theta }{2}%
\end{array}
\label{g28}
\end{equation}%
which imply%
\begin{equation}
\begin{array}{l}
z^{0}=t-r^{\prime }-4il\ln (\cos \frac{\theta }{2})\quad ,\quad
z^{1}=e^{i\varphi }\tan \frac{\theta }{2} \\ 
\\ 
z^{\widetilde{0}}=t+r^{\prime }-4il\ln (\cos \frac{\theta }{2})\quad ,\quad
z^{\widetilde{1}}=e^{-i\varphi }\tan \frac{\theta }{2}%
\end{array}
\label{g29}
\end{equation}%
The variable $\frac{t+r^{\prime }}{4l}$ is the $U(1)$ parameter $\tau $ and $%
\frac{r^{\prime }}{4l}$, $\theta $, $\varphi $ are related with the Euler
angles of $SU(2)$. We finally find the following LCR-tetrad%
\begin{equation}
\begin{array}{l}
\ell _{\mu }dx^{\mu }=\Lambda \lbrack dt-dr^{\prime }+4l\sin ^{2}\frac{%
\theta }{2}\ d\varphi ] \\ 
\\ 
n_{\mu }dx^{\mu }=N[dt+dr^{\prime }+4l\sin ^{2}\frac{\theta }{2}d\varphi ]
\\ 
\\ 
m_{\mu }dx^{\mu }=M[d\theta +i\sin \theta d\varphi ]%
\end{array}
\label{g30}
\end{equation}%
up to a tetrad-Weyl transformation. This tetrad is compatible with the Taub
metric\cite{Flahe}, therefore I call ($z^{\alpha },z^{\widetilde{\beta }}$)
"Taub" structure coordinates.

The corresponding closed self-dual null 2-forms, which are invariant under $%
t $ and $\varphi $ translations, have the form 
\begin{equation}
\begin{array}{l}
G^{\prime +}=\frac{C^{\prime }}{z^{1}}dz^{0}\wedge dz^{1}=\frac{C^{\prime }}{%
\sin \theta }\ell \wedge m \\ 
\\ 
\widetilde{G^{\prime }}^{+}=\frac{\widetilde{C^{\prime }}}{z^{\widetilde{1}}}%
dz^{\widetilde{0}}\wedge dz^{\widetilde{1}}=\frac{\widetilde{C^{\prime }}}{%
\sin \theta }n\wedge \overline{m}%
\end{array}
\label{g31}
\end{equation}%
and the static self-dual 2-form is

\begin{equation}
\begin{array}{l}
F_{j}-i\ast F_{j}:=-\frac{2C_{j}^{\prime }}{\sin \theta }\ell \wedge m-\frac{%
2C_{j}^{\prime \prime }}{\sin \theta }n\wedge \overline{m}= \\ 
\\ 
\qquad =-\frac{2(C_{j}^{\prime }+C_{j}^{\prime \prime })}{\sin \theta }%
(dt\wedge d\theta -i\sin \theta dr^{\prime }\wedge d\varphi +2l\cos \theta
d\theta \wedge d\varphi )- \\ 
\qquad -\frac{2(C_{j}^{\prime }-C_{j}^{\prime \prime })}{\sin \theta }(i\sin
\theta dt\wedge d\varphi -dr^{\prime }\wedge d\theta )%
\end{array}
\label{g32}
\end{equation}%
where the tetrad-Weyl factors are ignored. The source could come from the $%
d\theta \wedge d\varphi $ term. Therefore we assume $C_{j}^{\prime
}-C_{j}^{\prime \prime }=0$. But even then, if we integrate over the surface 
$t=const$ and $r^{\prime }=const$, we find zero too, because

\begin{equation}
\begin{array}{l}
\tiint (F_{j}-i\ast F_{j})=-4l(C_{j}^{\prime }+C_{j}^{\prime \prime })\tiint 
\frac{\cos \theta }{\sin \theta }d\theta \wedge d\varphi = \\ 
\qquad =-8\pi l(C_{j}^{\prime }+C_{j}^{\prime \prime })\underset{\varepsilon
\rightarrow 0}{\lim }\tint\limits_{\varepsilon }^{\pi -\varepsilon }d\ln
|\sin \theta |=0 \\ 
\end{array}
\label{g33}
\end{equation}%
Hence the "natural $U(2)$" LCR-manifold does not have any distributional
compact source, and no further restriction can be found on $(C_{j}^{\prime
}+C_{j}^{\prime \prime })$. This means that the electroweak vacuum does not
have a gluonic source.

\section{STANDARD\ MODEL\ DERIVATION}

\setcounter{equation}{0}

All the previous sections were devoted to the geometric consequences of the
fundamental LCR-structure on the tangent space of the spacetime. In the
present section I will describe how conventional "quantum field theory"
emerges as a mathematical targeted harmonic analysis through the BEGS
procedure. Electron and its neutrino are stable solitonic LCR-manifolds
which must be the time asymptotic free structures of any other LCR-surface
of $%
\mathbb{C}
^{4}$. This background geometric solitonic decay phenomenon is essentially
described by the Dyson expansion of the S-matrix into asymptotic free fields
identified with the elementary particles viewed as Poincar\'{e}
representations. In the context of BEGS formalism the Poincar\'{e} group and
Bogoliubov causality are assumed and applied to the expansion of the
S-matrix into representations of the Poincar\'{e} group. This procedure is
usually called "causal approach" which is essentially equivalent to the
ordinary QFT. The great advantage of the BEGS formalism is that it fully
"respects" the mathematical properties of the Schwartz distributions and the
order by order built up of the S-matrix expansion starting from the symmetry
properties of the free fields and using a BRS procedure for the elimination
of their unphysical modes as briefly described in the introduction. But the
Poincar\'{e} group and the elementary particles are axiomatically assumed
like in ordinary QFT. PCFT geometrically determines the Poincar\'{e} group
and all the "elementary particles" which appear in the SM, making the BEGS
procedure a kind of targeted harmonic analysis in the Hilbert-Fock space of
Poincar\'{e} group representations (precise free fields).

The convenient way to relate PCFT with the BEGS procedure is to consider the
representations of the background $SU(2,2)$ group in the context of the
generalized functions, which are necessary to be used in order to pass from
the basis of its maximal compact $S(U(2)\times U(2))$ subgroup to its affine
(non-compact) Poincar\'{e} subgroup. This is the subject of the first
subsection. We first start with the simple example of the analogous $SU(1,1)$
case and continue with the relation between the $S(U(2)\times U(2))$ Lie
algebra and the Poincar\'{e} Lie algebra. In the second subsection I will
describe how the results of order by order expansion of the "quantum" BEGS
procedure reconcile with the "classical" potentials (dressings) of the
"electron" LCR-structure. Besides I will point out how the BEGS procedure
provides the Einstein equations with a cosmological constant implying the
origin of dark energy already appeared through (\ref{e24}) in the geometric
level of PCFT. In the third subsection we reconciliate the disc singularity
of the dressings of the "free electron" LCR-structure with the point
singularity of "classical electron" potential implied by the BEGS procedure.

The assumption of the LCR-structure as the fundamental dynamical geometric
principle, contains all the observed particles as solitonic configurations.
It is clear that the identification criterion should be the distributional
singularities. Let me describe the case of every particle.

\textbf{Electron}: The free electron is the unique static LCR-structure with
the disc singularity, moving with a constant velocity. It is represented
with a massive Dirac free field, because of its fermionic gyromagnetic
ratio. Notice that the free field representation of the corresponding rigged
Hilbert space \emph{does not} contain all the surface properties. Even its
linear complex trajectory (recall that the free electron LCR-structure is a
ruled surface) and its disc singularity do not appear in the corresponding
free field rigged Hilbert space, on which QFT is built. Its distributional
singularity is the wavefront singularity implied by its free field
representation, like all the other waves.

\textbf{Neutrino}: The neutrino is the developable surface (consisting by
two $CP(3)$ planes) corresponding to the electron ruled surface. The one
plane has a complex trajectory while the other coincides with the trivial
one of the light-cone LCR-structure. It is represented as the left-hand part
of a massless Dirac field. Like in the electron case the simple Poincar\'{e}
representation (free field) does not contain the geometric properties. I
mention these points in order to point out that QFT is only a restricted
view of the geometry.

\textbf{Scaling is broken:} The massive static "electron" LCR-structure is
automorphic relative to the z-rotation and time-translation. The stationary
"neutrino" LCR-structure is automorphic relative to the z-rotation and the
light-cone translations. But \emph{no regular} solitonic LCR-structure
exists which is automorphic relative to an additional dilation
transformation. The Cartan subalgebra of the conformal $SU(2,2)$ group \emph{%
is not realized} in the solitonic sectors.

\textbf{Graviton}: The general LCR-structure defines a class of Einstein
metrics and the corresponding self-dual 2-forms. The Minkowski metric is
compatible to the "flat" LCR-structures of the Shilov boundary of the $%
SU(2,2)$ classical domain. Linearized deformations of this classical domain
define a symmetric tensor $h_{\mu \nu }$. In the case of the electron and
the neutrino LCR-structures it satisfies the wave equation with source their
singularities and the additional wavefront singularity of the free wave
equation. Recall that the distributional solutions inherit\cite{Horm} the
singularities of their source and the principal symbol of the PDE. This last
distributional singularity is identified with the spin-2 representation of
the (free) graviton.

\textbf{Electroweak gauge fields}: After fixing the tetrad-Weyl symmetry by
imposing the conditions $\Phi _{1}=1=-\Phi _{2}$ and $\Phi =|\Phi _{3}|$ of
the relative invariants we showed that the hermitian matrix (\ref{ew4})
coincides with a tetrad of the corresponding class of LCR-tetrads. It is the
SM gauge field with $\ell ^{\prime }$ the electromagnetic potential which
coincides with the electromagnetic field of the Kerr-Newman manifold. The
potentials\ describe 1-spin free fields (representations of the Poincar\'{e}
group) identified with the corresponding wavefront singularities.

\textbf{Higgs particle}: The remaining relative invariant component $\Phi
=|\Phi _{3}|$ is the remnant potential which plays the role of the Higgs
field.

\textbf{Gluons and quarks}: The gluonic field appears in PCFT through the
unique special PDE which is invariant under the tetrad-Weyl symmetry of the
fundamental geometric LCR-structure. No sources are permitted to be included
in the initial special PDE. For every leptonic generation a corresponding
quark generation emerges as sources in explicit distributional solitonic
solutions of the special symmetric PDE.

You may have noticed that we did not say anything yet for the field
equations. We have not introduced neither the Einstein equations nor the $%
U(2)$ gauge field equations. Only the existing free fields were introduced.
The field equations are essentially implied\cite{Sch2} by the BRS algorithm
applied in the causal approach.

\subsection{From SU(2,2) down to Poincar\'{e} rigged Hilbert space}

Lindblad and Nagel\cite{LiNa} studied the semisimple group $SU(1,1)$ and
showed that its compact subgroup $U(1)$ with the discrete basis determines
the space of test functions of the $SU(1,1)$ rigged Hilbert space and they
computed the eigenfunctions (Schwartz distributions) of its non-compact
generators. In its bounded realization%
\begin{equation}
\begin{array}{l}
E_{B}=%
\begin{pmatrix}
1 & 0 \\ 
0 & -1%
\end{pmatrix}%
\quad ,\quad Y=%
\begin{pmatrix}
Y^{0} \\ 
Y^{1}%
\end{pmatrix}%
=:%
\begin{pmatrix}
Y^{0} \\ 
wY^{0}%
\end{pmatrix}
\\ 
\\ 
Y^{\dag }E_{B}Y=\overline{Y^{0}}Y^{0}(1-\overline{w}w)=0%
\end{array}
\label{s1}
\end{equation}%
$SU(1,1)$ has the generators%
\begin{equation}
\begin{array}{l}
J_{0}=\frac{1}{2}%
\begin{pmatrix}
1 & 0 \\ 
0 & -1%
\end{pmatrix}%
\ ,\quad J_{1}=\frac{1}{2}%
\begin{pmatrix}
0 & 1 \\ 
-1 & 0%
\end{pmatrix}%
\ ,\quad J_{2}=\frac{-i}{2}%
\begin{pmatrix}
0 & 1 \\ 
1 & 0%
\end{pmatrix}
\\ 
\\ 
\lbrack J_{0},J_{\pm }]=\pm J_{\pm }\quad ,\quad \lbrack
J_{+},J_{-}]=-2J_{0}\quad ,\quad J_{\pm }:=J_{1}\pm iJ_{2} \\ 
C_{2}=J_{0}^{2}-J_{1}^{2}-J_{2}^{2}%
\end{array}
\label{s2}
\end{equation}%
where $C_{2}$ is the Casimir invariant. The eigenstates of $J_{0}$ and $%
C_{2} $ define the following standard basis%
\begin{equation}
\begin{array}{l}
<j,m|j,m^{\prime }>=\delta _{mm^{\prime }} \\ 
\\ 
C_{2}|j,m>=j(j+1)|j,m>\quad ,\quad J_{0}|j,m>=m|j,m>%
\end{array}
\label{s3}
\end{equation}

In the unbounded realization%
\begin{equation}
\begin{array}{l}
E_{U}=\frac{1}{2}%
\begin{pmatrix}
1 & 1 \\ 
1 & -1%
\end{pmatrix}%
\begin{pmatrix}
1 & 0 \\ 
0 & -1%
\end{pmatrix}%
\begin{pmatrix}
1 & 1 \\ 
1 & -1%
\end{pmatrix}%
=%
\begin{pmatrix}
0 & 1 \\ 
1 & 0%
\end{pmatrix}
\\ 
\\ 
X=\left( 
\begin{array}{c}
X^{0} \\ 
-irX^{0}%
\end{array}%
\right) =\frac{1}{\sqrt{2}}%
\begin{pmatrix}
1 & 1 \\ 
1 & -1%
\end{pmatrix}%
\begin{pmatrix}
Y^{0} \\ 
wY^{0}%
\end{pmatrix}
\\ 
\\ 
X^{\dag }E_{U}X=2\overline{X^{0}}X^{0}\frac{(r-\overline{r})}{2i}=0%
\end{array}
\label{s4}
\end{equation}%
we find the real line, which is the boundary of the upper half-plane and the
Cayley transformation between the projective coordinates is%
\begin{equation}
\begin{array}{l}
r=i(I-w)(I+w)^{-1}\quad \Longleftrightarrow \quad w=(iI-r)(iI+r)^{-1} \\ 
\end{array}
\label{s5}
\end{equation}%
In this unbounded realization the $SU(1,1)$ group is decomposed into the
translation, dilation and conformal subgroups with the corresponding
generators%
\begin{equation}
\begin{array}{l}
P=%
\begin{pmatrix}
0 & 0 \\ 
1 & 0%
\end{pmatrix}%
\quad ,\quad D=\frac{i}{2}%
\begin{pmatrix}
1 & 0 \\ 
0 & -1%
\end{pmatrix}%
\quad ,\quad K=%
\begin{pmatrix}
0 & 1 \\ 
0 & 0%
\end{pmatrix}
\\ 
\\ 
\lbrack D,P]=-iP\quad ,\quad \lbrack D,K]=iK\quad ,\quad \lbrack K,P]=-2iD%
\end{array}
\label{s6}
\end{equation}%
which are related with the $J_{i}$ generators with the relations

\begin{equation}
\begin{array}{l}
J_{0}=\frac{1}{2}(P^{\prime }+K^{\prime })\quad ,\quad J_{1}=\frac{1}{2}%
(P^{\prime }-K^{\prime })\quad ,\quad J_{2}=-D^{\prime } \\ 
\multicolumn{1}{c}{\Updownarrow} \\ 
P^{\prime }=J_{0}+J_{1}\quad ,\quad K^{\prime }=J_{0}-J_{1}\quad ,\quad
D^{\prime }=-J_{2}%
\end{array}
\label{s7}
\end{equation}%
where the accents indicate that the non-compact generators are first
transferred to the compact coordinates. The parabolic (energy) generator $%
P^{\prime }=J_{0}+J_{1}$ has a continuous spectrum. Its remarkable feature
is that the discrete principal series $D_{j}^{+}$ corresponds to the
positive real line spectrum while $D_{j}^{-}$ corresponds to the negative
real line spectrum, because of the existence of the external automorphism

\begin{equation}
\begin{array}{l}
(J_{0}^{\prime },J_{1}^{\prime },J_{2}^{\prime })=(-J_{0},-J_{1},J_{2}) \\ 
\\ 
(P^{\prime \prime },K^{\prime \prime },D^{\prime \prime })=(-P^{\prime
},-K^{\prime },D^{\prime })%
\end{array}
\label{s8}
\end{equation}%
which is the "temporal reflection" in $su(1,1)$ algebra and apparently
commutes with dilation. Recall that the $D_{j}^{+(-)}$ is the basis of
functions on the boundary which analytically extend in the interior
(exterior) of the disc.

The existence of a maximal compact subgroup $e^{-iJ_{0}\varphi }$ defines
the countable basis $|j,m>$ such that $J_{0}|j,m>=|j,m>$, which determines
the Hilbert space%
\begin{equation}
\begin{array}{l}
H=\{x=\tsum\limits_{m}a_{m}|j,m>\ ;\quad ||x||=\sqrt{\tsum%
\limits_{m}|a_{m}|^{2}}<\infty \} \\ 
\end{array}
\label{s9}
\end{equation}%
which is the central part of the Gelfand triplet $S\subset H\subset
S^{\prime }$. $S$ is the space of "rapidly decreasing sequences"%
\begin{equation}
\begin{array}{l}
S=\{a=\tsum\limits_{m}a_{m}|j,m>\ ;\quad \underset{|m|\rightarrow \infty }{%
\lim }m^{n}a_{m}=0\ ,\forall n\} \\ 
\\ 
p_{n}(a)=\sqrt{\tsum\limits_{m}(m^{2}+1)^{n}|a_{m}|^{2}}%
\end{array}
\label{s10}
\end{equation}%
where the second line defines the infinite set of norms, which provide the
strong topology of $S$. Notice that $p_{0}(a)$ is the norm of the Hilbert
space. The distributions $S^{\prime }$ is the space of "slowly increasing
sequences"%
\begin{equation}
\begin{array}{l}
S^{\prime }=\{a^{\prime }=\tsum\limits_{m}a_{m}^{\prime }|j,m>\ ;\quad
\exists N\in 
\mathbb{N}
\ :\underset{|m|\rightarrow \infty }{\lim }m^{-N}a_{m}=0\} \\ 
\end{array}
\label{s11}
\end{equation}

Starting from the $SU(1,1)$ classical domain, the above abstract method
becomes precise with the Celeghini et. al.\cite{CeGaOlm} method. The
boundary $w\overline{w}=1$ of bounded realization of the $SU(1,1)$ classical
domain provides the discrete Fourier transform with $|j,m>=\frac{1}{\sqrt{%
2\pi }}e^{im\varphi }$. In the unbounded realization, the parabolic energy
generator $P$ is diagonalized. Its eigenstates belong in the distribution
space $S^{\prime }$ and their expansion in the countable basis has been
computed (\cite{LiNa}).

Celeghini et. al.\cite{CeGaOlm} defined the rigged Hilbert space using the
Laguerre functions%
\begin{equation}
\begin{array}{l}
M_{n}(y):=e^{-y/2}L_{n}(y)\quad ,\quad n=0,1,2,... \\ 
y\in \lbrack 0,\infty )%
\end{array}
\label{s12}
\end{equation}%
where $L_{n}^{\alpha }(y)$ are the Laguerre polynomials. The fundamental
operators of the Laguerre basis are%
\begin{equation}
\begin{array}{l}
YM_{n}(y):=yM_{n}(y),\quad (YD_{y})M_{n}(y):=y\frac{d}{dy}M_{n}(y),\quad
NM_{n}(y):=nM_{n}(y) \\ 
\end{array}
\label{s13}
\end{equation}%
with%
\begin{equation}
\begin{array}{l}
K_{\pm }=\pm (YD_{y})+N+I-\frac{Y}{2}\quad ,\quad K_{3}=N+\frac{1}{2}I \\ 
\\ 
K_{\pm }M_{n}(y)=\sqrt{(n+\frac{1}{2}\pm \frac{1}{2})(n+\frac{1}{2}\pm \frac{%
1}{2})}M_{n+1}(y) \\ 
K_{3}M_{n}(y)=(n+\frac{1}{2})M_{n}(y) \\ 
\\ 
\lbrack K_{3},K_{\pm }]=\pm K_{\pm }\quad ,\quad \lbrack K_{+},K_{-}]=-2K_{3}
\\ 
C_{2}=K_{3}^{2}-\frac{1}{2}\{K_{+},K_{-}\}=-\frac{1}{4}I%
\end{array}
\label{s14}
\end{equation}%
and we find the unbounded operators%
\begin{equation}
\begin{array}{l}
Y=-(K_{+}+K_{-})+2K_{3} \\ 
\\ 
(YD_{y})=\frac{1}{2}(K_{+}-K_{-})%
\end{array}
\label{s15}
\end{equation}%
The relation of the compact support test functions with the precise
continuous basis index $y$ is 
\begin{equation}
\begin{array}{l}
|n>\ =\tint\limits_{0}^{\infty }dyM_{n}(y)|y>\quad \Leftrightarrow \quad
|y>\ =\tsum\limits_{n=0}^{\infty }M_{n}(y)|n> \\ 
\end{array}
\label{s16}
\end{equation}%
The eigenstate state $|y>$ is a distribution because it is a "slowly
increasing sequence"%
\begin{equation}
\begin{array}{l}
|y>\ =\tsum\limits_{n=0}^{\infty }M_{n}(y)|n>\ ;\quad \exists N\in 
\mathbb{N}
\ :\underset{n\rightarrow \infty }{\ \lim }n^{-N}M_{n}(y)=0\} \\ 
\end{array}
\label{s17}
\end{equation}

In the present case of $SU(2,2)$ group, its maximal compact subgroup is $%
S(U(2)\times U(2))$. The 15 $SU(2,2)$ generators are%
\begin{equation}
\begin{array}{l}
H_{0}=\frac{1}{2}%
\begin{pmatrix}
I & 0 \\ 
0 & -I%
\end{pmatrix}%
\quad ,\quad H_{j}=\frac{1}{2}%
\begin{pmatrix}
\sigma ^{j} & 0 \\ 
0 & 0%
\end{pmatrix}%
\quad ,\quad H_{3+j}=\frac{1}{2}%
\begin{pmatrix}
0 & 0 \\ 
0 & \sigma ^{j}%
\end{pmatrix}
\\ 
H_{7}=\frac{1}{2}%
\begin{pmatrix}
0 & I \\ 
-I & 0%
\end{pmatrix}%
\quad ,\quad H_{8}=\frac{1}{2}%
\begin{pmatrix}
0 & iI \\ 
iI & 0%
\end{pmatrix}
\\ 
H_{8+j}=\frac{1}{2}%
\begin{pmatrix}
0 & \sigma ^{j} \\ 
-\sigma ^{j} & 0%
\end{pmatrix}%
\quad ,\quad H_{11+j}=\frac{1}{2}%
\begin{pmatrix}
0 & i\sigma ^{j} \\ 
i\sigma ^{j} & 0%
\end{pmatrix}%
\end{array}
\label{s18}
\end{equation}%
in the bounded realization. $\sigma ^{j}$ are the Pauli matrices.

In the unbounded realization the conventional generators are%
\begin{equation}
\begin{array}{l}
P^{\mu }=%
\begin{pmatrix}
0 & 0 \\ 
\sigma ^{\mu } & 0%
\end{pmatrix}%
\quad ,\quad K^{\mu }=%
\begin{pmatrix}
0 & \sigma ^{\mu } \\ 
0 & 0%
\end{pmatrix}
\\ 
S^{j}=%
\begin{pmatrix}
\sigma ^{j} & 0 \\ 
0 & \sigma ^{j}%
\end{pmatrix}%
\quad ,\quad B^{j}=i%
\begin{pmatrix}
\sigma ^{j} & 0 \\ 
0 & -\sigma ^{j}%
\end{pmatrix}
\\ 
D=\frac{i}{2}%
\begin{pmatrix}
I & 0 \\ 
0 & -I%
\end{pmatrix}%
\end{array}
\label{s19}
\end{equation}%
where $S^{i}=\frac{1}{2}\epsilon _{ijk}M_{jk}$ and $B^{j}=M_{0j}$. The
relations between the generators of $SU(2,2)$ in their bounded and unbounded
realization are 
\begin{equation}
\begin{array}{l}
P^{\prime 0}=H_{0}+H_{7}\quad ,\quad P^{\prime j}=H_{j}-H_{3+j}+H_{8+j} \\ 
K^{\prime 0}=H_{0}-H_{7}\quad ,\quad K^{\prime j}=H_{j}-H_{3+j}-H_{8+j} \\ 
S^{\prime j}=H_{j}+H_{3+j}\quad ,\quad B^{\prime j}=H_{11+j}\quad ,\quad
D^{\prime }=H_{8}%
\end{array}
\label{s20}
\end{equation}

The generators of the maximal compact subgroup $S(U(2)\times U(2))$ are $%
H_{0}$, $H_{j}$ and $H_{3+j}$, which determine the test functions for the
rigged Hilbert space of $SU(2,2)$. But the non-existence of the automorphic
LCR-structure relative to z-rotation, time translation and dilation implies
that the $SU(2,2)$ symmetry is broken down to the Poincar\'{e} symmetry.
Hence the harmonic analysis must be restricted to the rigged Fock space of
the representations of the Poincar\'{e} group (wavefront singularities)
which appear to the well defined LCR-manifolds of the above list.

Using the Celeghini et. al.\cite{CeGaOlm} method, the orthogonal functions
are based on the Hermite polynomials 
\begin{equation}
\begin{array}{l}
\Psi _{n}(x):=\frac{e^{-x^{2}/2}}{\sqrt{2^{n}n!\sqrt{\pi }}}H_{n}(x)\ ,\quad
n=0,1,2,...\ ,\quad x\in (-\infty ,+\infty ) \\ 
\\ 
\tint\limits_{-\infty }^{+\infty }\Psi _{n}(x)\Psi _{m}(x)dx=\delta _{nm}\
,\quad \tsum\limits_{n=0}^{\infty }\Psi _{n}(x)\Psi _{n}(x^{\prime })=\delta
(x-x^{\prime })%
\end{array}
\label{s21}
\end{equation}%
We define the operators%
\begin{equation}
\begin{array}{l}
X\Psi _{n}(x):=x\Psi _{n}(x),\quad D_{x}\Psi _{n}(x):=\frac{d}{dx}\Psi
_{n}(x),\quad N\Psi _{n}(x):=n\Psi _{n}(x) \\ 
\\ 
a:=\frac{1}{\sqrt{2}}(X+D_{x})\quad ,\quad a^{\dag }=\frac{1}{\sqrt{2}}%
(X-D_{x}) \\ 
a\Psi _{n}(x)=\sqrt{n}\Psi _{n-1}(x)\quad ,\quad a^{\dag }\Psi _{n}(x)=\sqrt{%
n+1}\Psi _{n+1}(x)%
\end{array}
\label{s22}
\end{equation}%
It is well known that%
\begin{equation}
\begin{array}{l}
\lbrack a,a^{\dag }]=I\quad ,\quad \lbrack N,a^{\dag }]=a^{\dag }\quad
,\quad \lbrack N,a]=-a \\ 
\\ 
a\Psi _{n}(x)=\sqrt{n}\Psi _{n-1}(x)\quad ,\quad a^{\dag }\Psi _{n}(x)=\sqrt{%
n+1}\Psi _{n+1}(x)%
\end{array}
\label{s23}
\end{equation}%
is the Heisenberg algebra. The discrete $|n>$ and the continuous basis $|x>$
are related with the relation 
\begin{equation}
\begin{array}{l}
|n>\ =\tint\limits_{-\infty }^{\infty }dx\Psi _{n}(x)|x>\quad
\Leftrightarrow \quad |x>\ =\tsum\limits_{n=0}^{\infty }\Psi _{n}(x)|n> \\ 
\end{array}
\label{s24}
\end{equation}

In our case of the 3-dimensional space we have to start with the
3-dimensional Hermite function being the product of the three Hermite
functions (\ref{s21}). They span the vector space $S$ of rapidly decreasing
functions in $%
\mathbb{R}
^{3}$. It is dense in the Hilbert space of square integrable functions $%
L^{2}(%
\mathbb{C}
)$. The functionals on $S$ determine the generalized functions $S^{\prime }$
in the basis $|\overrightarrow{x}>$. For the following details I refer the
reader to the chapter 4 of the classical book of Bogoliubov, Logunov and
Todorov\cite{BoLoTo}.

\subsection{BEGS formulation of the standard model}

My claim that the geometry prevails and that the "quantum" SM S-matrix is a
targeted harmonic expansion into precise representations of the Poincar\'{e}
group has to clarify how the dressing potentials of an LCR-structure become
the radiation fields used in the BEGS procedure. We will recall the
formulation of the Lienard-Wiechert radiating potentials by an accelerating
charge, studying the LCR-structure based on a ruled surface with a general
(accelerating) complex trajectory (\ref{c75}). Then the homogeneous
coordinates are 
\begin{equation}
\begin{array}{l}
X^{nj}=%
\begin{pmatrix}
\lambda ^{A1} & \lambda ^{A2} \\ 
-i\xi _{a}(z^{0})\sigma _{A^{\prime }A}^{a}\lambda ^{A1} & -i\xi _{a}(z^{%
\widetilde{0}})\sigma _{A^{\prime }B}^{a}\lambda ^{B2}%
\end{pmatrix}
\\ 
\\ 
\lambda ^{Aj}=%
\begin{pmatrix}
1 & -z^{\widetilde{1}} \\ 
z^{1} & 1%
\end{pmatrix}%
\end{array}
\label{s25}
\end{equation}%
where $z^{\alpha },z^{\widetilde{\beta }}$ are two solutions of the
following equations 
\begin{equation}
\begin{array}{l}
\det [(r_{a}-\xi _{a}(\tau ))\sigma ^{a}]=0\quad ,\quad \tau _{1,2}=\xi
^{0}=r^{0}\mp \sqrt{(r^{i}-\xi ^{i}(\tau _{1,2}(r^{b})))^{2}} \\ 
\\ 
(r_{a}-\xi _{a}(z^{0}))\sigma _{A^{\prime }A}^{a}\lambda ^{A1}=0=(r_{a}-\xi
_{a}(z^{\widetilde{0}}))\sigma _{A^{\prime }A}^{a}\lambda ^{A2} \\ 
\end{array}
\label{s26}
\end{equation}%
Notice that if the two roots $\tau _{1,2}$ are different, the corresponding $%
\lambda ^{Aj}$ are also different and they provide the natural "retarded"
and "advanced" intersection points of the line $r^{a}$ with two sheets of
the algebraic surface of $CP(3)$. This becomes more clear in the case of
zero algebraic gravity ($\func{Im}r^{a}=0$) and $\xi ^{a}(\tau )=\xi
_{R}^{a}(\tau )+i\xi _{I}^{a}(\tau )$ with the vectors $\xi _{R}^{a}(\tau )$%
, $\xi _{I}^{a}(\tau )$ real analytic. Then up to 1st order approximation
relative to $\frac{1}{c}$ implies 
\begin{equation}
\begin{array}{l}
z^{\prime 0}\simeq t-\frac{1}{c}\sqrt{(x^{i}-\xi _{R}^{i}(t)-i\xi
_{I}^{i}(t))^{2}} \\ 
z^{\prime \widetilde{0}}\simeq t+\frac{1}{c}\sqrt{(x^{i}-\xi
_{R}^{i}(t)-i\xi _{I}^{i}(t))^{2}}%
\end{array}
\label{s27}
\end{equation}%
Apparently the trajectory (source) singularity of the LCR-structure occurs
at the space curve 
\begin{equation}
\begin{array}{l}
(x^{i}-\xi _{R}^{i}(t))^{2}=(\xi _{I}^{j}(t))^{2} \\ 
\\ 
\tsum\limits_{i=1}^{3}(x^{i}-\xi _{R}^{i}(t))\xi _{I}^{i}(t)=0%
\end{array}
\label{s28}
\end{equation}%
which is a moving disc with radius $\xi _{I}^{i}(t)$ around the real
trajectory $\xi _{R}^{i}(t)$, which may be viewed as an extension of ring
singularity of the "free electron" LCR-structure. I think it is clear that
these geometric calculations indicate the emergence of the radiation fields
but they turn out to be very difficult to be computed.

Detailed calculations for the derivation of the SM are presented in the
second book\cite{Sch2} of Scharf, using the nilpotent charge $Q$ (\ref{i11})
as described in the introduction. This procedure is valid if no anomalies
appear, which is the case in the $S(3)\times SU(2)\times U(1)$ SM. Recall
that the choice of the lagrangian of a field theory respected the rule not
to include higher order derivatives. Einstein proposed his field equations
for the metric respecting this rule taking up to second order derivatives
including the cosmological constant term. After the Pais-Uhlenbeck work\cite%
{Pais} it became clear that higher order derivatives may generate negative
norm (unphysical) modes. The BRS algorithm of Scharf and coworkers
essentially replaces the above ad hock assumption following the opposite
way. It starts from the free wave field and it eliminates order by order all
the interaction terms which could create unphysical modes. In the case of
the spin-2 field $h^{\mu \nu }$ considered as an operator valued
distribution describing the spin-2 Poincar\'{e} group representation through 
\begin{equation}
\begin{array}{l}
\lbrack h^{\alpha \beta }(x),h^{\mu \nu }(y)]=-ib^{\alpha \beta \mu \nu
}D(x-y) \\ 
\\ 
b^{\alpha \beta \mu \nu }:=\frac{1}{2}(\eta ^{\alpha \mu }\eta ^{\beta \nu
}+\eta ^{\alpha \nu }\eta ^{\beta \mu }-\eta ^{\alpha \beta }\eta ^{\mu \nu
})%
\end{array}
\label{s28a}
\end{equation}%
where $D(x)$ is the mass-zero Jordan-Pauli distribution. The nilpotent gauge
charge is 
\begin{equation}
\begin{array}{l}
Q:=\tint\limits_{x^{0}=t}d^{3}x(\partial _{\beta }h^{\alpha \beta })%
\overleftrightarrow{\partial _{0}}u_{\alpha } \\ 
\end{array}
\label{s28b}
\end{equation}%
and the ghosts fields $u^{\mu }$ and $\widetilde{u}^{\nu }$ satisfy the
anticommutation relation 
\begin{equation}
\begin{array}{l}
\{u^{\mu }(x),\widetilde{u}^{\nu }(y)\}=i\eta ^{\mu \nu }D(x-y) \\ 
\end{array}
\label{s28c}
\end{equation}%
The variations of the fundamental fields are 
\begin{equation}
\begin{array}{l}
d_{Q}h^{\mu \nu }=-\frac{i}{2}(\partial ^{\mu }u^{\nu }+\partial ^{\nu
}u^{\mu }-\eta ^{\mu \nu }\partial _{\alpha }u^{\alpha }) \\ 
\\ 
d_{Q}u^{\mu }=0\quad ,\quad d_{Q}\widetilde{u}^{\mu }(y)\}=i\partial _{\nu
}h^{\mu \nu }%
\end{array}
\label{s28d}
\end{equation}%
Applying the invariance (up to a total derivative) under the above
variations to the possible self-interactions of the spin-2 field and after
very long calculations the group of Scharf found that up to third order the
terms of the S-matrix coincide with those of the Hilbert-Einstein action
including a cosmological constant. The interested reader must study the
original works of the group or the books of Scharf for more details. The
first important point for the present case is that the compact U(2)
universe, indicated by the background geometry (\ref{e24}), is also found
from the BEGS harmonic analysis. The second important point is the
cancellation of the anomalies for the precise Weinberg-Salam model implied
by PCFT. An analogous BRS algorithm was used to prove that in the
2-dimensional PCFT (Polyakov action) (\ref{i1}) that the conformal aomaly
vanishes for the well known 26-dimension of the vector space of $X^{\mu }$.
Notice that in the present 4-dimensional PCFT the gluonic gauge field (\ref%
{g1}) corresponds to the $X^{\mu }$ field of the 2-dimensional PCFT. Hence
the physically observed $SU(3)$ unitary gluonic group may be considered as a
consequence of the cancellation of the axial anomaly.

I have already pointed out that the Poincar\'{e} group of PCFT is the proper
orthochronous group of special relativity (Minkowski space). The spatial and
temporal reflections are \emph{external automorphisms} of its Lie algebra,
i.e. they do not exponentiate into elements of the group like (\ref{s8}) in
the $SU(1,1)$ example. In PCFT there are two additional discrete external
transformations. The asymmetry between the left $X^{n1}$ and the right $%
X^{n1}$ homogeneous coordinates in the case of reducible Kerr polynomials
(related to the chirality) and the ($z^{\alpha },z^{\widetilde{\beta }}$)$%
\leftrightarrow $($\overline{z^{\alpha }},\overline{z^{\widetilde{\beta }}}$%
) transformation related to the particle-antiparticle transformation. All
these four discrete transformations are also permitted in the BEGS harmonic
analysis. In the following subsection I discuss the possible manifestation
of the electron ring singularity in the BEGS procedure.

\subsection{On self-consistency conditions}

The perturbative approach permits the definition of general dynamical
variables through the generating functional introduced considering the
formal existence of a "classical" current $J(x)$ for every field $\phi (x)$
of the action. The generating functional $Z_{0}(J)$ and the connected
generating functional $Z_{c}(J)$ are

\begin{equation}
\begin{array}{l}
Z_{0}(J)=<0|T[\exp \{i\tint (L_{I}(x)+\phi (x)J(x))d^{4}x\}]|0> \\ 
\\ 
Z_{c}(J)=-i\ln [Z_{0}(J)]%
\end{array}
\label{s29}
\end{equation}%
The general and the connected Green functions are defined taking (formal)
functional derivatives of the generating functionals. Through this formal
procedure, the symmetries of the action become Ward identities for the Green
functions. The anomalies appear as disagreements between the formal and the
exact (quantum) computations.

Any field $\phi (x)$ defines a generating field $\Phi (x;J)$\ and the
Legendre transformation

\begin{equation}
\begin{array}{l}
\Phi (x;J)=\frac{\delta Z_{c}(J)}{\delta J(x)} \\ 
\\ 
Z_{c}(J)\rightarrow W(\Phi )=Z_{c}(J)-\tint \Phi (x;J)J(x)d^{4}x%
\end{array}
\label{s30}
\end{equation}%
In the context of the Bogoliubov-Shirkov notation [\cite{BoShir}, Chap. VII]

\begin{equation}
\begin{array}{l}
\Phi (x;g)=-\frac{\delta H(x;g)}{\delta J(x)}=\frac{-i}{g(x)}(\frac{\delta S%
}{\delta J(x)}\overset{\ast }{S})|_{J=0} \\ 
\\ 
H(x;g):=i(\frac{\delta S(g)}{\delta g(x)}\overset{\ast }{S}(g))%
\end{array}
\label{s31}
\end{equation}%
where $H(x;g)$ is the "quantum" hamiltonian of the system. Hence the causal
approach formalism permits the computation of a classical potential with the
following formula

\begin{equation}
\begin{array}{l}
\phi (x;1)=-\frac{\delta E(J)}{\delta J(x)}|_{J=0}=\frac{-i}{<S>}\overset{%
\ast }{\Phi }_{1}(\frac{\delta S}{\delta J(x)}\overset{\ast }{S})\Phi
_{1}|_{J=0} \\ 
\\ 
\Phi _{1}=(2\pi )^{\frac{3}{2}}a_{\nu }^{+}(\overrightarrow{k})\Phi _{0}%
\end{array}
\label{s32}
\end{equation}%
where $\Phi _{1}$ is the one-electron state. Notice that the elementary
particle has the same initial and final energies and their creation and
annihilation operators are outside the time ordering. The physical intuition
is that we use the classical current $J(x)$ as a sensor of the potential
generated by a particle.

Hence the BEGS procedure permits the computation of the electromagnetic
dressing of the electron with the following first order term

\begin{equation}
\begin{array}{l}
A_{1\mu }(x;1)\simeq \frac{-i}{2}\overset{\ast }{\Phi }_{1}\frac{\delta 
\widehat{S}_{2}(J)}{\delta J^{\mu }(x)}\Phi _{1}|_{J^{\mu }=0} \\ 
\\ 
\widehat{S}_{2}(J)=\tint T((L_{I}(x_{1})+A_{\nu }(x_{1})J^{\nu
}(x_{1}))(L_{I}(x_{2})+A_{\nu }(x_{2})J^{\nu }(x_{2}))[dx]%
\end{array}
\label{s33}
\end{equation}%
which becomes

\begin{equation}
\begin{array}{l}
A_{1}^{\mu }(x)\simeq -e\tint D_{0}^{c}(x-y)\overset{\ast }{\Phi }%
_{1p^{\prime }}:\overline{\psi _{e}}(y)\gamma ^{\mu }\psi _{e}(y):\Phi
_{1p}d^{4}y \\ 
\\ 
\Phi _{1p}=(2\pi )^{\frac{3}{2}}\overset{\ast }{a_{\nu }^{+}}(%
\overrightarrow{p})\Phi _{0}%
\end{array}
\label{s34}
\end{equation}%
This potential is singular at the point $\overrightarrow{x}=0$, but the
electromagnetic dressing of the "electron" LCR-structure (\ref{l19a}) in
cartesian coordinates\ 
\begin{equation}
\begin{array}{l}
A=\frac{qr^{3}}{4\pi (r^{4}+a^{2}(x^{3})^{2})}(dx^{0}-\frac{rx^{1}-ax^{2}}{%
r^{2}+a^{2}}dx^{1}-\frac{rx^{2}+ax^{1}}{r^{2}+a^{2}}dx^{2}-\frac{x^{3}}{r}%
dx^{3}) \\ 
\\ 
dF=0\quad ,\quad d\ast F=-\ast j_{e}%
\end{array}
\label{s35}
\end{equation}%
is singular at the entire disc with radius $a$, which is a non-local
interaction. This incompatibility may be solved from the observation\cite%
{Cart} that the Kerr-Newman manifold attributed to the "electron"
LCR-structure has spin $ma$. Notice that the expansion of the
electromagnetic dressing relative to $a$, has first term (\ref{s34}). But
because of the relation $a=\frac{h}{2m}$, the higher order terms of (\ref%
{s35}) should emerge from the loop diagrams. It would be interesting to
check it for the electrogravity potentials (\ref{l19a}).

\section{DISCUSSION}

\setcounter{equation}{0}

The very interesting result of the present work is that by simply replacing
the Einstein metric $g_{\mu \nu }$ on the tangent space of spacetime with
the Frobenius integrable LCR-structure (about) all current phenomenology in
elementary particle physics emerges. The solitonic static quadratic
LCR-structure is the static electron with electroweak, "Higgs" and
gravitational potentials which generate wavefront singularities in the
accelerating case. Its corresponding neutrino is also computed. Quantum
field theory emerges by simply applying the BEGS procedure to the rigged
Hilbert-Fock space of the Poincar\'{e} representations of the above
wavefront singularities. Besides it seems that in PCFT the only static
(non-baryonic) particle is the electron LCR-structure. Hence there is no
place for the existence of weakly interacting massive particles (WIMPs),
which is compatible with current observations. It is apparent that geometry
prevails as expected by Einstein, implying that all the assumptions of the
standard model are provided by the geometry.

In the context of astrophysical observations the formulation of the
mathematical problem changes. Recall that algebraic gravity of matter
emerges from a deformation of the SU(2,2) classical domain with a
Kerr-Schild ansatz. In the present case the asymptotic space must be the
deSitter space. Recall that the tetrd-Weyl transformations and the
corresponding $\rho _{ij}^{\prime }=f_{ij}\rho _{ij}$ transformation show
that the LCR-structure does not uniquely determine neither the metric of
spacetime nor the metric of the ambient Kaehler manifold. But in the case of
the vacuum LCR-structure it is natural to assume the symmetric ambient
metric with curvature

\begin{equation}
\begin{array}{l}
\widehat{R}_{\alpha \beta \gamma \delta }=R_{0}(a_{\alpha \gamma }a_{\beta
\delta }-a_{\alpha \delta }a_{\beta \gamma }) \\ 
\end{array}%
\end{equation}%
which implies that the Gauss equation 
\begin{equation}
\begin{array}{l}
R_{ijkl}=\tsum\limits_{\sigma =5}^{8}e_{\sigma }(\Omega _{\sigma |ik}\Omega
_{\sigma |jl}-\Omega _{\sigma |il}\Omega _{\sigma |jk})+\widehat{R}_{\alpha
\beta \gamma \delta }y_{,i}^{\alpha }y_{,j}^{\beta }y_{,k}^{\gamma
}y_{,l}^{\delta } \\ 
\\ 
y_{,ij}^{\alpha }+\{_{\mu \nu }^{\alpha }\}_{a}y_{,i}^{\mu }y_{,j}^{\nu
}=:\tsum\limits_{\sigma =5}^{8}e_{\sigma }\Omega _{\sigma |ij}\xi _{\sigma
|}^{\alpha }\quad ,\quad a_{\alpha \beta }y_{,i}^{\alpha }\xi _{\sigma
|}^{\alpha }=0%
\end{array}%
\end{equation}%
gives the following asymptotic form of the spacetime curvature 
\begin{equation}
\begin{array}{l}
R_{ijkl}\simeq \tsum\limits_{\sigma =5}^{8}e_{\sigma }(\Omega _{\sigma
|ik}\Omega _{\sigma |jl}-\Omega _{\sigma |il}\Omega _{\sigma
|jk})+R_{0}(g_{ik}g_{jl}-g_{il}g_{jk}) \\ 
\end{array}%
\end{equation}%
where $g_{ik}$ is the first fundamental form (metric),$\ \Omega _{\sigma
|ik} $ are the four second fundamental forms and $\xi _{\sigma |}^{\alpha }$
are the four normal vectors to the spacetime being formally embedded in the
ambient complex manifold. The notation of the Eisenhart book\cite{Eisen} is
used.\ The emergence of the deSitter spacetime is apparent, but there is an
additional term implied by the second fundamental forms of the \emph{%
flatprint} LCR-manifold of matter (stars, galaxies, etc). Explicit use of
this formula into the astrophysical computations with the typical flatprint
embedding%
\begin{equation}
\begin{array}{l}
z^{0}=t-r+ia\cos \theta \quad ,\quad z^{1}=e^{i\varphi }\tan \frac{\theta }{2%
} \\ 
\\ 
z^{\widetilde{0}}=t+r-ia\cos \theta \quad ,\quad z^{\widetilde{1}}=\frac{r+ia%
}{r-ia}e^{-i\varphi }\tan \frac{\theta }{2}%
\end{array}%
\end{equation}
of a rotating star or galaxy could be a test of PCFT.

\emph{Acknowledgements}: I am grateful to P. Grang\'{e} for very useful
discussions on the BEGS procedure, to S. Papadopoulos for discussions on
hadronic scattering and to T. Papakostas for discussions\ on general
relativity.

\newpage

(*) IEP, Ministry of Education, Tsoha 36, Athens, 11521, Greece (email:
ragiadak@gmail.com)

\end{document}